\documentclass[a4paper,notitlepage,11pt]{article} 
\usepackage{amsfonts}
\usepackage{amssymb}
\usepackage{amsmath}
\usepackage{bm}
\usepackage{graphicx}
\usepackage{multirow}
\usepackage{color}
\usepackage{graphics}
\usepackage{cite}
\usepackage{enumitem}
\usepackage{caption}

\def\one{{\hbox{1\kern-.8mm l}}}

\newcommand{\beq}{\begin{equation}}
\newcommand{\eeq}{\end{equation}}
\def\be#1\ee{\begin{align}#1\end{align}}
\newcommand{\ov } {\over }

\def\vk{{\vec k}}
\def\ein{{\eta_{\text{in}}}}
\newcommand{\ri}[1]{\mbox{\text{{\tiny{\rm #1}}}}}
\newcommand{\rii}[2]{\mbox{\text{{\tiny{\rm #1}}}\hspace{-1.0pt}\text{{\tiny{\rm #2}}}}}
\newcommand{\riii}[3]{\mbox{\text{{\tiny{\rm #1}}}\hspace{-1.0pt}\text{{\tiny{\rm #2}}}\hspace{-1.0pt}\text{{\tiny{\rm #3}}}}}
\def\stiny{\fontsize{4pt}{4pt}\selectfont}
\def\sstiny{\fontsize{2pt}{2pt}\selectfont}
\begin{document}
\begin{titlepage}
\bigskip
\begin{flushright}
\end{flushright}
\begin{center}
\vskip 2cm
\Large{Signatures of very high energy physics in the squeezed limit of the
  bispectrum 
  }
\end{center}
\vskip 1cm
\begin{center}
\large{Diego Chialva \\ 
      {\it Universit\'e de Mons, Service de M\'ecanique et gravitation, Place
       du Parc 20, 7000 Mons, Belgium} \\
        \tt{diego.chialva@umons.ac.be}}

\end{center}
\date{}

\pagestyle{plain}

\begin{abstract}

We investigate the signatures in the squeezed
limit of the primordial scalar bispectrum due to
modifications of the standard theory at high energy. In particular, we
consider the cases of modified dispersion relations and/or modified initial
quantum state (both in the Boundary Effective Field Theory and in the
New Physics Hyper-Surface formulations)
. Using the in-in formalism we study in details
the squeezed limit of the contributions to the bispectrum from all
possible cubic couplings in the effective 
theory of single-field inflation.
We find general features such as enhancements and/or non-local 
shape of the non-Gaussianities, which are relevant, for example, for
measurements of the halo bias and which distinguish these scenarios from the
standard one (with Bunch-Davies vacuum as initial state and standard
kinetic terms).
We find that the signatures change according to the 
magnitude of the scale of new physics, and therefore several
pieces of information regarding high energy physics could be obtained
in case of detection of these signals, especially bounds on the scales
of new physics.  

\end{abstract}

\end{titlepage}

\tableofcontents

\setcounter{section}{0}

\section{Introduction}

The non-Gaussian features of the primordial cosmological perturbations are going to
become one of the most important experimental and theoretical tools for
investigating the early history of the Universe
\cite{Komatsu:2009kd}. The 
scalar bispectrum, coming from the three-point correlator of the
comoving curvature perturbation $\zeta$, is the leading contribution
to the scalar non-Gaussianities in the perturbative theory. 

In recent times, several reasons of interests  have emerged for investigating     
its so-called {\em squeezed limit}, when 
one of the three momenta it depends on in Fourier space is much smaller than
the other two: $k_1 \ll k_S, k_{2, 3} \simeq k_S$.
In fact, large-scale probes of the primordial 
non-Gaussianities like the halo bias are expected to be soon competing
with or even surpass in accuracy those of the CMBR
\cite{Desjacques:2010jw, Dalal:2007cu, Matarrese:2008nc, Slosar:2008hx,
  Shandera:2010ei, Desjacques:2009jb, Wagner:2010me, Smith:2010gx},
and this 
prompts for a better understanding 
of their sensitivity on the details of the bispectrum and 
the theoretical models. 

Furthermore, studying the squeezed limit provides powerful results capable of
distinguishing (hopefully falsify) classes of models. This is
particularly true for single-field models (or more generally, for
models where there are only adiabatic perturbations). All
single-field models (adiabatic perturbations) with 
Bunch-Davies vacuum and standard kinetic terms (we call this the
{\em standard scenario} from now on) predict
the so-called {\em local} form of the bispectrum and a
very low level of non-Gaussianities in the limit
\cite{Maldacena:2002vr, Creminelli:2004yq, Creminelli:2011rh}. 
Remarkably, measurements of the halo bias
could discriminate cases with different underlying assumptions
\cite{Verde:2009hy, Schmidt:2010gw}: for example models with 
modified kinetic terms whose bispectrum template is maximized on {\em
  equilateral} configurations, or with initial state different than the
Bunch-Davies, in which case the proposed template is maximized on {\em folded}
configurations \cite{Meerburg:2009ys}.

In this paper, we want to look at the field theory cosmological predictions
for the squeezed limit of the bispectrum for single-field slow-roll inflationary
models, investigating the effects of 
two very simple modifications of the standard scenario:
\vspace{-0.12cm}
\begin{itemize}[leftmargin=0.8cm,itemsep=0.12cm,parsep=0.0cm]
\setlength{\parindent}{0.2cm}
\item[1)] the presence of higher derivative  
corrections to the action that are unsuppressed at a certain physical
scale $\Lambda$ 
and modify the dispersion relations of the perturbation fields
\cite{Martin:2000xs, Martin:2002kt, Martin:2003kp,
  Lemoine:2001ar}\footnote{As in \cite{Chialva:2011iz},
our scenario differs from that considered in \cite{Seery:2005wm,
  Chen:2006nt}, where the only 
modification to the field equations was 
a change in the speed of sound (the modifications at the level of the
Lagrangian depended only on first order 
derivatives of the fields). In particular, we include also derivatives
of higher order, and the effect on the field 
equations is more profound. In many scenarios
with time-varying speed of 
sound the Maldacena's condition \cite{Maldacena:2002vr,
  Creminelli:2004yq} on the squeezed limit of the bispectrum 
holds, see \cite{Creminelli:2011rh, RenauxPetel:2010ty, 
  Burrage:2011hd, Khoury:2008wj}. Similarly, see \cite{Creminelli:2011rh}, in the
adiabatic ekpyrosis scenario of \cite{Khoury:2009my,
  Khoury:2011ii}. Some contrived model (non-attractor background)
\cite{Baumann:2011dt} do 
not satisfy the relation \cite{Creminelli:2011rh}.},  
\item[2)] the choice of an initial state other than the Bunch-Davies
vacuum\footnote{The halo bias and the squeezed limit for this case
  have been studied in the literature mostly by
  using the {\em template} of \cite{Meerburg:2009ys} and not the
  actual field theoretical results for the bispectrum.
  We will show that this appreciably changes the
  outcomes (this was noted also in \cite{Wagner:2011wx}, which
  however still employed the template and did not study in details the field
  theoretical results). After the appearance of this article (August 2011),
  we were signalled reference \cite{Agullo:2010ws}, which also makes remarks, 
  focusing on the presence of enhancements,
  about the squeezed limit in a 
  particular formalism (density matrix approach) with a
  specific initial condition for the fields (akin to the one called BEFT in
  section \ref{revmodinvac}, which explicitly breaks scale
  invariance) and dealing  
  only with one specific cubic coupling (the interaction
  (\ref{HIcubic})). See also \cite{Ganc:2011dy} 
  which analyses the same scenario.
  Also in these articles the study of the general field
  theoretical results was not undertaken and part of the signatures
  were missed (see our sections \ref{bispsqueezlimsec} and
  \ref{signaturessec}).}, reflecting the effects of different
physics at higher  
energies/earlier times as proposed in \cite{Danielsson:2002kx,
Danielsson:2002qh,
Easther:2002xe, Schalm:2004qk,
Schalm:2004xg,
Nitti:2005ym} (we consider both the BEFT and the NPHS scenarios, see
section \ref{revmodinvac} for a brief review). 
\end{itemize}
\vspace{-0.12cm}
There are in fact several reasons, see section \ref{generalarguments}, to
expect that in such scenarios the  
standard result for the squeezed limit of the bispectrum will not
occur. 
 
The description of the physics at high energies is largely unknown,
and, at the same time, there are various different models of
single-field inflation. It is therefore important to derive general
results, as indeed was in the spirit of \cite{Creminelli:2004yq, 
  Creminelli:2011rh}. We will therefore deal with this topic
adopting the same phenomenological approach and scenarios of
\cite{Meerburg:2009ys, Holman:2007na, Chialva:2011iz}, not assuming
specific detailed high-energy models, but still obtaining quantitative
as well as qualitative results. 
We will present in more details the approach and the
scenarios in section \ref{revmodscen}.
It is also important to stress that we consider the
{\em realistic squeezed 
limit}, where $k_1$ is small but yet nonzero, as defined in \cite{Verde:2009hy,
  Schmidt:2010gw, Creminelli:2011rh}, in view of possible future
observations. Finally, we study the contribution from all possible
cubic couplings in the effective action for single-field inflation
\`a-la Weinberg \cite{Weinberg:2008hq}.

Beside \cite{Meerburg:2009ys, Holman:2007na} (which studied the
bispectrum, but not in the squeezed limit), 
modified initial states have been proposed and studied in \cite{Danielsson:2002kx,
Danielsson:2002qh, Easther:2002xe, Schalm:2004qk,
Schalm:2004xg,Nitti:2005ym, Danielsson:2002mb, Einhorn:2003xb} and
many successive papers. The 
scenarios that we consider are 
well-defined phenomenologically and formally, as investigated by
a large number of works, see for example \cite{Danielsson:2002mb,
Einhorn:2003xb}. Modified initial conditions for the perturbations can
derive for example from the preceding evolution of the universe (see
the examples of \cite{Vilenkin:1982wt, Dey:2011mj}) or from
integrating out heavy fields \cite{Shiu:2011qw, Jackson:2010cw}. On
the other hand, modified dispersion relations occur in a 
variety of models, such as certain realization
of quantum gravity, like the Ho\v{r}ava-Lifshitz model
\cite{Horava:2009uw}, or also in braneworld
models \cite{Chung:1999zs, Chung:1999xg, Chung:2000ji,
Csaki:2000dm, Dubovsky:2001fj}, and are of interest for the studies
of Lorentz violation \cite{Mattingly:2005re,
Jacobson:2005bg}. They occur as well in
the effective theory when the fields propagate with small speed of sound 
\cite{Baumann:2011su, Tolley:2009fg}.

The outline of the paper is as follows:
we begin in section
\ref{formalismnotation} with a very brief presentation of the necessary
formalism, followed by a short review of the scenarios we consider in
section \ref{revmodscen}. 
We then study the bispectrum in the squeezed
limit.
First, we provide a
general discussion of the (expected) results and their physical interpretation
in section \ref{generalarguments}.

We then present the technical
computation of the bispectrum in the 
squeezed limit in section \ref{bispsqueezlimsec}. For the reader's
advantage, we proceed in three
steps. First, in sections \ref{mincoupcubsec} and
   \ref{highdercubsec}, we discuss in full details two
   examples of 
   contributions from two well-known cubic coupling (with and without
   higher-derivatives). These example are meant to help the reader to
   quickly grasp the features of the more general computations and
   results later on. 

Second, in
   section \ref{gencoupsec}, we move to the general analysis and study the 
   contributions to the bispectrum in the 
   squeezed limit from all allowed couplings in the effective theory
   (single-field).

As a third step, armed with these general results, in section
\ref{signaturessec} we 
analyse the features of the 
various contributions, individuating the leading ones and
providing the predictions for the bispectrum in the squeezed limit for
the modified scenarios we have considered.

Finally, we briefly comment on the
consequences of our results on the halo bias in section \ref{halobiassec},
and conclude in \ref{conclusions}. Some useful equations and material,
which would have made heavy the main text, have been moved to 
the appendices.

\section{Formalism and notation}\label{formalismnotation}
\vspace{-0.2cm}
Let us first introduce the basic notation and elements of
the formalism of cosmological perturbation theory, for more details
see \cite{Maldacena:2002vr}. Scalar perturbations in
single-field slow-roll inflationary models are efficiently
parametrized by the comoving curvature perturbation\footnote{We follow
  the conventions in  
  \cite{Maldacena:2002vr, Meerburg:2009ys, 
   Holman:2007na, Chialva:2011iz}; other authors call this perturbation
  $\mathcal{R}$ to distinguish it from the uniform energy-density
  curvature perturbation.} 
\vspace{-0.2cm}
 \beq
  \zeta(\eta, \vec x) = \int {d^3 k \ov (2\pi)^{{3 \ov 2}}}
    \zeta_{\vk}(\eta) e^{i \vk\cdot \vec x}. 
 \eeq

The slow-roll parameters are defined as 
$\epsilon = {\dot \phi^2 \ov 2 H^2 M_{\text{Planck}}^2}, \; 
 \eta_{_\text{sl}} = -{\ddot \phi \ov \dot \phi H} + 
 {\dot \phi^2 \ov 2 H^2 M_{\text{Planck}}^2}$, where $H$ is the Hubble
rate. $\phi$ is the inflaton (background), and 
$a(\eta) \sim -{1 \ov H\eta} \; (\eta < 0)$ is the scale factor at zeroth 
order in slow-roll. Throughout the paper, dots (primes) indicate
derivatives with respect to cosmic time 
$t$ (conformal time $\eta$).

The two-point function is defined as
 \beq \label{twopointdefgen}
  \langle \zeta(\vk_1)\zeta(\vk_2) \rangle
       = (2\pi)^3\delta^{(3)}(\vk_1+ \vk_2)\, P(k_1) , \qquad k_1 = |\vk_1| 
       \, .
 \eeq
and, following \cite{Babich:2004gb}, we define the bispectrum $B$ as
\beq\label{shape-func}
  \langle
  \zeta_{\vk_1}(\eta)\zeta_{\vk_2}(\eta)\zeta_{\vk_3}(\eta)\rangle
   = (2\pi)^3 \delta^{(3)}({\textstyle \sum\limits_i} \vk_i) B(\vk_1,\vk_2,\vk_3, \eta) \, ,
\eeq
where, at leading order in the perturbative expansion,
 \beq \label{threepointzeta}    
  \langle \zeta(\eta, \vec x_1)\zeta(\eta, \vec x_2)\zeta(\eta, \vec x_3)\rangle =
  -2 \text{Re}\left( \int^\eta_\ein d \eta' i
  \langle\psi_{\text{in}}|\zeta(\eta, \vec x_1)\zeta(\eta, \vec x_2)\zeta(\eta, \vec x_3)
  H_{(I)}(\eta')|\psi_{\text{in}}\rangle\right)\, .
 \eeq
Here, $\ein$, $|\psi_{\text{in}}\rangle$ are the initial
time and state, and $H_{(I)}$ is the interaction
Hamiltonian.

A key ingredient in the computation of correlators is the Whightman
function
 \beq \label{Whightman} 
  (2\pi)^3\delta^{(3)}(\vk_1+ \vk_2)\,G_{\vk}(\eta,\eta') \equiv 
  \langle \zeta_{\vk_1}(\eta)\zeta_{\vk_2}(\eta') \rangle =
  (2\pi)^3\delta^{(3)}(\vk_1+ \vk_2)\,{H^2 \ov \dot{\phi}\!^{2}}
  {f_k(\eta) \ov a(\eta)}{f_k^{\ast}(\eta') \ov a(\eta')}
 \eeq
 where $f_{\vk}(\eta), f_{\vk}^*(\eta)$ are two linearly independent
 solutions of the field equation
 \beq \label{eqofmo}
   f_\vk'' + \left(\omega(\eta,\vk)^2- {z'' \ov z} \right) f_\vk = 0
    \qquad z = {a\dot{\phi} \ov H} \, ,
 \eeq
such that, once quantized,
 \beq
  \zeta_{\vk}(\eta) = {f_{\vk}(\eta) \ov z} \hat a_\vk^\dagger +
  {f_{\vk}^*(\eta) \ov z} \hat a_\vk \, .
 \eeq
Imposing the standard commutation relation on the operators
$\hat a_{_\vk}^{\text{\,\,\tiny $\dagger$}}, \hat a_{_\vk}$, entails a certain normalization for the
Wronskian of $f_{\vk}^*(\eta), f_{\vk}(\eta)$. Different choices for
$f_{\vk}^*(\eta), f_{\vk}(\eta)$ correspond to different choices of
initial state \cite{Maldacena:2002vr}. In equation 
(\ref{eqofmo}), $\omega$ is the (comoving) frequency as read from the
effective action: we will limit ourselves to the isotropic case,
where $\omega$ depends only on $k\equiv|\vk|$, dropping the arrow symbol.

\section{Brief review of the scenarios under
  consideration}\label{revmodscen} 
 
To ease the
discussion of the results for the three-point function and make the
paper self-contained, let us recap
here the main features of the single-field scenarios we will deal
with. For more details see \cite{Martin:2000xs, Martin:2002kt,
Martin:2003kp, Lemoine:2001ar, Schalm:2004qk, Schalm:2004xg,
Nitti:2005ym,  Danielsson:2002kx,Danielsson:2002qh, Easther:2002xe,
Meerburg:2009ys, Holman:2007na, Chialva:2011iz}.

\subsection{The standard scenario}

In the standard scenario, as defined in the introduction, the
kinetic terms are standard, the dispersion relation is the usual 
Lorentzian one and the mode functions $f_{\vk}(\eta)$ are determined by
choosing the Bunch-Davies vacuum, so that
 \beq \label{basicstandard}
   \omega(\eta,k)^2 = k^2 \, , \qquad
   f_{k}(\eta) = {\sqrt{\pi} \ov 2} e^{i{\pi \ov 2}\nu+i{\pi \ov 4}}\, \sqrt{-\eta} H^{(1)}_\nu(-k\eta) \, ,
 \eeq
where $H^{(1)}$ is the first Hankel's functions with $\nu = {3 \ov 2}
+ {1-n_s \ov 2}$, and the 
equations and solutions are extrapolated up to $\ein = -\infty$ where
the initial state has been chosen as the empty adiabatic vacuum
(Bunch-Davies). At leading order in slow-roll the spectral index
reads $n_s=1-6\epsilon+2\eta_{\text{sl}}$ and the late time
behaviour of the two-point function is
 \beq \label{twopointstand}
  P_{_\text{st}}(k) \underset{\eta \to 0}{\sim} 
   {H^2 \ov 4 M_{_\text{Planck}}^2 \epsilon k^3}
 \eeq
where the suffix $_\text{st}$ indicates that we are considering the standard case.

\subsection{Modified dispersion relations}\label{revmoddisprel}

Higher-derivative corrections to the action are
deeply rooted in the well-established effective field theory
framework.
In fact, one expects that at certain times at least some of the
higher derivative corrections to the action cannot be treated
perturbatively, because, since the metric scale factor $a(\eta)$ is
rapidly decreasing back in time, the  
suppressing factors, given by powers of ratios such as
${p \ov \Lambda} = {k \ov a(\eta)\Lambda}$, where $\Lambda$ is the
relevant high-energy scale, are not small any 
more.
These no longer small corrections can lead to modified dispersion
relations. As recalled in the introduction, such possibility
appears in a variety of 
scenarios, see for example those in \cite{Horava:2009uw, Chung:1999zs,
Chung:1999xg, Chung:2000ji, Csaki:2000dm, Dubovsky:2001fj, Mattingly:2005re,
Jacobson:2005bg, Baumann:2011su, Tolley:2009fg}.

As we said, we adopt a phenomenological approach and consider {\em generic}
modifications to the dispersion relation due to higher-derivative
corrections to the 
effective action, encoding them in a function $F(-{p \ov \Lambda})$ by
writing the comoving frequency as 
   \beq \label{powerexpfreq}  
    \omega(\eta, k) =  a(\eta) \omega_{_\text{phys}}(p) 
      =  a(\eta) p \;  F\left(-{p \ov \Lambda}\right) 
      = k \;  F\left({H \ov \Lambda}k\eta\right) \, ,
    \qquad F(x \to 0) \to 1 \, ,
    \qquad {H \ov \Lambda} \ll 1 \, .
   \eeq

In \cite{Chialva:2011iz, Ashoorioon:2011eg}, it has been shown that
if the modified 
dispersion relations always satisfy the WKB conditions
(adiabaticity) at early times (that is, WKB is violated only when the
dispersion is in the standard linear regime $\omega \simeq k$ as
usual) the bispectrum is similar to that of the standard
scenario, and so it will be its squeezed limit. This is easily
understood, since the particle production is practically absent in
those cases. 

Instead, \cite{Chialva:2011iz}
has shown that the bispectrum is particularly sensitive to
modifications that lead to a violation of the WKB conditions
(adiabaticity) at early times, where particle production is more
substantial. We will therefore focus on those scenarios.

\begin{figure}[t]
\centering
\includegraphics*[width=210pt, height=140pt]{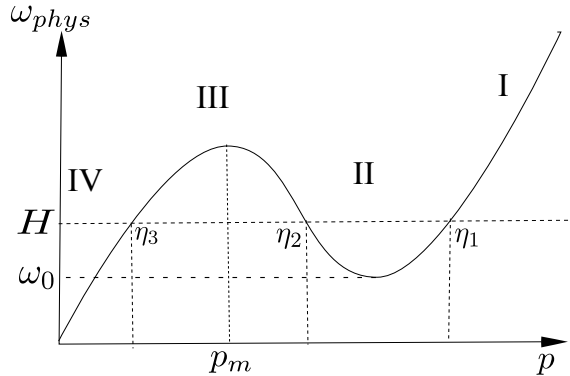}
\caption{Generic example of dispersion relation with violation of WKB at early times.}
\label{INOUTdispersion}
\end{figure}

The generic shape of a dispersion relation with WKB violation
once at early times is depicted in figure
\ref{INOUTdispersion}. We stress that we are not proposing any specific
model, which would be a strong assumption on the high energy physics:
we consider the most generic function $F\left(-{p \ov \Lambda}\right)$
in equation (\ref{powerexpfreq}), 
with the only constraint that adiabaticity is once violated at
early times (it is easy to extend our analysis to multiple
violations). We are in fact interested in the general consequences on
the bispectrum from this scenario. Similar phenomenological approach
and generic dispersion
relations have been considered and discussed, studying the spectrum,
for example in 
the references \cite{Martin:2000xs, Martin:2002kt, Martin:2003kp,
  Lemoine:2001ar}, see also \cite{Mattingly:2005re, Jacobson:2005bg}. 

A general treatment of the field equation (\ref{eqofmo}) in these cases
yields the following solution \cite{Chialva:2011iz}:   
 \beq \label{piecewisesolution}
 f_k(\eta) = 
  \begin{cases}
  \varsigma_k \, u_1(\eta, k) & 
     \qquad {\rm I} : \; \eta < \eta^{(k)}_{\ri I} \\
  B_1 \, \mathcal{U}_1(\eta, k) + B_2 \, \mathcal{U}_2(\eta, k) 
    & \qquad {\rm II} : \; \eta^{(k)}_{\ri I} < \eta < \eta^{(k)}_{\rii II} \\
  \alpha^{^\text{mdr}}_k \, u_1(\eta, k) + \beta^{^\text{mdr}}_k \, u_2(\eta, k)
    & \qquad {\rm III} : \; \eta^{(k)}_{\rii II} < \eta < \eta^{(k)}_{\riii III} \\
  D_1 \, \mathcal{V}_1(\eta, k) + D_2 \,\mathcal{V}_2(\eta, k)
  & \qquad {\rm IV} : \; \eta^{(k)}_{\riii III} < \eta  
 \end{cases}
 \eeq
where $u_{1, 2}(\eta, k)$ can be very well approximated using the
WKB method, while $\mathcal{V}_{1, 2}(\eta, k)$ and
$\mathcal{U}_{1, 2}(\eta, k)$ need to be found by other means, as
the WKB conditions are not satisfied~\footnote{Some different general
  approximations are possible in this case, see \cite{Chialva:2011iz} and the appendix
\ref{solutionmoddisprel}. With particular choices
  of time coordinate, also the WKB approximation can still be applied sometimes,
  see \cite{Martin:2002vn}.}. 

As known, given equation 
(\ref{eqofmo}), the times
$\eta^{(k)}_{{\ri I}, {\rii II}, {\riii III}}$
where the WKB approximation first fails are in proximity of the turning
points of $V(\eta, k)^2 \equiv \omega(\eta,k)^2- {z'' \ov z}$, see  
\cite{BenderOrszagMathMeth, Martin:2002vn}. They depend on $k$. When
clear from the context we will omit the label $^{(k)}$. Looking at  figure
\ref{INOUTdispersion}, these times will be in proximity of $\eta_{1, 2, 3}$
\cite{Chialva:2011iz, Martin:2000xs, Martin:2002kt, Martin:2003kp,
  Lemoine:2001ar}. 
The coefficients $D_{1, 2}, B_{1, 2}, \alpha^{^\text{mdr}}_k,
\beta^{^\text{mdr}}_k$ are obtained 
by asking for the continuity of the 
function and its first derivative, and by imposing the Wronskian condition
$\mathcal{W}\{f, f^*\} = -i$ in order to have the standard commutation
relations in the quantum theory. The label ``mdr'' indicates that we
are considering the scenario of modified dispersion relations.

The error made by using these approximations can be made small at will
by going to higher order in the approximation, see 
for example \cite{Chialva:2011iz} or
\cite{BenderOrszagMathMeth}. The 
techniques are standard in cosmology and allow
controlled global approximations of the
equations solutions and/or Green functions.

The details of the solution can be found in
\cite{Chialva:2011iz}, and are briefly reviewed in the appendix
\ref{solutionmoddisprel}. Some of the salient features needed here are
that: 
\begin{itemize}
 \item backreaction constraints the interval $[\eta_{\ri I},
   \eta_{\rii II}]$ of WKB violation to be very small, that is 
   \beq \label{intervWKBviol} 
    \Delta = {\eta_{\ri I} -\eta_{\rii II} \ov \eta_{\ri I}} \ll 1 \, ,
   \eeq
 \item expanding for small $\Delta$, we obtain generically 
    (i.e., independently of the detailed form of $\omega(\eta, k)$)
  \beq \label{alphabetaIII}
   \alpha^{^\text{mdr}}_k = \biggl(\!\!1 -i{V(\eta_{\ri I}, k)\,\eta_{\ri I} \ov 2} {\mathcal{Q} \ov V^2}\biggr|_{\eta_{\ri I}} \Delta 
     + \mathcal{O}(\Delta^2)\!\!\biggr) \, \varsigma_k \, , \qquad
   \beta^{^\text{mdr}}_k = \biggl(\!\!i{V(\eta_{\ri I}, k)\,\eta_{\ri I} \ov 2} {\mathcal{Q} \ov V^2}\biggr|_{\eta_{\ri I}} \Delta  
     + \mathcal{O}(\Delta^2)\!\!\biggr)e^{-{2i\Lambda \ov H}\Omega_{_{\text{{\tiny F}}}\!|_{\eta_{_{\ri I}}}}} \, \varsigma_k ,
  \eeq
  where $V(\eta, k)^2 \equiv \omega(\eta,k)^2- {z'' \ov z}$. Here,
  ${\mathcal{Q} \ov V^2}\bigr|_{\eta_{\ri I}}$ is the order 1 factor
  signalling the violation of the WKB conditions at $\eta_{\ri I}$, and
  $\mathcal{Q}$ is reported in equation (\ref{WKBparameteratorderfour}) in appendix 
  \ref{solutionmoddisprel}.
  The magnitude of $|\beta^{^\text{mdr}}_k|$ is constrained by backreaction. 
  We will discuss more of the $k$-dependence and the
  magnitude of these coefficients in section \ref{constraintsonBog}, 
 \item  we choose $\varsigma_k =1$ picking up the usual adiabatic vacuum.
\end{itemize}

At late time the two-point function does not differ much from the
standard one, since the Bogoliubov coefficient is quite constrained by
backreaction (see section \ref{constraintsonBog}). This is good for
the agreement with the observations on the (late 
time) spectrum.
At leading order in $\epsilon$, $\beta^{^\text{mdr}}_{k}, \Delta$,
 \beq \label{twopointmoddisprel}
  P_{_\text{mdr}}(k) \underset{\eta \to 0}{\sim} 
  {H^2 \ov 4 M_{_\text{Planck}}^2\epsilon k^3}
  \biggl(1+2 \, \text{Re}(\beta^{^\text{mdr}}_{k}) \biggr) \, .
 \eeq
In the case of the bispectrum 
it will be important the behaviour of the Whightman function at
earlier times, which shows more relevant modifications, see
appendix \ref{Whightmanappendix}. 

\subsection{Modified initial state (BEFT and NPHS scenarios)}\label{revmodinvac}

Let us review the two implementations of the modified initial
state approach that have been first proposed in
\cite{Schalm:2004qk,
Schalm:2004xg,
Nitti:2005ym, Danielsson:2002kx,
Danielsson:2002qh,
Easther:2002xe}, reflecting the effects of
the physics at higher energies/earlier times than inflation in
setting the initial/boundary condition for the perturbation
fields.
 \begin{itemize}
  \item BEFT approach: one
    follows an effective theory approach by  
    fixing the boundary conditions for the fields at 
    the finite time $\eta_c$ (beginning of inflation)
    independently of the modes $k$, through a 
    Boundary Effective Field Theory (BEFT), which accounts for the high energy
    physics via renormalization \cite{Schalm:2004qk, Schalm:2004xg,
    Nitti:2005ym}. This kind of
    boundary condition strongly breaks scale invariance, being imposed
    for all modes $k$ at the same boundary time. The (cutoff)
    scale $\Lambda$, proper of
    the effective action formulation, is not related to the boundary time  
    $\eta_c$ \cite{Schalm:2004qk,
Schalm:2004xg,
Nitti:2005ym}.
  \item NPHS approach: consider the effective theory valid up to  
   a certain energy scale $\Lambda$, and choose the boundary  
   conditions for the solution to the field equation at the New
   Physics Hyper-Surface (NPHS) corresponding to when the physical
   momentum reaches that cutoff scale \cite{Danielsson:2002kx,
Danielsson:2002qh,
Easther:2002xe}.  
   In this case,  the cutoff is imposed in a 
   scale-invariant way via the condition ${k \ov a(\eta_c)} =
   \Lambda$, so that the time $\eta_c$ when the initial state is 
   picked is $k$- and $\Lambda$- dependent. At zero slow-roll order,
   $\eta_c = -{\Lambda \ov k H}$. 
 \end{itemize}
To spare notation, we use the symbol $\eta_c$ for both the NPHS and BEFT cases
and make clear in the context which approach we are following.
In both cases, we stress that the initial condition is
fixed when the modes $k$ are well within the horizon, that is, for
$\eta_c$'s such that $|k\eta_c| \gg 1$.

The physical motivation why such modified initial states can be chosen
is the fact that inflation and cosmological 
perturbation theory are effective theories valid below a certain energy
scale (if the theory was valid at all scales, the only sensible choice
would be the so-called Bunch-Davies vacuum) \cite{Danielsson:2002mb,
Einhorn:2003xb}. Many
studies have proven 
that these initial states and the perturbation theory defined upon them are
well-defined also in the formal sense (for example, see
\cite{Danielsson:2002mb,
Einhorn:2003xb}). 
As in \cite{Meerburg:2009ys, Holman:2007na}, when studying these
scenarios we adopt a phenomenological approach.

A modified initial condition may
present both a Gaussian part and an intrinsic non-Gaussian one. As we
will discuss in section \ref{generalarguments}, using the result of 
\cite{Porrati:2004dm,
Porrati:2004gz} -- see also the comment in \cite{Meerburg:2009ys} -- one finds 
that the leading corrections to the bispectrum in the squeezed limit due to  
intrinsic non-Gaussianities of the initial condition, if present, would be in line
with the standard result, with a local form and a suppressed amplitude,
because of backreaction and lack of cumulation with time. 

The Gaussian part of the initial
condition, which modifies the Whightman functions of the perturbation
theory, will instead turn out to be responsible for new and more
dominant features of the squeezed limit.
As for this part, both in the BEFT and NPHS cases the solution
of the field equation is of the Bogoliubov form 
\cite{Danielsson:2002kx,
Danielsson:2002qh,
Easther:2002xe, Schalm:2004qk,
Schalm:2004xg,
Nitti:2005ym, Meerburg:2009ys, Holman:2007na}:
 \beq \label{solutionmodinitstat}
  f_{k}(\eta) = \alpha_k^{\text{mis}} \sqrt{-\eta} H^{(1)}_\nu(-k\eta) 
    + \beta_k^{\text{mis}} \sqrt{-\eta} H^{(2)}_\nu(-k\eta) \, , 
 \eeq
where $\alpha_k^{\text{mis}}, \beta_k^{\text{mis}}$ are determined by
the specific boundary conditions 
imposed on the solution at the time $\eta_c$, and by
the Wronskian condition, which translates into 
$|\alpha_k^{\text{mis}}|^2 - |\beta_k^{\text{mis}}|^2 = 1$.
The Bogoliubov coefficients depend on the mode $k$, the cutoff
scale and the time $\eta_c$ (these latter are related in the NPHS
case, but not in the BEFT, as we said). The label ``mis'' indicates
that we are dealing with the case of modified initial state.
 
At late time the two-point function does not differ much from the
standard one, because of the smallness of $\beta_k^{^\text{mis}}$ due
to backreaction constraints (see section \ref{constraintsonBog}).
At leading order in slow-roll and $\beta_k^{^\text{mis}}$
 \beq \label{twopointmodvac}
  P_{_\text{mis}}(k) \underset{\eta \to 0}{\sim} 
  {H^2 \ov 4 M_{_\text{Planck}}^2 \epsilon k^3}
  \biggl(1+2 \, \text{Re}(\beta_k^{\text{mis}}e^{^\text{-$i$Arg($\alpha_k^{\text{mis}}$)}}) \biggr) \, .
 \eeq
Once again, for the bispectrum
it will be important the behaviour of the Whightman
function at earlier times, which shows more relevant differences, see appendix
\ref{Whightmanappendix}.    

\subsection{General constraints on Bogoliubov
  coefficients}\label{constraintsonBog} 

The magnitude and scale ($k$)
dependence of the coefficients
$\beta_k^{\text{mdr}}, \beta_k^{\text{mis}}$  
entering respectively equations (\ref{twopointmoddisprel})
and (\ref{twopointmodvac})
are determined in full details by the
specific model or boundary condition giving rise
to the modified dispersion relation  or modified initial state. However, to
be in accordance with observations and to avoid backreaction 
stopping the slow-roll inflationary evolution, these coefficients
are subject to a series of phenomenological constraints, see for
example \cite{Holman:2007na, Chialva:2011iz}:  
\begin{itemize}
 \item the observations on the spectral index show that
   $-k{d\log(k^3 P(k)) \ov dk}$ is small, so that $\beta_k^{\text{mdr}},
   \beta_k^{\text{mis}}$ must be slowly varying with $k$ at the
   observed scales;
 \item backreaction imposes two further constraints:
  \begin{itemize}
   \item the total energy density must be finite. In general this
     demands that $|\beta_k^{\text{mdr}}|, |\beta_k^{\text{mis}}|$ decay
     faster than $k^{-2}$ at large $k$;
   \item preserving the slow-roll inflationary evolution enforces the
     constraint \cite{Holman:2007na, Chialva:2011iz}  
     \beq \label{betaconstraintnobackreaction}
      |\beta_{k}^{\text{mdr}}|, |\beta_k^{\text{mis}}| \leq 
        \sqrt{\epsilon |\mu|} {H M_{_\text{Planck}} \ov \Lambda^2}
      \, , \qquad \mu \equiv \eta_{_\text{sl}}-\epsilon \,.
     \eeq 
  \end{itemize} 
\end{itemize}
In the following we will keep indicating the scale dependence of 
these coefficients with the label $_k$. We will also continue
distinguishing the modified initial state scenario from that with
modified dispersion relations using the labels ``mis'' and ``mdr''.

\section{Bispectrum in the squeezed limit: general
  arguments}\label{generalarguments} 

We will now discuss the three-point
function.
Before presenting the detailed analysis, it is useful to
outline the points that make the difference with the standard
scenario in the cases we consider. 

To make the section self-contained, let us first recall briefly that
the scalar bispectrum is a three-point correlator for the scalar 
perturbation field evaluated at a late time after horizon exit of the
observed modes. It is 
calculated on an initial state defined at an 
initial (conformal) time $\eta_{\text{in}}$, which is then evolved up
to the time the bispectrum is 
evaluated at, and back to $\eta_{\text{in}}$ (in-in formalism, we are
using the interaction picture). Its leading perturbative formula has been
presented in equation (\ref{threepointzeta}). 

The standard result for the squeezed limit $k_1 \ll k_{2, 3} \sim
k_{S}$ follows from very simple 
arguments \cite{Maldacena:2002vr, Creminelli:2004yq,
  Creminelli:2011rh}: 
\begin{itemize}[leftmargin=0.4cm,itemsep=0.0cm,parsep=0.0cm]
\setlength{\parindent}{0.2cm}
 \item[{\em a)}]  for an initial Bunch-Davies vacuum state and a
   standard comoving
   Lorentzian dispersion relation
   the non-Gaussianities are essentially generated at horizon exit,
   and thus the most relevant part of the time evolution is from horizon
   exit of the modes until the time the bispectrum is evaluated at; 
 \item[{\em b)}]  since $k_1 \ll k_{2, 3}$ in the squeezed limit, the
   horizon-exit time for the perturbations depending on $k_{2,3}$ occurs
   much later than the one for the perturbation depending on
   $k_1$. The latter then acts as a background  
     for the other perturbations, shifting their horizon-exit
     time. This leads to a certain dependence of the result on
     $k_1$ (dubbed ``local'') and on the spectral index (which is
     related to the shift of horizon-exit time). It is called the
     Maldacena's consistency relation \cite{Maldacena:2002vr, Creminelli:2004yq,
  Creminelli:2011rh}.
\end{itemize}
\vspace{-0.2cm}
These pieces of physical information are encoded in the form of
the Whightman functions (the two-point functions, see section
\ref{formalismnotation}) of 
the standard scenario. Every higher-order correlator is written
in terms of them because of Wick's theorem. In particular, their form is
such that the time integral from $\eta_{\text{in}}$ to $\eta \sim 0$
coming from the time evolution in the bispectrum formula
(\ref{threepointzeta}) is dominated
by the upper limit of integration (see sections
\ref{bispsqueezlimsec}, \ref{secstadardmincubcoup}). This is how
condition {\em a)} is encoded mathematically.  

It is therefore natural to expect that if the Whightman functions are
modified, in particular in a way such that the time-integral in the 
bispectrum formula picks up other important contributions violating
conditions {\em a)} and/or {\em b)}, then the
squeezed limit will be different from the standard one. 
This is what occurs in the scenarios of modified dispersion
relation and of modified initial state/condition. 

At this point one could think that if the scenarios allow modifications
to the Whightman functions,
one would completely loose general predictivity on the squeezed 
limit. However, we will show that this is not the case: new general
physical arguments based on the concepts of ``particle
creation/content''\footnote{As
  well-known, the concept of particle is not 
  well-defined on time evolving background. Approximate concepts with respect
to comoving observers have a standard use in connection with
adiabaticity, see \cite{BirrelDavies, Mottola:1984ar,
Parker:1968mv, Parker:1969au, Parker:1971pt, Fulling}. Recall that adiabaticity in this
context concerns the time evolution of certain quantities (usually
the effective frequency/mass, from the {\em quadratic} part of the
action \cite{BirrelDavies, Mottola:1984ar, Parker:1968mv,
  Parker:1969au, Parker:1971pt, Fulling}) and is independent and
conceptually different from 
non-Gaussianity.}, interference and accumulation 
in time, take the place of 
those operating in the standard scenario.

'Particles' arise in different ways in the two scenarios we consider,
see section \ref{revmodscen}. In the case of modified
initial state the initial condition for the 
field mode functions is fixed at $\eta_{\text{in}}$ when the modes are well within
the horizon (adiabaticity is satisfied), but the
physics at times/scales 
{\em preceding} $\eta_{\text{in}}$ can be parametrized and
interpreted in terms of what we will call the 'particle content' of
the state. The idea is that the physics at scale higher than inflation
has generated an initial state that is not the adiabatic vacuum, but an
excited one \cite{Vilenkin:1982wt, Dey:2011mj, Shiu:2011qw,
  Jackson:2010cw} -- hence its nonzero energy and particle content. In
the case of modified dispersion relation, instead, 
the time evolution of the dispersion relation can
lead to 'particle 
production' even if the initial state is
the standard empty adiabatic vacuum \cite{Chialva:2011iz,
  Ashoorioon:2011eg}.   

It is the presence of this early energy and
'particle' content that generates the additional
non-Gaussianities. 
Of course, this content is severely constrained by
backreaction on inflation, as we will review. 

The effects of
particle content/creation will be stronger at early times (before dilution by
cosmic expansion). 
The fact that the
time-evolution integral extends to such early times explains why the
bispectrum (and in principle all 
higher-order correlators) are particularly sensitive to these
modifications to the Whightman functions.
Indeed, the spectrum is 
affected by them as well, but much less so, because it does not involve an
integration over time and, especially for observational reasons, it 
is computed at late times, after the horizon exit of the
perturbation, where the modifications are negligible.

The modifications of the Whightman functions will affect all
correlators and lead to different results, compared to the standard
case, even for the simplest couplings.
The new features in the squeezed limit will concern scale dependence
as well as magnitude (enhancements). 
Indeed, although particle creation is 
certainly strongly constrained by backreaction,
it can lead to interference and phase cancellation in
the integrand of the time integral in the bispectrum, giving rise to
enhancements. 

As we will see, the greatest effect occurs when the largest
contributions to the 
oscillating phase of the integrand cancel out so that the suppression
due to oscillations is strongly reduced. This happens
when the  early
particle content for the 
perturbations depending on the largest modes $k_{2,3}$ is relevant and
interference occurs among them.
At those times,
the perturbation depending on $k_1 \ll k_S$, initially subhorizon, may
or may not be superhorizon yet. As we will see, this condition will
depend on the magnitude of $k_1, k_S$ and the scale of
the new physics. There can then occur two different cases.

If the perturbation depending on $k_1$ is not already superhorizon at
those times, 
we do not expect (and do not get indeed) the same $k_1$-dependence
(local shape) as in the standard 
result, because that is entirely determined by the superhorizon
condition. 

If instead the $k_1$-perturbation was already superhorizon,
we obtain a local shape and an effect due to the shift of the horizon
crossing time for the other perturbations as in the standard
scenario. However, also in this case 
we have a new result, because the overall
amplitude of the bispectrum does not match the standard one, and can
also be enhanced. This is a
consequence of the
additional non-Gaussianities generated by the particle
content/creation at early times for the perturbations depending on
$k_{2, 3}$, see sections \ref{bispsqueezlimsec} and 
\ref{signaturessec}. 

In the case of modified initial conditions, there can also be
non-Gaussianities intrinsic to the initial condition. However, 
their effects are subdominant with respect to those due
to the modifications of
the Whightman functions.
In fact, in the squeezed limit the contribution of intrinsic
non-Gaussianities is in line with the standard one: local form and
very suppressed amplitude (because of backreaction). 

This can be seen in various ways. First of all, the
general results for the leading contributions 
to the bispectrum from intrinsic 
non-Gaussianities of the initial conditions were calculated in
\cite{Porrati:2004dm,
Porrati:2004gz}, using the BEFT 
formalism --see also the comment in \cite{Meerburg:2009ys}--. By
taking the squeezed limit of those results one obtains an outcome in
line with the standard one.
One can also argue that the contribution of
non-Gaussianities intrinsic
to the initial condition is negligible from general arguments.
Indeed, those non-Gaussianities, already strongly constrained by backreaction,
are nonzero only at the initial  
time. Thus, their contributions lack 
the integration over time. This makes them
subdominant with respect to the contributions due to the modified Whigthman functions
(Gaussian part), where interference and
time accumulation occur and enhance the result, as we will see.

Generally predictable features of the squeezed limit
in the modified scenarios indeed appear precisely because
the most important new effects are played by the modifications of the
Whightman functions and not by specific couplings or peculiar initial
intrinsic non-Gaussianities. The results are then dominated (we 
will see in what measure) by the general features of
particle content/creation, interference, accumulation and
sub-/superhorizon evolution. This permits to
constrain and possibly falsify entire 
classes of models as the differences between the specific single-field
slow-roll models enter the subleading corrections.

We are now going to investigate these features of the squeezed limit of the
bispectrum by  performing the analysis at the rigorous level of the
field theory description using the in-in formalism. We perform a thorough analysis
from this point of view, providing general
result for all single-field models of inflation within the scenarios
of section \ref{revmodscen}, and studying all cubic couplings that arise in
an effective theory formalism \`a-la Weinberg \cite{Weinberg:2008hq}.

In the case of modified initial state, we will also show that the
field theoretic result 
for the squeezed limit is very different from those obtained in
previous studies, see \cite{Verde:2009hy, Schmidt:2010gw}, which used
the folded 
template proposed for CMBR analysis in \cite{Meerburg:2009ys}.
This disagreement could have been anticipated, since
the standard evaluators (cosine and fudge factor) for the matching
between the template 
\cite{Meerburg:2009ys} and theoretical 
prediction \cite{Holman:2007na} indicate that the two
depart more and more for large $k_L\eta_c$, where $k_L$ is the
largest momentum and $\eta_c$ is the time when the 
boundary condition picking up the initial state is
imposed, see \cite{Meerburg:2009ys}. Even more 
importantly, the template does not depend on the scale $\eta_c$,
and therefore taking the squeezed limit in the full result is different
than taking it in the template, because of the 
presence of distinct scales.

\section{Bispectrum in the squeezed limit: technical
  analysis}\label{bispsqueezlimsec} 

We present here the most technical part of the paper, where we
calculate the contributions to the squeezed limit of the
bispectrum.
We subdivide our presentation in three parts. 
\begin{itemize}
 \item The first two parts (sections \ref{mincoupcubsec} and
       \ref{highdercubsec}) consist of two detailed examples (for
       cubic couplings with and
       without higher derivatives), to illustrate in details what are
       the differences between the scenarios we 
       discuss and the standard one. 

       Let us stress
       that at this point we are not setting apart the interactions
       in these examples as special or dominant compared to all the other 
       possible ones. We choose them
       simply because they are two well-known and well-studied cubic
       interactions\footnote{However, the squeezed limit of their
         contributions to the bispectrum has not been studied in the
       modified scenarios we consider.},
       see \cite{Maldacena:2002vr, Meerburg:2009ys, Holman:2007na,
         Chialva:2011iz, Creminelli:2003iq}, and thus the discussion
       should be easier to follow for the reader. Only after the general
       analysis has been performed in section \ref{gencoupsec} we will come back to
       the question whether these, or other, interactions play a predominant role
       (see section \ref{signaturessec}). 
 \item In section \ref{gencoupsec} we then deal with the full general
   analysis of the squeezed limit of the bispectrum,
   considering all possible cubic interactions in the effective action
   for the inflaton.
\end{itemize}
The results we obtain will be then fully analysed and
discussed in section \ref{signaturessec}. We will find out what are the
leading features of the bispectrum in the squeezed limit for the
modified scenarios and what are the prediction for each specific
modified scenario.

 \subsection{Example 1: Minimal coupling cubic interaction}\label{mincoupcubsec}

We begin by studying the example of the cubic interaction
 \cite{Maldacena:2002vr}: 
 \beq \label{HIcubic}
  H_{(I)} = -\int d^{3}x \, a^3 \, ({\dot \phi \ov H})^4 \, {H \ov M_{_\text{Planck}}^2}
    \, \zeta_c'^2 \partial^{-2} \zeta'_c \, .
 \eeq
We call this interaction the ``minimal coupling cubic
interaction'' because it is already present 
in the simplest case of a scalar field (inflaton) minimally coupled to
gravity \cite{Maldacena:2002vr}.
 
We have followed the
practice of \cite{Maldacena:2002vr} writing this 
interaction in terms of the field redefinition\footnote{In
  \cite{Holman:2007na}, $\zeta_c$ is simply 
  written as $\zeta$, see their equation (3.11). The same is done in
  many other papers, such as \cite{Meerburg:2009ys}.}
 \beq \label{fieldredefzetac} 
  \zeta = \zeta_c + {1 \ov 8} {\dot \phi^2 \ov H^2 M_{_\text{Planck}}^2} \zeta_c^2
    + {1 \ov 4} {\dot \phi^2 \ov H^2 M_{_\text{Planck}}^2}
    \partial^{-2} (\zeta_c \partial^2 \zeta_c) 
    + {1 \ov 2} {\ddot \phi \ov \dot \phi H} \zeta_c^2.
 \eeq
The two-point function and the quadratic part of the action are
the same for $\zeta$ and $\zeta_c$ \cite{Maldacena:2002vr}.

Using the definitions of the slow-roll parameters, the bispectrum of
$\zeta$, which is the relevant one for observations, is
related to the three-point function for $\zeta_c$ as
 \begin{multline} \label{threepointallconserved}
   \langle\zeta_{\vk_1}(\eta)\zeta_{\vk_2}(\eta)\zeta_{\vk_3}(\eta)\rangle =
    \langle\zeta_{c, \,\vk_1}(\eta)\zeta_{c, \,\vk_2}(\eta)\zeta_{c, \,\vk_3}(\eta)\rangle 
    \\
    + (2\pi)^3 \delta^{(3)}({\textstyle \sum\limits_i} \vk_i)\sum_{\substack{i=1 \\ \text{mod 3}}}^3
    \Bigl(\epsilon \bigl({3 \ov 2} + {k_{i+1}^2+k_{i+2}^2 \ov 2 k_{i}^2}\bigr)
     - \, \eta_{_\text{sl}} \Bigr)  
     P(k_{i+1}) \, P(k_{i+2}) \, ,
 \end{multline}
where
 \beq \label{threepoint}    
  \langle \zeta_c(\eta, \vec x_1)\zeta_c(\eta, \vec x_2)\zeta_c(\eta, \vec x_3)\rangle =
  -2 \text{Re}\left( \int^\eta_\ein d \eta' i
  \langle\psi_{\text{in}}|\zeta_c(\eta, \vec x_1)\zeta_c(\eta, \vec x_2)\zeta_c(\eta, \vec x_3)
  H_{(I)}(\eta')|\psi_{\text{in}}\rangle\right)\, ,
 \eeq
and where $\eta \sim 0$ is a late time when all modes $k_i$
are outside the horizon.

As we know from the review in sections \ref{revmoddisprel},
\ref{revmodinvac} the modifications affecting the second line
of (\ref{threepointallconserved}) via the 
spectra $P(k_i)$'s for the modified scenarios at late time are very
suppressed. Hence, we focus on the connected 
contribution (\ref{threepoint}). 
From (\ref{HIcubic}), (\ref{threepoint})
{\small \begin{multline} \label{bispectrumcub}
  \!\!\!\!\langle\zeta_{c,\,\vk_1}(\eta)\zeta_{c,\,\vk_2}(\eta)\zeta_{c,\,\vk_3}(\eta)\rangle \!=\! 
   2 \text{Re} \biggl(\!-i (2\pi)^3 \delta^{(3)}({\textstyle \sum\limits_i} \vk_i)\biggl({\dot{\phi} \ov H}\biggr)^4
   \!\!{H \ov M_{_\text{Planck}}^2}\!\! \int^\eta_{\eta_{\text{in}}} d\eta' {{a(\eta')^3} \ov k_3^2} \prod_{i=1}^3
  \partial_{\eta'}G_{k_i}(\eta, \eta')
    + \text{permutations}\!\biggr)
 \end{multline}}
We will now present the results for the
standard and modified scenarios. 
Our notation is as follows. We will write the
result for the three-point function
in powers of $|\beta_{k_i}|$. Up to linear order, 
 \beq \label{connectedmoddisrel}
  \langle\zeta_{c, \,\vk_1}\zeta_{c, \,\vk_2}\zeta_{c, \,\vk_3}\rangle
  =
  \delta_0 \langle\zeta_{c, \,\vk_1}\zeta_{c, \,\vk_2}\zeta_{c, \,\vk_3}\rangle +
  \delta_\beta \langle\zeta_{c, \,\vk_1}\zeta_{c, \,\vk_2}\zeta_{c, \,\vk_3}\rangle.
 \eeq
We then write
 \beq \label{deltaconnectedmoddisrel}
  \negthinspace\delta_0 \langle\zeta_{c, \,\vk_1}\zeta_{c, \,\vk_2}\zeta_{c,\,\vk_3}\rangle
   \!=\!  A(k_1, k_2, k_3) \delta F_0(k_1, k_2, k_3) \, ,
  \quad
  \delta_\beta \langle\zeta_{c, \,\vk_1}\zeta_{c, \,\vk_2}\zeta_{c, \,\vk_3}\rangle      
   \!=\! A(k_1, k_2, k_3) \delta F_1(k_1, k_2, k_3), \negmedspace
 \eeq
where for the standard scenario
 \beq \label{standarddF0dF1}
  \delta F_0^{_\text{standard}} = 1, \qquad 
  \delta F_1^{_\text{standard}} = 0,
 \eeq 
so that
$A(k_1, k_2, k_3)$ is indeed the standard three-point function for
$\zeta_c$.

This representation of the three-point function is useful because in
this way one will read from $\delta F_0$ and $\delta F_1$ 
the corrections to the standard result, respectively of
order $|\beta_{k_i}|^0$ and $|\beta_{k_i}|^1$, in the modified
scenarios, due to the new high 
energy physics.

\subsubsection{Example 1: standard scenario (review)}\label{secstadardmincubcoup}

In the standard scenario it is $\eta_{_\text{in}} = -\infty$, and the
Whightman functions are obtained from (\ref{Whightman}),
(\ref{basicstandard}). We list them in
appendix \ref{Whightmanappendix}. Inserting them in 
equation (\ref{bispectrumcub}), one finds the integral over
time\footnote{\label{interactingvacuumprescription}The 
  path of integration in time must be chosen
  such that the oscillating piece of the integrand
  becomes exponentially decreasing for $\eta \to -\infty$. This corresponds to taking the
  vacuum of the interacting theory \cite{Maldacena:2002vr}.}
 \beq \label{BunchDaviesBispectrumIntegralMincub}
  \int^{^{\eta\sim 0}}_{_{\eta_\text{in}}} d \eta'
   e^{i(k_1+k_2+k_3)\eta'} \, (k_1+k_2+k_3) = \, -i,
 \eeq
where we have inserted a factor of $k_t \equiv k_1+k_2+k_3$ for later
comparison with the modified scenarios, see  equations
(\ref{correctioncoeffmodvacmincub}),
(\ref{threepointFourierinteg}), and to make the integral
dimensionless. The result from (\ref{bispectrumcub}) is then
\cite{Maldacena:2002vr}
 \beq \label{standardfactor}
  A(k_1, k_2, k_3) \equiv 
  \langle\zeta_{c, \,\vk_1}\zeta_{c, \,\vk_2}\zeta_{c, \,\vk_3}\rangle_{_\text{st}} =
  4 (2\pi)^3 \delta^{(3)}({\textstyle \sum\limits_i} \vk_i){H^6 \ov \dot{\phi}^2 M_{_\text{Planck}}^2}
  {k_1^2k_2^2k_3^2 \ov \prod_{i=1}^3 (2k_i^3)} \sum_l {1 \ov k_t k_l^2} \, ,
  \quad k_t = \sum_{i=1}^3k_i.
 \eeq
Adding the disconnected contribution in equation 
(\ref{threepointallconserved}) to obtain the bispectrum for $\zeta$,
and taking the squeezed limit $k_1 \ll k_S, \, k_{2, 3} \sim k_S$,
finally \cite{Maldacena:2002vr}
 \beq \label{squeezedlimitA}
  \langle\zeta_{\vk_1}\zeta_{\vk_2}\zeta_{\vk_3}\rangle_{_\text{st}}
  \underset{k_1 \ll k_S}{=}
   (2\pi)^3 \delta^{(3)}({\textstyle \sum\limits_i} \vk_i)(1-n_s) P_{_\text{st}}(k_1) P_{_\text{st}}(k_S) \, .
 \eeq
at leading order in ${k_1 \ov k_S} \ll 1$ \cite{Maldacena:2002vr,
  Creminelli:2004yq, Creminelli:2011rh}.
The result (\ref{squeezedlimitA}) shows Maldacena's consistency condition.
The bispectrum in the form (\ref{squeezedlimitA}) is in the
so-called local form \cite{Maldacena:2002vr,
  Creminelli:2004yq, Creminelli:2011rh}. 
 
\subsubsection{Example 1 with modified initial state}\label{corbispmincubmodvac}

As the computations are simpler in this case, we present it first.
We have already discussed before that possible intrinsic
non-Gaussianities of the initial 
condition lead to suppressed and standard-looking results (see section
\ref{generalarguments} and \cite{Porrati:2004dm,
Porrati:2004gz, Meerburg:2009ys}).
We focus here on the Gaussian part of the initial condition/state, which is
responsible for the form of the Whightman functions, and will appear to 
give rise to much more important and dominant modifications to the
squeezed limit of the bispectrum. 

It is straightforward to calculate the contribution to the bispectrum
from (\ref{HIcubic}) for generic $k_1, k_2, k_3$ in this scenario. The
result has been first presented in 
\cite{Holman:2007na}. Let 
us quickly review it. We insert the relevant Whightman
functions, see appendix \ref{Whightmanappendix}, in
 (\ref{bispectrumcub}) and
write the result in the form (\ref{connectedmoddisrel}),
(\ref{deltaconnectedmoddisrel}) using (\ref{standardfactor}). We also
define $k_t \equiv k_1+k_2+k_3$, obtaining \cite{Holman:2007na} 
  \beq  \label{correctioncoeffmodvacmincub}
   \delta F_0 \sim 1 \, ,\qquad 
   \delta F_1 \! = \!\! {\textstyle \sum\limits_{j=1}^3} \delta F_1^{^{(j)}} \!\! = \!
      {\textstyle
        -\!\!\sum\limits_{j=1}^3}\text{Re}\bigl[i\beta^{^\text{mis*}}_{k_j} \!\!k_t\!\!
      \int^{\eta\sim 0}_{\eta_c} \!\!\!\!\!\!d \eta'
      {\textstyle e^{i(\sum_{h\neq j}k_h-k_j)\eta'}} \bigr]\!\!
   = \! -{\textstyle \!\sum\limits_{j=1}^3\!} 
     {k_t{\textstyle \text{Re}\bigl[\beta^{^\text{mis*}}_{k_j} \!\bigl(1\!-\!e^{i (\sum_{h \neq j} k_h-k_j)\eta_c}\bigr)\!\bigr]} \ov \text{{\small$\sum\limits_{h \neq j}$}} k_h\text{-}k_j}  
     \, .
   \eeq

While $\delta F_0$ matches the standard result, $\delta F_1$ leads to
very different outcomes. The presence of a nonzero 
$\delta F_1$ part is a consequence of the negative-frequency component
of the Whightman functions. Note 
the finite-time lower limit of integration at
 $\eta_{_\text{in}} = \eta_c$ for the integral\footnote{The initial
  conditions for the perturbations $\zeta_{k_1}, \zeta_{k_2}, \zeta_{k_3}$
  are fixed respectively at 
  $\eta_{_c}^{(k_{1})}, \eta_{_c}^{(k_{2})}, \eta_{_c}^{(k_{3})}$ when
  modes are well within the horizon ($|k_i\eta_{_c}^{(k_{i})}| \gg 1$,
  $i=1, 2, 3$). In the BEFT case the initial time is the same for all
  modes ($\eta_c$ is independent of $k$, that is,
  $\eta_{_c}^{(k_{1})}=\eta_{_c}^{(k_{2})}=\eta_{_c}^{(k_{3})} = \eta_c$).  
  In the NPHS case, instead, the initial times can be
  different as they depend on the different wavenumbers, therefore the
  overlap of the perturbations (and so their interaction) is nonzero
  only after the latest of the initial times. In the
  squeezed limit, this time is $\eta_{_c}^{(k_{2})} \sim
  \eta_{_c}^{(k_{3})} \sim \eta_{_c}^{(k_{S})}$.
  These points are also explained in section \ref{secpredicscena}.
  We will often neglect the $^{(k_{S})}$ label to avoid cluttering of
  formulas.}. 

Starting from this result, we
move now to the novel part of the analysis and study the squeezed
limit ($k_1 \ll k_S, \, k_{2, 3} \sim k_S$) for this example of interaction.
It appears from (\ref{correctioncoeffmodvacmincub}) that $\delta F_1$ is
the sum of three contributions. 
By taking the squeezed limit, we find that the contribution proportional to
$\beta_{k_1}$ is very small: at leading order
 \beq \label{deltaFmv1cub}
   \delta F_1^{^{(1)}} 
      \underset{k_1 \ll k_S}{=}  
      -\text{Re}\bigl[\beta^{^\text{mis*}}_{k_1} \bigl(1-e^{i 2 k_S\eta_c}\bigr)\bigr] \ll 1 
     \, .
 \eeq

We consider then the contributions proportional to $\beta_{k_2},
\beta_{k_3}$ in (\ref{correctioncoeffmodvacmincub}), where the
perturbations depending on the 
large momenta in the squeezed limit are in opposition of phase 
(that is, $k_j = k_2$ or $k_3$). These contributions will be
much larger than (\ref{deltaFmv1cub}). Indeed, expressing 
$k_{h \neq \{1, j\}}$ in terms of $k_{1}$ and $k_j$ as 
 \beq \label{khmomentumconservation}
  k_{h \neq \{1, j\}} =
   (k_1^2+k_j^2+2k_1k_j\cos{\theta_{j}})^{{1 \ov 2}}
 \eeq 
using momentum conservation, and expanding in 
${k_1 \ov k_j} \sim {k_1 \ov k_S} \ll 1$, one obtains from
(\ref{correctioncoeffmodvacmincub})  
\vspace{-0.2cm}
 \beq \label{argKmdvsqueezed}
  \text{for $j =2$ or $3$} \qquad \sum_{h \neq j}k_h - k_j
   \underset{k_1 \ll k_{2, 3}}{\simeq} k_1 (1 + \cos{\theta_{j}})
   \equiv k_1 \, v_{_{\theta_{\text{\stiny $j$}}}}\, .
 \eeq
so that at leading order
 \beq \label{dF1squeezedgeneralmincub} 
 \delta F_1^{^{(j=2, 3)}}\!\!\!\! \underset{k_1 \ll k_S}{=}\!\! 
     - {k_1+k_2+k_3 \ov k_1 \, v_{_{\theta_{\text{\stiny $j$}}}}}  
      \text{Re}\bigl[\beta^{^\text{mis*}}_{k_j} \bigl(1-e^{i k_1 \,\eta_c v_{_{\theta_{\text{\stiny $j$}}}}}\bigr)\bigr]\, .
 \eeq
It can be easily checked (for example by plotting) 
that the one in (\ref{argKmdvsqueezed}) is an almost perfect
approximation for all values of $\theta_j$ already for 
${k_1 \ov k_j} \sim {k_1 \ov k_S} \sim 0.1$. 

The form of the correction $\delta F_1^{^{(j=2, 3)}}$, then,
depends on the interplay between $k_1$,
$\eta_c$ and $v_{_{\theta_{\text{\stiny $j$}}}}$. 
Recalling that we
consider the realistic limit, where $k_1$ is 
small but not zero, there are two asymptotic
possibilities where the result can be evaluated most explicitly. From
(\ref{dF1squeezedgeneralmincub}), at leading order
 \be \label{modvacGreatProb} 
   1)\; |k_1\eta_c v_{_{\theta_{\text{\stiny $j$}}}}| & \gg 1 \quad 
   \Rightarrow \qquad
    \delta F_1^{^{(j=2, 3)}}\!\!\!\! \underset{k_1 \ll k_S}{=}\!\!
    -2 {k_S \ov k_1}v_{_{\theta_{\text{\stiny $j$}}}}^{-1} \, 
       \text{Re}\bigl[\beta^{^\text{mis*}}_{k_S}\bigr]
            \\
     \label{modvacSmallProb}
   2)\; |k_1\eta_c v_{_{\theta_{\text{\stiny $j$}}}}| & \ll 1 \quad
   \Rightarrow \qquad
   \delta F_1^{^{(j=2, 3)}} \underset{k_1 \ll k_S}{=} -2 k_S \eta_c \,
      \text{Im}\bigl[\beta^{^\text{mis*}}_{k_S} \bigr] \, ,
 \ee
where in the case 1) we have eventually taken into account that in standard
observables such as the halo bias, see section \ref{halobiassec}, the
final result involves an 
integration/average over quantities such as $\theta_j, k_S$ and so the
terms with large oscillations average to zero.

We see that  (\ref{modvacGreatProb}), (\ref{modvacSmallProb})
dominate over (\ref{deltaFmv1cub}) because of the large
factors ${k_S \ov k_1}$ and $k_S \eta_c$. This result is due to
the interference leading to phase cancellation in the
integrand of (\ref{correctioncoeffmodvacmincub}), and the cumulative
effect due to the time integration. Comparing 
to (\ref{BunchDaviesBispectrumIntegralMincub}), we see that the
latter is instead ineffective in the standard scenario, where the
result is dominated by the upper limit of integration (late time). It
is also clear why intrinsic 
non-Gaussianities of the initial condition would have
a different form and be subdominant: 
in that case there is no time-integration, as intrinsic
non-Gaussianities are nonzero at the initial time only.
 
Thus, the leading correction to the squeezed limit of the three-point
function for nonzero $\beta_{k}$ in the case of the ``minimal cubic
coupling'' (\ref{HIcubic}), neglecting subleading corrections 
in ${k_1 \ov k_S} \ll 1$, reads
 \beq \label{consistrelmodvac}
  \delta_\beta\langle\zeta_{c, \,\vk_1}\zeta_{c, \,\vk_2}\zeta_{c, \,\vk_3}\rangle^{^\text{mis}}_{_\text{min cub}} 
  \equiv A(k_1, k_2, k_3) \, \delta F_1(k_1, k_2, k_3)
      \underset{k_1 \ll k_S}{=} (2\pi)^3 \delta^{(3)}({\textstyle \sum\limits_i} \vk_i) \, \,  
       \mathcal{B}^{^\text{mis}}_{_\text{min cub}} \, P_{_\text{st}}(k_1) P_{_\text{st}}(k_S) \, ,
 \eeq
with
 \beq \label{Bfactmodvacthreepoint}
  \mathcal{B}^{^\text{mis}}_{_\text{min cub}} = \sum_{j=2}^3\mathcal{B}^{^\text{mis, min cub}}_{^{(j)}} , \qquad
  \mathcal{B}^{^\text{mis, min cub}}_{^{(j)}} =
  \begin{cases} 
  -4 \, \epsilon \,{k_S \ov  k_1} \,v_{_{\theta_{\text{\stiny $j$}}}}^{-1}
       \text{Re}\bigl[\beta^{^\text{mis*}}_{k_S} \bigr] &
       \text{if $|k_1\eta_c v_{_{\theta_{\text{\stiny $j$}}}}| \gg 1$} \\
  -4 \, \epsilon \,  k_S \eta_c \,
      \text{Im}\bigl[\beta^{^\text{mis*}}_{k_S} \bigr] &
       \text{if $|k_1\eta_c v_{_{\theta_{\text{\stiny $j$}}}}| \ll 1$}
  \end{cases}
 \eeq

For observational purposes, for example concerning the halo bias, one
is interested in the magnitude 
of these corrections and in their dependence
on the probed large scale $k_1^{-1}$. The $k_1$-dependence of the
leading contributions 
(\ref{consistrelmodvac}), (\ref{Bfactmodvacthreepoint}) is fully
determined, since $\beta^{^\text{mis}}_{k_S}$ depends on
$k_S$ and not $k_1$. The magnitude
will have to be estimated and bound using the phenomenological
constraints reviewed in section \ref{constraintsonBog}. 

In section \ref{signaturessec} we will discuss
in details the features of this result (and of the more general
results we will obtain later on), and see if the different models of
modified initial condition
favour case 1) or 2) in (\ref{modvacGreatProb}), (\ref{modvacSmallProb}).  
For the moment, the main evident differences compared to the standard
scenario are as follows. For
$|k_1\eta_c v_{_{\theta_{\text{\stiny $j$}}}}| \ll 1$ the result has the same
dependence on $k_1$ and $1-n_s \sim \epsilon$ as the standard one,
but with a new amplitude factor $|k_S\eta_c| \gg 1$. Instead, for 
$|k_1\eta_c v_{_{\theta_{\text{\stiny $j$}}}}| \gg 1$, the
$k_1$-dependence of this and the standard result are different.

\subsubsection{Example 1 with modified dispersion relations}\label{corrbispmoddisprel} 

The contribution to the bispectrum in this case was first
computed in \cite{Chialva:2011iz} for generic $k_1, k_2, k_3$. Before
discussing the squeezed 
limit, we briefly review the computation,
addressing the reader to \cite{Chialva:2011iz} for more 
details.
 
The three-point function (\ref{bispectrumcub}) in this scenario is
computed using the relevant Whightman functions listed in appendix
\ref{Whightmanappendix}. The latter ones are obtained
using the field solution (\ref{piecewisesolution}) in the general definition
(\ref{Whightman}). Let us also recall that the error
induced by the approximations  
in solving the equation for the mode functions is
calculable and can be made small at will by going to higher order in
the approximations. The results for the bispectrum
are robust against that (small) error, see \cite{Chialva:2011iz}. 

The time integral in the bispectrum (\ref{bispectrumcub}) can then be
divided into the different intervals of validity of the piecewise
solutions (\ref{piecewisesolution}).
In \cite{Chialva:2011iz}, it was shown that 
the leading contribution to the bispectrum occurs when
the Whightman functions have support in the intervals {\rm IV} and {\rm
  III}, see figure \ref{INOUTdispersion}.
Other contributions are indeed suppressed by $\Delta$, defined in equation
(\ref{intervWKBviol}), or by a rapidly decaying
integrand\footnote{Recall that the prescription in footnote
  \ref{interactingvacuumprescription} must be followed also here.},
see \cite{Chialva:2011iz}.  
We concentrate, therefore, only on
the leading contributions from intervals {\rm IV} and {\rm III}.

It is convenient to write the result in terms of the variables 
 \beq \label{yxvariables}
  y \equiv -{p_{_\text{max}}(\eta) \ov \Lambda} = {H \ov \Lambda}k_{_\text{max}}\eta \, , \qquad
  x_i \equiv {k_i \ov  k_{_\text{max}}} \, .
 \eeq
In particular, when taking the squeezed limit it will be $k_{_\text{max}} \approx
k_{2,3} \approx k_S$ and so in these variables the limit reads $x_1\ll
1, \, x_{_{2, 3}} \simeq x_{_S} \simeq 1$. Since $y$ is a rescaled time 
coordinate, sometimes we will call it simply ``time'' in this section.

By inserting in equation 
(\ref{bispectrumcub}) the relevant Whightman functions, listed in
appendix \ref{Whightmanappendix}, and writing the result in the form
(\ref{connectedmoddisrel}), 
(\ref{deltaconnectedmoddisrel}) using (\ref{standardfactor}), the
corrections $\delta F_{_{m=\{0, 1\}}}$ read \cite{Chialva:2011iz} 
 \beq \label{threepointFourierinteg}
  \delta F_{_{m=\{0, 1\}}}(x_{1}, x_{2}, x_{3}, y)\!\! = \!\!
  \sum_{j=1}^3\text{Re} \biggl[(\beta_{k_j}^{^\text{mdr*}})^{m}
  {\Lambda \ov H}\int^{y}_{y_{_{\rii II}}}\!\! dy' x_t 
  g(\{x_{h \neq j}\}, x_j, y', m) 
  \,  e^{i{\Lambda \ov H}S_0(\{x_{h \neq j}\}, x_j, y', m)}\biggr] \, ,
 \eeq
where $x_t \equiv \sum_{i=1}^3 x_i$, and
$g(\{x_{h \neq j}\}, x_j, y', m)$ is reported in equation
(\ref{gmoddisprel}) of appendix
\ref{compbispmoddisprel}.
$S_{0}$ is given by\footnote{Observe that $\omega(x, y) \equiv xF(x y)$ is
dimensionless.} 
 \beq \label{phaseintegcorrect}
  S_{0}(\{x_{h \neq j}\}, x_j, y', m) = 
  \int^{y'}  dy''
  \biggl(\sum_{h \neq j}\omega(x_h\,, y'')+(-1)^m\omega(x_j\,, y'')\biggr) \, ,
  \quad h, j \in \{1, 2, 3\} \, , 
 \eeq

The limits of integration for the variable $y'$ in
(\ref{threepointFourierinteg}) have been discussed in 
\cite{Chialva:2011iz}, which we refer the reader to for the details,
while reporting here the results. The upper limit is  
$y \approx 0$ because the bispectrum is evaluated at late time
$\eta \approx 0$, 
while the lower limit $y_{_{{\rii II}}}$ is determined by
the smallest among 
$\eta^{(k_{{i=1, 2, 3}})}_{_{{\rii II}}}$, which bound region {\rm III} for the
three momenta $k_{_{i=1, 2, 3}}$, see equation (\ref{piecewisesolution}). 
As shown in \cite{Chialva:2011iz}, it is
$y_{_{{\rii II}}} =  {H \ov \Lambda} k_{_\text{max}} 
\eta^{(k_{_\text{max}})}_{_{{\rii II}}} \approx -1$~\footnote{This is indeed
straightforward: the corrections to the linear  
dispersion relation at the time $\eta^{(k)}_{_{{\rii II}}}$ that
separates region {\rm III} and 
{\rm II}  must be quite important in order to drive the
frequency close to the turning point (see figure
\ref{INOUTdispersion}). As these corrections grow as powers
of $-{p(\eta) \ov \Lambda}$, see (\ref{powerexpfreq}), it must then be
$|p(\eta^{(k)}_{_{{\rii II}}}) \Lambda^{-1}| \sim
1$. Furthermore, since $p(\eta^{(k)}_{_{{\rii II}}}) =
-k\eta^{(k)}_{_{{\rii II}}} H$, the relevant   
boundary time $\eta^{(k)}_{_{{\rii II}}}$ (the smallest, as we recalled) is
the one for the largest 
$k$, which is indeed $k_{_\text{max}}$, and thus,
from the definition of $y$, it follows that
$|y_{_{{\rii II}}}| = |p_{\text{max}}(\eta^{(k_{_\text{max}})}_{_{{\rii II}}}) \,
\Lambda^{-1}| \sim 1$.}. Observe also that, in the squeezed limit, 
 $|x_1 y| \leq |x_1 y_{_{{\rii II}}}| \ll 1$ 
$\forall |y| \leq |y_{_{{\rii II}}}|$, so effectively 
$F({\textstyle {H \ov \Lambda}}k_1\eta) = F(x_1y) \simeq 1$ and 
$\omega(x_1y) \simeq x_1$.

We move now to the novel part of the analysis and study the squeezed
limit.
The integral
in (\ref{threepointFourierinteg}) can be conveniently written as the
sum of two parts, respectively when the Whightman functions depending
on $k_{2, 3}$ have support in their regions {\rm IV} or {\rm III}.  

We begin studying $\delta F_{_{m=0}}$. In region {\rm IV} 
$\zeta_{k_{2}}$ and $\zeta_{k_{3}}$ are
superhorizon, and the contribution to the bispectrum is straightforward
to calculate using the Whightman functions listed in
appendix \ref{Whightmanappendix}. 
In region {\rm III}, instead, the
integral in equation (\ref{threepointFourierinteg}), which is a typical
Fourier integral, is dominated by the contributions of points 
where the strong suppression (${\Lambda \ov H} \gg 1$) from the
oscillating phase of the integrand is reduced. This happens for
stationary points of $S_0$ and boundary points \cite{Erdelyi,
  Chialva:2011iz}.  

But in the case of $\delta F_{_{m=0}}$ there are no stationary
points. Indeed, from
(\ref{phaseintegcorrect}) we see that for 
$m=0$ the first
derivative of $S_0(x_{1, 2, 3}, y', m=0)$ is the sum
of the positive-defined frequencies and thus is always
positive and never zero \cite{Chialva:2011iz}. Then, in this case the
Fourier integrals can be 
approximated for ${\Lambda \ov H} \gg 1$ by 
integrating by parts (see \cite{Erdelyi} and section 4.1.2 in
\cite{Chialva:2011iz}). 

Putting together all the contributions and using the information in appendix
\ref{compbispmoddisprel},
we find that in the squeezed limit, neglecting subleading corrections, 
 \beq \label{mincubbispcaseLargedF0}
  \delta F_0 \underset{\substack{k_1 \ll k_S \\ (x_1 \ll 1)}}{=}
   1 - {1 \ov \omega(1, -1)}
   \text{Im}\bigr[g(x_1, x_2, x_3, -1, 0)
    \,  e^{i{\Lambda \ov H}S_0(x_1, x_2, x_3, -1, 0)}\bigl]\Bigr|_{\substack{\!\!\!\! x_1 \simeq 0 \\ x_{_{2, 3}} \simeq x_{_S}\simeq 1}}
  \simeq 1 \, ,
 \eeq
where in the last passage, as
before, we have taken into account the averaging to zero of the rapid 
oscillations in observables such as the
halo bias. As we see, comparing to the first one of (\ref{standarddF0dF1}), this
contribution does not lead to very significant 
corrections to the standard local form of the bispectrum.

We expect a more interesting behaviour for what concerns $\delta F_{_{m=1}}$. 
We start by discussing the contribution from region {\rm IV} of
$\zeta_{k_{2,3}}$ (the region where $\zeta_{k_{2,3}}$ are
superhorizon). In the squeezed limit, that contribution
is very much subdominant. This is because those 
modes become superhorizon only at 
$\eta_S \simeq -k_{2, 3}^{-1} \simeq -k_{_\text{max}}^{-1}$, that is,
using (\ref{yxvariables}), for   
$y_S = -{H \ov \Lambda} \sim 0$.
This time is then very close to the
late time $y \sim 0$ that the bispectrum is evaluated at. Computing then
the contribution of region {\rm IV} from
(\ref{threepointFourierinteg}), we find that it is  of order $\beta$,
hence very small. 

We turn now to the contribution from
region {\rm III} of $\zeta_{k_{2,3}}$.  
From (\ref{threepointFourierinteg}) and
(\ref{phaseintegcorrect}) for $m=1$, we see that
in the squeezed limit $x_1 \ll 1$ the dominant
contributions are those where 
the mode function in the negative-energy branch 
is the one depending on $k_2$ or $k_3$. The reason is that in those cases the
suppression from the oscillations of the integrand is very reduced, because the
perturbations depending on the large momenta are in opposition of
phase and the overall phase of the integrand becomes very small. Indeed,  
 \begin{multline} \label{S0integrandsqueezlimit}
   \!\!\!\!\shoveleft{\text{for $j=2$ or $3$:}} \, \\
   \!\!\!\!\shoveleft{S_0(\{x_{h \neq j}\}, x_j, y', 1) =  \!\!\int^{y'} \!\!\!\!  dy''
  \biggl(\sum_{h \neq j}\omega(x_h\,, y'')-\omega(x_j\,, y'')\biggr)} 
   \!\! \underset{x_1 \ll 1}{\simeq}  \!\!
  x_1 \!\!\int^{y'} \!\!\!\!  dy'' 
  \bigl(1 +\partial_x\omega(x, y'')\big|_{x_S}\cos{\theta_{j}}\bigr) +
  \mathcal{O}(x_1^2) \\
  = x_1\, y'(1+F(y'))\cos(\theta_j)+\mathcal{O}(x_1^2)\quad
  \equiv \; x_1 \, \tilde{v}_{_{\theta_{\text{\stiny $j$}}}}\!(y')\, , \hspace{4.2cm}
 \end{multline}  
using equations (\ref{powerexpfreq}), (\ref{khmomentumconservation})
and $x_s \simeq 1$ ($\mathcal{O}$ is Landau's big-O symbol).

Inserting (\ref{S0integrandsqueezlimit}) back in equation
(\ref{threepointFourierinteg}), 
the integral has then two asymptotic regimes \cite{Erdelyi}, depending on
the magnitude of the integrand's phase factor ${\Lambda \ov  H}
\,x_1$. Using the information listed in appendix   
\ref{compbispmoddisprel}, table \ref{behaviourtable}, we find at leading
order
 \beq \label{mincubbispcaseLarge} 
   \!\!\!\!\delta F_1   =
  {\textstyle\sum\limits_{j=2}^3}
  \begin{cases}   
    \!\!\left. 
   \begin{aligned}
   & \!\! 
  \left[\! 
   \begin{aligned}
   & {\textstyle\!\text{-}{2 \ov x_1} 
   \,{\text{Re} \Bigl[\beta_{k_S}^{^\text{mdr*}}\Bigr] \ov 1+\cos(\theta_j)}}
    && {\textstyle\text{{\small if
    $v_{_{\theta_{\text{\stiny $j$}}}}\!>\!{H \ov \Lambda}{1 \ov x_1}$}}} \\
   &  
   {\textstyle \bigl({\Lambda \ov H}\bigr)^{^{\!\!\text{{\tiny$1\text{-}{1 \ov \kappa+1}$}}}} 
   {2\Gamma({1 \ov \kappa+1}) \ov x_{_1}^{{1 \ov \kappa+1}}}\;   
   \!\!{\text{Im} \Bigl[\beta_{k_S}^{^\text{mdr*}} e^{i{\pi \ov 2}{\text{{\tiny$\text{sign}(\!F^{\text{{\stiny(}}\!\kappa\!\text{{\stiny)}}}\!)$}} \ov \kappa\!+\!1}}\Bigr] \ov (\kappa+1) |F^{\text{{\stiny(}}\kappa\text{{\stiny)}}}|^{{1 \ov \kappa+1}}}
   }
    && {\textstyle \text{{\small if
    $v_{_{\theta_{\text{\stiny $j$}}}}\!<\!{H \ov \Lambda}{1 \ov x_1}$}}}
   \end{aligned} \!\right]
   {\textstyle  -   {2 \ov x_1}
   {\text{Im} \Bigl[\beta_{k_S}^{^\text{mdr*}}g(\text{-}1)e^{i{\Lambda \ov H} x_1\tilde{v}_{_{\theta_{\text{\stiny $j$}}}}\!(\text{-}1)}\Bigr] \ov \tilde{v}_{_{\theta_{\text{\stiny $j$}}}}^{(1)}(\text{-}1)}
    }
    \\
   & \\
   & \quad\quad\quad\quad  \quad\quad\quad\quad  \quad\quad\quad\quad  
   {\textstyle+
    {\textstyle\sum\limits_{\{y_*\}}}
   \bigl({\Lambda \ov H}\bigr)^{^{\!\!1\text{-}{\mu_* \ov \nu_*}}}\,
   {1 \ov x_{_1}^{~{\!\!{\mu_* \ov \nu_*}}}} 
   \text{Re}\Bigl[\beta_{k_S}^{^\text{mdr*}} \mathcal{I}_{_{\theta_{\text{\stiny $j$}}}}^{(\mu_*, \nu_*)}(y_*)
   e^{i{\Lambda \ov H} x_1 \tilde{v}_{_{\theta_{\text{\stiny $j$}}}}\!(y_*)}
   \Bigr]
    } 
    \end{aligned} \right\}
    &  \text{if} \, {\Lambda \ov H} x_{_1} \!\gg \!1   \vspace{0.2cm}\\
     & \\
    2 {\Lambda \ov H}
    \,\text{Im}\,
    \bigl[\beta_{k_S}^{^\text{mdr*}}(\mathcal{O}(1) + \mathcal{O}(x_1))\bigr] 
    &  \text{if} \, {\Lambda \ov H} x_{_1} \!\ll \!1
  \end{cases}
 \eeq
where we have defined
 \beq \label{amplcoeffstatpointmincub}
  \mathcal{I}_{_{\theta_{\text{\stiny $j$}}}}^{(\mu_*, \nu_*)}(y_*) = 
   {\Gamma({\mu_* \ov \nu_*}) \ov \nu_* \bigl({|\tilde{v}_{_{\theta_{\text{\stiny $j$}}}}^{(\nu_*)}(y_*)|}\bigr)^{{\mu_* \ov \nu_*}}}
   g^{\text{\stiny $(\mu_*\text{-}1)$}}(y_*)
   ({\textstyle\sum\limits_{l=0}^1}(\text{-}1)^{l(\mu_*\text{-}1)}
   e^{(\pm 1)^{l} i{\pi \ov 2}{\mu_* \ov \nu_*}\text{sign}(\tilde{v}_{_{\theta_{\text{\stiny $j$}}}}^{\text{{\stiny(}}\nu_*\text{{\stiny)}}}(y_*))}),
 \eeq
and the $\mathcal{O}(1)$ coefficient in the last line of
(\ref{mincubbispcaseLarge}) is reported in detail in equation
(\ref{Oonemincub}), appendix \ref{compbispmoddisprel}.

Equation (\ref{mincubbispcaseLarge}) is the complete
leading-asymptotic result of the
integral. 
It is a complicated
formula, which we write for
illustrative purposes, but the final result will be simpler 
and neat.
We have introduced some new notation, which we explain here below as
well as in appendix \ref{compbispmoddisprel}.

First of all, let us discuss the contribution from 
the possible presence of stationary points, which appear in the second line of
(\ref{mincubbispcaseLarge}), case 
${\Lambda \ov H} x_1 \!\gg \!1$. We have indicated those points by
$\{y_*\}$, and their orders of stationariness for the function
(\ref{S0integrandsqueezlimit}) are given by the numbers
$\{\nu_*\}$. In general terms, they are also
zeros of orders $\{\mu_*\text{-}1\}$ for the function $g$ in
(\ref{threepointFourierinteg})\footnote{The coefficients of the leading
  behaviour of the functions $\tilde{v}_{_{\theta_{\text{\stiny $j$}}}}(y)$,  
$g(\{x\}, y)$ in the squeezed limit
in proximity of a
stationary point $y_*$ have been written as
$\tilde{v}_{_{\theta_{\text{\stiny $j$}}}}^{(\nu_*)}(y_*)$, 
$g^{\text{\stiny $(\mu_*\text{-}1)$}}(y_*)$, see table \ref{behaviourtable} in appendix
\ref{compbispmoddisprel} and
(\ref{amplcoeffstatpointmincub}). 
In particular, if $\nu_*$ is an integer, then $\tilde{v}_{_{\theta_{\text{\stiny $j$}}}}^{(n)} =
n!^{-1}\,\partial_{y}^n\tilde{v}_{_{\theta_{\text{\stiny $j$}}}}$
(similarly for $\mu_*$ and $g^{\text{\stiny $(\mu_*\text{-}1)$}}$). 
The sign $\pm$ in the phases in (\ref{amplcoeffstatpointmincub})
depends on 
$\tilde{v}_{_{\theta_{\text{\stiny $j$}}}}\!(y)-\tilde{v}_{_{\theta_{\text{\stiny $j$}}}}\!(y_*)$ 
being even or odd under $(y-y_*) \to -(y-y_*)$,  
for $y \sim y_*$.}.

The presence of stationary points is
obviously 
model-dependent. However, their contributions will be 
suppressed in the final observables such as the halo bias, because of
the averaging to zero of the large oscillations due to their phase,
visible in (\ref{mincubbispcaseLarge}). Thus, at
the end we can neglect them, and they do not spoil the generality of
the final results we will obtain.

We are left then with the other contributions in
(\ref{mincubbispcaseLarge}), which are always present \cite{Erdelyi}.
Note that in the case of the contribution from the (nearly) folded
configurations (the one for
$v_{_{\theta_{\text{\stiny $j$}}}} \equiv 1\!+\!\cos(\theta_j) < {H
  \ov \Lambda}{1 \ov x_1}$ in (\ref{mincubbispcaseLarge})), one finds
that the scaling
is determined by the lowest-order
correction to the dispersion relation (see appendix  
\ref{compbispmoddisprel}, table \ref{behaviourtable}). That is, $F^{(\kappa)}\sim
\mathcal{O}(1)$ and $\kappa$ in (\ref{mincubbispcaseLarge}) enter the expansion
(\ref{powerexpfreq})
 \beq \label{kappafirstdispcorrec}
  \omega_{_\text{phys}}(p) \sim p(1 + \,\bigl({p \ov \Lambda}\bigr)^{\kappa} F^{(\kappa)}+\cdots ) \,,
 \eeq     
thus they capture the leading correction to the dispersion relation of the
specific models.

Finally, from (\ref{connectedmoddisrel}), using
(\ref{deltaconnectedmoddisrel}), (\ref{standardfactor}) 
(\ref{mincubbispcaseLargedF0}), (\ref{mincubbispcaseLarge}) , the leading
correction to the standard consistency relation due to nonzero
$\beta_{k}$ in the case of the ``minimal cubic coupling''
(\ref{HIcubic}) reads  
 \beq \label{consistrelmoddisprelII}
  \delta_\beta\langle\zeta_{c, \,\vk_1}\zeta_{c, \,\vk_2}\zeta_{c, \,\vk_3}\rangle^{^\text{mdr}}_{_\text{min cub}} 
   \underset{k_1 \ll k_S}{=}
   (2\pi)^3 \delta^{(3)}({\textstyle \sum\limits_i} \vk_i) \; \mathcal{B}^{^\text{mdr}}_{_\text{min cub}} \, \, P_{_\text{st}}(k_1) P_{_\text{st}}(k_S)
  \, ,
 \eeq
 \beq \label{BfactmdrII}
  \mathcal{B}^{^\text{mdr}}_{_\text{min cub}} = {\textstyle\sum\limits_{j=2}^3}
  \begin{cases}
  \left[\! 
   \begin{aligned}
   & {\textstyle\!-4 \,\epsilon \, {k_S \ov k_1} \,
     {1 \ov 1+\cos(\theta_j)}\text{Re} \Bigl[\beta_{k_S}^{^\text{mdr*}}\Bigr]}
    && {\textstyle\text{{\small if $1\!+\!\cos(\theta_j)>{H \ov \Lambda}{1 \ov x_1}$}}} \\
   & 
    {\textstyle 4 \,\epsilon\bigl({\Lambda \ov H}\bigr)^{\!\!\text{{\tiny$1\!-\!{1 \ov \kappa+1}$}}} 
   \bigl({k_S \ov k_1}\bigr)^{\!\!\text{{\tiny${1 \ov \kappa+1}$}}}
   \Gamma(\!{1 \ov \kappa+1}\!)
   {\text{Im} \Bigl[\beta_{k_S}^{^\text{mdr*}} e^{i{\pi \ov 2}{\text{{\tiny$\text{sign}(\!F^{\text{{\stiny(}}\!\kappa\!\text{{\stiny)}}}\!)$}} \ov \kappa\!+\!1}}\Bigr] \ov (\kappa+1) |F^{\text{{\stiny(}}\kappa\text{{\stiny)}}}|^{{1 \ov \kappa+1}}}
    }
    && {\textstyle \text{{\small if $1\!+\!\cos(\theta_j) < {H \ov \Lambda}{1 \ov x_1}$}}}
   \end{aligned} \!\right]
    & \text{if ${\Lambda \ov H}x_1 \gg 1$} \\
    & \\
   \; \; 4\, \epsilon \,
   \, {\Lambda \ov H}
    \,\text{Im}\, \bigl[\beta_{k_S}^{^\text{mdr*}}\mathcal{O}(1) \bigr]
    & \text{if ${\Lambda \ov H}x_1 \ll 1$}   
  \end{cases},
 \eeq
where we have accounted for the averaging to zero of the rapid
oscillations in (\ref{mincubbispcaseLarge}) in observables such as the
halo bias. Recall that the $\mathcal{O}(1)$ coefficient
in the last line is specified in (\ref{Oonemincub}) appendix
\ref{compbispmoddisprel}. 

Deferring comments to section \ref{signaturessec}, for the
moment we just note that, once again, at leading order the dependence on $k_1$ of 
the result is fully specified, and there appear enhancement factors
such as ${\Lambda \ov H}$ and ${k_S \ov k_1}$ thanks to interference
and accumulation in time. Finally, the only relevant sensitivity to the
specific models, a part from $\beta_{k_S}^{\text{mdr*}}$,  is due
to $F^{(\kappa)}\sim \mathcal{O}(1)$ and $\kappa$, which capture the
leading correction to the dispersion relation of the 
specific models, and thus are interesting for detection.

The final result (\ref{consistrelmoddisprelII}), (\ref{BfactmdrII}) is
very similar to the one obtained in the modified 
initial state case. This points at the fact that, indeed, at leading
order the squeezed limit is dominated by the generic
features common to both scenarios (such as particle content/creation,
interference effects). 
The only difference is that in the
modified initial state case, the folded configurations (given by
$\sum_{h \neq j}k_h-k_j=0$, that is, $v_{_{\theta_{\text{\stiny $j$}}}}=0$) 
yield a contribution corresponding to $\kappa \to \infty$ in 
(\ref{BfactmdrII}), since the phase of the integrand and all
its derivatives are identically zero, see the second line in
(\ref{Bfactmodvacthreepoint}). 
In the case of modified dispersion relations, instead, the true
analogous of the folded configuration (which would be 
$\sum_{h \neq j}\omega(x_h\,, y')-\omega(x_j\,, y')=0$ for some $x$'s
but {\em for all $y'$'s}) does not occur\footnote{It can be easily
  verified at least if
  $\omega$ is given by elementary functions and/or power series.} and
$\kappa$ is finite.

\subsection{Example 2: Quartic derivative interaction}\label{highdercubsec}

In this section, we present our second detailed example
before passing to the more general discussion
in section \ref{gencoupsec}. We consider
the higher-derivative cubic interaction coming from the correction
$\mathcal {L}_{_\text{HDI}} =\sqrt{-\text{det}G} 
{g \ov 8 \Lambda^4}((\nabla \Phi)^2)^2$ to
the effective action of the inflaton $\Phi$, where $G$ is the metric, $g$ the
coupling.

Expanding in perturbations and converting in terms of $\zeta$ as
explained in details in \cite{Holman:2007na, Chialva:2011iz}, from
$\mathcal {L}_{_\text{HDI}}$ one obtains the cubic term 
 \beq \label{hamilthigder}
  {H}_{(I)}
    = -\int d^3x a {g \, \dot{\phi}^4 \ov 2 H^3 \Lambda^4} \zeta' 
    \bigl({\zeta'}^2-\bigl(\partial_i \zeta \bigr)^2\bigr) \, .
 \eeq

Because of its origins from the term $\mathcal {L}_{_\text{HDI}}$,
which contains quartic power of derivatives, we call this
cubic interaction ``quartic derivative interaction''.
It is allowed to treat these interactions perturbatively also when the
physical momenta $p = a^{-1} k$ are such that ${p \ov \Lambda}$ is not
small any more, because, compared to the operators changing the
dispersion relation, they are further suppressed by 
additional powers of the perturbation fields.

In the standard scenario, the contributions to the bispectrum from
higher derivative interactions are suppressed 
in the squeezed limit by powers of ${k_1 \ov k_S}$
\cite{Creminelli:2011rh}. It is  
interesting, then, to study if one finds the same suppressed
contributions and the same overall scale dependence in the scenarios of modified
initial state and modified dispersion relations. 
In the standard single-field slow-roll scenario, 
reference \cite{Creminelli:2003iq} has discussed that the one in
(\ref{hamilthigder}) is the 
lowest-dimensional operator in the class of operators
that is the most important for non-Gaussianities. Nonetheless, in the
standard scenario its contribution is very suppressed in the
squeezed limit, as we have said.

Using equations (\ref{threepointzeta}), the bispectrum for the
coupling (\ref{hamilthigder}) reads 
 \be \label{threepointhighder}
  \!\!\!\!\langle\zeta_{\vk_1}(\eta)\zeta_{\vk_2}(\eta)\zeta_{\vk_3}(\eta)\rangle \!\! 
   & = 2\text{Re}\biggl[-i (2\pi)^3
  \delta^{(3)} \bigl({\textstyle \sum\limits_i} \vec{k}_i\bigr) {g \dot{\phi}^4 \ov 2H^3 \Lambda^4}
  \!\!\int_{\eta_{in}}^{\eta} \!\!d\eta' \, a 
  \bigl(\partial_{\eta} G_{k_1}(\eta, \eta') \partial_{\eta} G_{k_2}(\eta, \eta')\partial_{\eta} G_{k_3}(\eta, \eta') + \nonumber \\
  & \qquad + \vk_1\cdot \vk_2 G_{k_1}(\eta, \eta') G_{k_2}(\eta, \eta')\partial_{\eta} G_{k_3}(\eta, \eta'))+\text{permutations}
   \biggr] \, ,
 \ee
where, again, $\eta \sim 0$ is a very late time when all modes $k_i$
are outside the horizon.

\subsubsection{Example 2 with modified initial state}\label{sectcubhigdermodvac}

In this scenario the leading contribution to
the bispectrum from (\ref{threepointhighder}) for generic 
external momenta has been computed in \cite{Holman:2007na,
  Meerburg:2009ys}. Writing (\ref{threepointhighder}) in the same form
as (\ref{connectedmoddisrel}), 
(\ref{deltaconnectedmoddisrel}), using (\ref{standardfactor}), the
main modifications compared to the standard result come from $\delta
F_1$, which reads
 \be \label{modvachigderintbisp}
   \delta F_1(k_1, k_2, k_3, \eta \simeq 0) = \text{Re} 
     \biggl[i C {k_S^3 \ov k_1^2 k_2^2 k_3^2}\sum_{j=1}^3 \beta^{^\text{mis*}}_{{k}_j}  \int_{\eta_c}^{0} d\eta' e^{i(\sum_{h \neq j}k_h-k_j)\eta'}
    P_j(k_1,k_2,k_3,\eta')+{\text c.c.} \biggr],
 \ee
with
 \beq \label{defCfact}
  C \equiv   {x_t \, \prod_{l=1}^3 x_l^2 \ov \sum_{r > v} x_r^2x_v^2}
  {g H^2 M_{_\text{Planck}}^2 \ov 4 \Lambda^4} \, ,
 \eeq
 \begin{multline}
   P_j(k_1,k_2,k_3,\eta') =-k_t\prod_{i=1}^3 (\sum_{h \neq i}k_h-k_i) 
    - i\eta'
    \biggr((k_{j+1}^2-k_{j}^2-k_{j+2}^2)k_{j+1}^2(k_{j+2}-k_{j})+
    (k_{j+2}^2-k_{j}^2-k_{j+1}^2)
     \\  
     \qquad\qquad k_{j+2}^2(k_{j+1}-k_{j}) +
    (k_{j}^2-k_{j+1}^2-k_{j+2}^2)k_{j}^2(k_{j+1}+k_{j+2})
    \biggl)
      +  (-\eta')^2  (\sum_{h \neq j}k_h-k_j)(\prod_{i=1}^3 k_i)(k_t^2-4k_{j+1}k_{j+2}) \, ,
 \end{multline}
where $j$ is defined modulo 3. The factor $C$ has been written in terms of the
variables $x_i$ defined in equation (\ref{yxvariables}).

Now we study the squeezed limit.
As in section \ref{corbispmincubmodvac}, we see from (\ref{modvachigderintbisp})
that in that
limit the dominating contributions to the integral
occur for $j = 2, 3$ in equation (\ref{modvachigderintbisp}), when the
perturbations depending on the large wavenumbers are in opposition of
phase, so that the overall phase of the integrand is small and the
suppression due to the oscillations 
is reduced. It is then straightforward to
see that the
largest contributions come from the terms of order $\eta$ and $\eta^2$ in
$P_j$. Indeed, their integration yields 
 \be \label{oneeta}
  j & =2, 3 \qquad \int_{\eta_c}^{0} d\eta' e^{i(\sum_{h \neq j}k_h-k_j)\eta'}\eta' = 
   \text{{\small ${1 + e^{i k_1\eta_c v_{_{\theta_{\text{\stiny $j$}}}}}(-1+i k_1 \eta_c v_{_{\theta_{\text{\stiny $j$}}}}) \ov k_1^2 v_{_{\theta_{\text{\stiny $j$}}}}^2}
   \equiv {I_1^{^{(j)}} \ov k_1^2 v_{_{\theta_{\text{\stiny $j$}}}}^2}$}} \, ,
   \\
  \label{twoeta}
  j &=2, 3 \qquad \int_{\eta_c}^{0} d\eta' e^{i(\sum_{h \neq j}k_h-k_j)\eta'}\eta'^{2} = 
   \text{{\small $i{2 + e^{i k_1\eta_c v_{_{\theta_{\text{\stiny $j$}}}}}(-2+2 i k_1\eta_c v_{_{\theta_{\text{\stiny $j$}}}}+ k_1^2\eta_c^2 v_{_{\theta_{\text{\stiny $j$}}}}^2) \ov k_1^3 v_{_{\theta_{\text{\stiny $j$}}}}^3}
   \equiv {I_2^{^{(j)}} \ov k_1^3 v_{_{\theta_{\text{\stiny $j$}}}}^3}$}}  \, ,
 \ee
so that at leading order
 \be
  \delta F_1 & \underset{k_1 \ll k_S}{=}
   \sum_{j=2}^3 {2 g H^2 M_{_\text{Planck}}^2 \ov \Lambda^4}  
    {k_S \ov k_1} v_{_{\theta_{\text{\stiny $j$}}}}^{-2} 
   \biggl(-\text{Im}\bigl[ \, \beta_{k_j}^{^\text{mis*}} I_2^{^{(j)}}\, \bigr]
   - \cos(\theta_{j})\text{Re}\bigl[ \, \beta_{k_j}^{^\text{mis*}}I_1^{^{(j)}} \, \bigr]\biggr)
 \ee

Looking at (\ref{oneeta}), (\ref{twoeta}), we find again two
asymptotic behaviours for the correction to the 
squeezed limit of the three-point function depending on whether
$|k_1\eta_c v_{_{\theta_{\text{\stiny $j$}}}}| \gg 1$ 
or $|k_1\eta_c v_{_{\theta_{\text{\stiny $j$}}}}| \ll 1$. Therefore, at leading order
we obtain, for the ``quartic derivative'' coupling (\ref{hamilthigder}),
 \beq \label{consistrelmodvachigderint}
  \delta_\beta\langle\zeta_{\vk_1}\zeta_{\vk_2}\zeta_{\vk_3}\rangle^{^\text{mis}}_{_\text{quart der}} 
      \underset{k_1 \ll k_S}{=} (2\pi)^3 \delta^{(3)}({\textstyle \sum\limits_i} \vk_i) \, \,  
       \mathcal{B}^{^\text{mis}}_{_\text{quart der}} \, P_{_\text{st}}(k_1) P_{_\text{st}}(k_S) \, ,
 \eeq
with
 \begin{gather}
  \mathcal{B}^{^\text{mis}}_{_\text{quart der}} =
   \label{BfactmodvacthreepointHigDerSinglej}
  {\textstyle \sum\limits_{j=2}^3}
  \begin{cases} 
   -4 \, \epsilon \,{g H^2 M_{_\text{Planck}}^2 \ov \Lambda^4}  
    {k_S \ov k_1}  v_{_{\theta_{\text{\stiny $j$}}}}^{-2}
    (2+\cos(\theta_{j}))
    \text{Re}\bigl[ \, \beta_{k_S}^{^\text{mis*}} \, \bigr] 
    &   \text{if $|k_1\eta_c v_{_{\theta_{\text{\stiny $j$}}}}| \gg 1$} \\
   2 \, \epsilon \, {g H^2 M_{_\text{Planck}}^2 \ov \Lambda^4}
    {k_1 \ov k_S} (k_S \eta_c)^{2}
    \Bigl({2 \ov 3} 
    \, {k_1 \ov k_S} (k_S \eta_c)  v_{_{\theta_{\text{\stiny $j$}}}}
    \,\text{Im}\bigl[ \, \beta_{k_S}^{^\text{mis}}\,\bigr]\!+\!
     \cos(\theta_{j})
    \,\text{Re}\bigl[ \, \beta_{k_S}^{^\text{mis}}\,\bigr]\Bigr)
    &   \text{if $|k_1\eta_c v_{_{\theta_{\text{\stiny $j$}}}}| \ll 1$}  \, .
  \end{cases}
 \end{gather}
In the first of (\ref{BfactmodvacthreepointHigDerSinglej}), 
we have again taken into
account the averaging of large oscillations to zero in observables
such as the halo bias. 
Recall that $|k_S \eta_c| \gg 1$ (in the NPHS scenario, in particular, 
$|k_S \eta_c|= {\Lambda \ov H}$). We comment on the actual magnitude and
scale dependence of these corrections to the bispectrum in section \ref{signaturessec}.

\subsubsection{Example 2 with modified dispersion relations}\label{corrbispmoddisprelhigderint}

In this scenario the three-point scalar correlator from
(\ref{threepointhighder}) for generic $k_1, k_2, k_3$ 
was computed in \cite{Chialva:2011iz}. We briefly review that computation.
The dominant 
contribution occurs when the Whightman functions have support in
regions {\rm IV/III}, see (\ref{piecewisesolution}), as 
shown in \cite{Chialva:2011iz} and reviewed in section
\ref{corrbispmoddisprel}. 
Inserting in equation (\ref{threepointhighder}) the relevant Whightman
functions listed in appendix 
\ref{Whightmanappendix}, and writing the result as in
(\ref{connectedmoddisrel}), 
(\ref{deltaconnectedmoddisrel})  using (\ref{standardfactor}), with
the variables defined in 
(\ref{yxvariables}), one obtains that the leading modification
compared to the standard result is \cite{Chialva:2011iz}
 \beq \label{enhanchigder} 
  \delta F_{1}
  =  -\text{{\small $\sum_j$}} \text{Re}\biggl[ \, 
  C \, \biggl({\Lambda \ov H}\biggr)^3 \, \beta_{k_j}^{^\text{mdr*}}
  \int_{y_{\rii II}}^{y} dy' y'^{2} \,
  e^{i{\Lambda \ov H} S_{0}(\{x_h\}, x_j, y', 1)} \,  
  q(\{x_h\}, x_j, y')\biggr]  \qquad h \neq j \, , 
 \eeq
where the limits of integration
are those presented in section  \ref{corrbispmoddisprel}. The
function $q$ is reported in equation 
(\ref{qfactenhanchigder}) of appendix \ref{compbispmoddisprel},
while $S_0$ and $C$ are defined respectively in equation
(\ref{phaseintegcorrect}) and (\ref{defCfact}). 
 
Now we study the squeezed limit. As we see, the correction
(\ref{enhanchigder}) is again given by a Fourier 
integral. Once again, the suppression
from the oscillations of the integrand is reduced when the
perturbations depending on $k_2, k_3$ are in opposition of phase. Then, 
$S_0({\{x_{1, 2(3)}\}, x_{3(2)}, y', 1})$ behaves as in equation
(\ref{S0integrandsqueezlimit}). The integral
can be solved by asymptotic techniques, and we obtain at leading order
 \beq 
  \delta_\beta\langle\zeta_{\vk_1}\zeta_{\vk_2}\zeta_{\vk_3}\rangle^{^\text{mdr}}_{_\text{quart der}}
   \underset{k_1 \ll k_S}{=} 
  (2\pi)^3 \delta^{(3)}({\textstyle \sum\limits_i} \vk_i) \mathcal{B}^{^\text{mdr}}_{_\text{quart der}} \, P_{_\text{st}}(k_1) P_{_\text{st}}(k_S)
  \nonumber
 \eeq 
with (recalling appendix \ref{compbispmoddisprel} and
table \ref{behaviourtable}) 
 \be \label{BfactmoddisrelHigDer}
  \mathcal{B}^{^\text{mdr}}_{_\text{quart der}} & =
   -4\epsilon {g H^2 M_{_\text{Planck}}^2 \ov \Lambda^4} \\
    & \quad \times{\textstyle\! \sum\limits_{j=2}^3}
  \begin{cases}
  \left[\! 
   \begin{aligned}
   & {\textstyle\! 
     {k_S \ov k_1} \,
     {\textstyle{\Gamma(3)+\cos(\theta_j) \ov (1+\cos(\theta_j))^{2}}}
     \text{Re}\bigl[ \, \beta_{k_S}^{^\text{mdr*}} \bigr]}
    && {\textstyle\text{{\small if
    $v_{_{\theta_{\text{\stiny $j$}}}}\!\!>\!\!{H \ov \Lambda}{1 \ov x_1}$ }}} \\
   & 
   {\textstyle {\Lambda \ov H}
   \bigl({\Lambda \ov H}
   {k_1 \ov k_S}\bigr)^{^{\!\!\text{{\tiny$1\!\!-\!\!{2 \ov \kappa\!+\!1}$}}}}
   {\Gamma(\!{\kappa+3 \ov \kappa+1}\!) \ov (\kappa+1)}  
   {2 \, \text{Im}\bigl[ \, \beta_{k_S}^{^\text{mdr*}}F^{\text{{\stiny(}}\text{{\tiny$\kappa$}}\text{{\stiny)}}} \,e^{i{\pi\text{{\tiny$(\kappa\!\!+\!\!3)$}} \ov 2\text{{\tiny$(\kappa\!\!+\!\!1)$}}}{\text{{\tiny$\text{sign}(\!F^{\text{{\stiny(}}\!\kappa\!\text{{\stiny)}}}\!)$}}}} \bigr] \ov |F^{\text{{\stiny(}}\text{{\tiny$\kappa$}}\text{{\stiny)}}}|^{{\kappa+3 \ov \kappa+1}}}
   }
    && {\textstyle \text{{\small if
    $v_{_{\theta_{\text{\stiny $j$}}}}\!\!<\!\!{H \ov \Lambda}{1 \ov x_1}$ }}}
   \end{aligned} \!\right]   
    & \!\!\text{if} \; {\Lambda \ov H} x_1 \gg 1 \\
    & \\ 
   {1 \ov 2} {k_1 \ov k_S} \,
   \bigl({\Lambda \ov H}\bigr)^2  \, 
    \Bigl({k_1 \ov k_S} \,
    {\Lambda \ov H} \, 
    \bigl(\mathcal{O}(1)+{2 \ov 3}\cos(\theta_j)\,\bigr)\text{Im}\bigl[ \, \beta_{k_S}^{^\text{mdr*}} \bigr]
    +\cos(\theta_j)\text{Re}\bigl[\beta_{k_S}^{^\text{mdr*}}\bigr]\Bigr)
    & \!\! \text{if} \; {\Lambda \ov H} x_1 \ll 1
  \end{cases}, \nonumber
 \ee
where we have again taken into account the averaging to zero of the
strongly oscillating terms in final observables as the
halo bias, and the $\mathcal{O}(1)$ term in the last line is specified
in (\ref{Oonemodquart}), appendix \ref{compbispmoddisprel}. 
Equation (\ref{BfactmoddisrelHigDer}) matches
(\ref{BfactmodvacthreepointHigDerSinglej}); to see that, note that
in the modified initial state scenario, as we said, the contribution
from the folded
configuration corresponds to $\kappa \to \infty$, and that the dominating
term in the second line of (\ref{BfactmodvacthreepointHigDerSinglej}) is 
the one proportional to $\text{Re}[\beta_{k_S}^{\text{mis}}]$,
since $|k_1\eta_c v_{_{\theta_{\text{\stiny $j$}}}}| \ll 1$.

\subsection{General analysis for all cubic couplings in the effective
  action}\label{gencoupsec} 

So far we have presented two examples to illustrate in details the kind of
modifications that would occur in the squeezed limit of the bispectrum
in the modified scenarios we
consider. We now want to analyse the squeezed limit
for all possible cubic coupling arising in a generic
effective action for the inflaton.

Let us first summarize the key-points concerning the violation of
the standard result for the bispectrum in the squeezed limit, which
have emerged from the detailed examples discussed above: 
\vspace{-0.08cm}
\begin{itemize}[leftmargin=0.6cm,itemsep=0.2cm,parsep=0.0cm]
 \item the most important modification to the standard result are due to
   the differences in the Whightman functions between the standard and
   the modified scenarios
 \item non-Gaussianities can be enhanced\footnote{The 
   actual magnitude of the enhancements depends on the precise value
   of $|\beta_{k_S}|$ and the scales of ``new'' physics, and thus can
   only be estimated, as we will show in section \ref{signaturessec},
   in our phenomenological 
   approach.} in the squeezed limit by two effects: 
  \begin{itemize}
   \item interference, which reduces the suppression due
    to the oscillating phase of the integrand in the time integral
    in the bispectrum formula, and is strongest when the  
    perturbations depending on the large modes 
    $k_{2, 3} \sim k_S$ are in 
    opposition of phase, 
   \item accumulation in time, which leads to larger enhancement from the time
    integration when interactions
    scale with higher powers of ${1 \ov a(\eta')} \sim -\eta'$~\footnote{This is 
  quite intuitive, due to the importance of early times (large $|\eta'|$) around $\eta_c$ 
for the modified initial state scenario, or around the time of WKB
violation/particle creation 
$\eta_{\rii II} \sim -\Lambda k_S^{-1}H^{-1}$ for the modified dispersion one.
Indeed, if $|k_1\eta_c v_{_{\theta_{\text{\stiny $j$}}}}|\ll 1$ or  
${k_1 \ov k_S}{\Lambda \ov H} \ll 1$, the integral for the
bispectrum is approximately of the form 
$
\int^0_{\eta_{c,{\rii II}}} \!\! d\eta'\,\eta^{\prime n} \!\sim\! \eta_{_{c,{\rii II}}}^{n+1}$, and
one easily sees the growth with $n$. We will soon show a similar growth
for     
$|k_1\eta_c| \!\gg\! 1$, ${k_1 \ov k_S}{\Lambda \ov H} \!\gg\! 1$.}.
  \end{itemize} 
 \item if the squeezed mode and the time/scale of new physics are such that 
      $|k_1\eta_c^{(k_{2,3})}| \ll 1$~\footnote{We recall that the initial  
        conditions for each mode of the perturbations are set when they are
        subhorizon, so that it is always $|k_1\eta_c^{(k_1)}| \gg 1$, see section
        \ref{signaturessec}.}, or ${k_1 \ov k_S} {\Lambda \ov H} \ll 1$, the
      perturbation depending on $k_1$ is already outside the horizon
      when the effects of new physics are relevant for the perturbations 
      depending on $k_{2, 3}$ 
      (responsible for the maximum interference and enhancement, see previous point). 
      Then the result is of the
      local  form but with {\em enhanced non-Gaussianities} (because of
      the particle content/creation, interference,
      and accumulation in time for $\zeta_{k_2}, \zeta_{k_3}$)
 \item if instead $|k_1\eta_c^{(k_{2,3})}| \gg 1$ or 
   ${k_1 \ov k_S}{\Lambda \ov H}\gg 1$ (in a realistic squeezed limit $k_1 \neq 0$), the
      perturbation depending on $k_1$ is {\em inside the horizon} when
      the effects of the new physics on the $k_{2, 3}$-perturbations
      are relevant and the bispectrum in the squeezed limit is {\em
        not of the local form}. 
\end{itemize}

We want now to generalize our analysis beyond the two examples of interactions
discussed above. It is convenient to adopt a different gauge than the one we 
have used so far: in the new gauge it is $\zeta = 0$, and
the perturbations are accounted for by a ``matter field'' perturbation
$\varphi$.
This gauge is fully equivalent, of course, to the one used before, and it
is useful because it leads to a technically simpler expansion in
perturbations \cite{Maldacena:2002vr}. It also makes clear from the
  outset the actual order in slow-roll parameters of
  the cubic couplings, see \cite{Maldacena:2002vr}. In particular, we
  can choose $\varphi$ to  
be the inflaton perturbation, expanding the inflaton in background
plus perturbations
 \beq \label{inflatonexpansion}
  \Phi(x, t) = \phi(t) + \varphi(x, t) \, .
 \eeq  

The results in terms of the variable $\zeta$ are then obtained at the
end by performing a gauge
transformation as described in \cite{Maldacena:2002vr}:
 \beq \label{gaugechangezetavarphi}
  \zeta = - H {\varphi \ov \dot\phi} +
   \mathcal{O}\bigl(\epsilon^2, \varphi^2 \bigr) ,
 \eeq 
so that
 \beq 
  \langle\zeta\zeta\zeta\rangle = - {H^3 \ov \dot\phi^3} \langle\varphi\varphi\varphi\rangle+
   \Bigl[\langle\mathcal{O}\bigl(\epsilon^2, \varphi^2\bigr)\varphi\varphi\rangle
   +\text{permutations}\Bigr] .
 \eeq
In particular, it is enough to consider the gauge-transformation at
leading linear order, because the higher-order corrections will not
yield time-integrated contributions to the bispectrum at tree-level,
and, as it appears from the previous sections, it is the
time-integrated contribution that leads to the leading modified results (see
also  \cite{Meerburg:2009ys, Holman:2007na, Chialva:2011iz}).

The most generic effective action for the inflaton
$\Phi$ in
single-field slow-roll inflation has been presented in
\cite{Weinberg:2008hq}~\footnote{Another way 
  of writing the most generic effective 
  action for a ``matter scalar'' takes advantage of the so-called
  Stueckelberg trick using the Goldstone mode for the broken
  time diffeomorphisms in the unitary gauge, see
  \cite{Cheung:2007sv}.}. 
In a Lorentz invariant theory, the terms in the action will be function of
the Lorentz invariant objects $\Phi$,\,
$g^{\mu\nu}\partial_\mu\Phi\partial_\nu\Phi$,\,
$R^{\mu\nu}\partial_\mu\Phi\partial_\nu\Phi$,\, $\Box \Phi$, $R$. In
particular, those that are functions of $\Phi$ only, and not of the 
objects involving derivatives, constitute the scalar potential
$V(\Phi)$. Recall also that all second order time derivatives (and
first order derivatives of auxiliary fields) in higher order terms in
the action (more than quadratic in the field) have to be eliminated
using the equations of motion, see \cite{Weinberg:2008hq}. This applies
for example to $\Box \Phi$ and $R^{\mu\nu}$. In a Lorentz-broken
theory (such as it is the case with 
modified dispersion relations), there are also additional terms that are
not built up only from the Lorentz-invariant operators listed
above. These terms involve higher
derivatives of the fields that can be written as purely
spatial derivatives in a convenient system of
coordinates~\footnote{Lorentz-breaking always implies a privileged
  frame, the convenient system of coordinates is precisely the one
  adapted to this frame.}.

By expanding the effective action in the inflaton perturbation
(\ref{inflatonexpansion}), given the 
fundamental building blocks listed above (including the higher-derivative
Lorentz-breaking ones), and using conformal time,
the cubic couplings for the perturbation $\varphi$ have the most
general and schematic form 
 \be \label{gencubcoup} 
  \int d\eta d^3x \, a^4 \, {\lambda_{_{nms}} \ov \Lambda^{n+m+s-1}} 
    {\partial^n \varphi\ov a^n} {\partial^m \varphi \ov a^m} {\partial^s \varphi \ov a^s}.
 \ee
Thus, a cubic coupling is individuated by the integers $n, m, s$ and
by the coupling constant $\lambda_{_{nms}}$.

In equation (\ref{gencubcoup}) we have sloppily
indicated both time and space derivatives (comoving coordinates) with
the same symbol $\partial$ for notational simplicity and 
because it will simplify the counting of powers of momenta in the
correlators. However, we will distinguish more carefully the two kinds of
derivatives in the detailed calculations we are about to perform.
We recall again, however, that in the case of time derivatives, their 
order is at most 1 (as we said, higher orders must be eliminated by using the
equations of motion, see \cite{Weinberg:2008hq}).
Finally, the dimensionless coefficients $\lambda_{_{nms}}$ can depend on $\dot \phi$
and in general on the slow-roll parameters, and are constrained by backreaction.

Let us comment on the possible values of $n, m, s$ for the cubic
couplings (\ref{gencubcoup}).
In a Lorentz-invariant theory, given the Lorentz-invariant objects
listed above,
$n, m, s$ can be at most equal to 1~\footnote{Obviously, integrating by
parts, derivatives can be ``moved around'', in which case the more
appropriate condition is that $\text{max}(n+m+s) = 3$. What we will
say in the following is valid, of course, also for the terms obtained
by integration by parts.}, while in a
Lorentz-broken theory their values can be higher (for the higher
spatial derivatives couplings, in the convenient frame adapted to the
Lorentz breaking). 

Furthermore, observe that in a gauge-fixed theory, when solving for the
gravitational constraints (for example those related to the lapse $N$ and
shift-vector $N^i$ in 
the ADM formalism) there will also appear non-local
operators. In the gauge we are using, they arise because of the
solution\footnote{The standard notation $(\vec\partial)^{-2}$ for the
  inverse of $\text{{\small $\sum_j$}} \partial^j\partial_j$ is
  formal. It becomes clear after a Fourier transform.} 
$N^i\simeq \epsilon^{{1 \ov 2}}\partial^i
(\vec\partial)^{-2}\dot\varphi+\mathcal{O}(\epsilon^{{3 \ov 2}})
$, see \cite{Maldacena:2002vr}. Thus, when expanding in 
perturbations, $n, m, s$ can assume value -1 because of the
combination $(\vec\partial)^{-2}\dot\varphi$ in the leading order
contribution from $N^i$, relevant for the bispectrum. However, in the solution for $N^i$ 
this combination is acted upon by $\partial^i$, which increases by one
the value of $n+m+s$ and thus
``compensates'' the -1.

Moreover, $N^i$ enters the action only via the metric $g^{\mu\nu}$
or the extrinsic curvature. In the first case, the spatial index of
$N^i$ will result always 
contracted with a spatial derivative, as it can be seen from 
$g^{\mu\nu}\partial_\mu\Phi\partial_\nu\Phi$ considering that the shift
vector enters the components $g^{0i} = N^{-2}N^i$ and $g^{ij} =
N^{-2}N^iN^j$. Also when coming from the extrinsic curvature a
component $N^i$ will always be 
accompanied by a spatial derivative, because the extrinsic curvature is
defined as  
$K_{ij} = (2N)^{-1}(\dot h_{ij}-2{^{^{(3)}}}\nabla_{\{i} N_{j\}})=
 (2N)^{-1}(\dot h_{ij}-h_{ik}\partial_jN^k-h_{jk}\partial_iN^k-N^k\partial_kh_{ij})$, where 
$h_{ij}, {^{^{(3)}}}\nabla$ are the ADM three-metric and
relative covariant derivative.
These properties imply that in a coupling where originally $N^i$ was
present, the combination $\partial^i(\vec\partial)^{-2}\dot\varphi$ will
multiply at least one other factor with a spatial derivative, which
increases further the value of $n+m+s$ by one.

Couplings
with $n=m=s=0$ arise instead from the expansion of terms such as
 \beq
  V(\Phi) \to {V^{'''}(\phi) \ov 3!} \varphi^3 \, ,\qquad 
  \text{or} \quad \Bigl({\partial^r\Phi \ov a^r}\Bigr)^{u} \Phi^m \to 
   (\partial_t^r\phi)^{u}\phi^{m-3}\varphi^3 \, ,
 \eeq
and thus are higher-order in slow-roll
and would give a negligible
contribution to the bispectrum, see
\cite{Maldacena:2002vr,Weinberg:2008hq, Cheung:2007sv}. We will
neglect them in the following. Moreover, couplings of the form
${\lambda_{011} \ov \Lambda}\, \varphi
(\vec\partial\varphi)^2$, where $\vec\partial$ denotes spatial derivatives,
are reabsorbed (eliminated) by field redefinitions such as 
$\varphi \to \varphi+{\lambda_{011} \ov 2\Lambda}\varphi^2$. Finally, couplings
with a single time or spatial derivative are absent because of isotropy.

The two detailed examples of cubic couplings that we have analysed in
sections \ref{mincoupcubsec}, \ref{highdercubsec}
fall into the scheme we are describing. 
Indeed, once written as in
(\ref{gencubcoup}) using the relation between $\zeta$ and $\varphi$, they
are of the type (see also the comment at the end of section
\ref{gencoupmoddisprelsec}) 
 \be \label{classdetailedexamplesmincub}
  \text{minimal cubic} &: \qquad \{n, m, s\} = \{1, 1, -1\}, && 
  \lambda^{^\text{min cub}}_{_{11\text{-}1}}=\sqrt{2\epsilon} {H \ov M_{_\text{Planck}}} \\
     \label{classdetailedexamplesquartder}
  \text{quartic derivative}&: \qquad n=m=s=1, && 
   \lambda^{^\text{quart der}}_{_{111}}= \sqrt{\epsilon \ov 2}\, g {H M_{_\text{Planck}} \ov \Lambda^2}
   \, .
 \ee

From equations (\ref{gencubcoup}), (\ref{gaugechangezetavarphi}) and
(\ref{threepointzeta}) the contribution to the bispectrum for a generic
cubic coupling (\ref{gencubcoup}) reads
 \begin{multline} \label{genericcontrib}
  \langle\zeta_{\,\vk_1}\zeta_{\,\vk_2}\zeta_{\,\vk_3}\rangle_{_\lambda} \propto
  {\dot\phi^3 \ov H^3}{\lambda_{_{nms}} \ov \Lambda^{n+m+s-1}}\int^0_{\eta_i} d\eta' \; a(\eta')^{4-n-m-s} \;
  (\partial^n G)_{k_1} (\partial^m G)_{k_2} (\partial^s G)_{k_3} + \text{permutations} \\
   +\text{complex conjugate}, 
 \end{multline}  
where we have neglected the momenta conserving delta function,
$2\pi$ factors and
overall minus signs. We have used equation
(\ref{gaugechangezetavarphi}) to cast the 
result in terms of the perturbation $\zeta$ (the Whightman function in
(\ref{genericcontrib}) are those of the $\zeta$-variable). $(\partial^r G)_{k}$
indicates the Fourier transform of $\partial^r G$.

We will now study the dominant contributions to
(\ref{genericcontrib}) in the modified
scenarios. We recall that, as it has emerged from
the detailed examples, such contribution is due to the modification of
the Whightman 
functions and occurs {\em i)} when the interference effects are
stronger -- that is, for $\zeta_{k_2}, \zeta_{k_3}$ in
opposition of phase --, and {\em ii)} for the terms in the integrand
with higher powers of $\eta' \sim -{1 \ov a}$ and,
obviously, lower powers of $k_1$. In the following we will
not pay attention to numerical factors of order $\mathcal{O}(1)$ in
the formulas as they are irrelevant for our considerations. 

In section (\ref{signaturessec}), equipped with the results we are
about to obtain, we will finally discuss which contributions, from all
the possible couplings, determine the leading features of the bispectrum.

\subsubsection{Modified initial state}\label{gencoupmodvacsec}

The relevant Whightman functions go as, see appendix
\ref{Whightmanappendix},
 \beq
  (\partial_{\eta'} G^\pm)_k(0, \eta') = \pm{H^4 \ov \dot\phi^2} k^{2}\eta' {e^{\pm i k \eta'} \ov 2k^3}, \quad
  (\partial_{i}^{r\geq 0} G^\pm)_k(0, \eta')=\pm {H^4 \ov \dot\phi^2}(\pm ik_i)^{r}(1\mp ik\eta') {e^{\pm i k \eta'} \ov 2k^3},
 \eeq
so that when they are inserted in (\ref{genericcontrib}) we find the leading
contribution linear in $\beta_{k_i}$
 \begin{multline} \label{corrmodvacgen}
  \!\!\!\!\delta_\beta\langle\zeta_{\vk_1}\zeta_{\vk_2}\zeta_{\vk_3}\rangle^{^\text{mis}}_{_\lambda}
  \!\!\propto {\lambda_{_{nms}} \ov \Lambda^{n+m+s\text{-}1}}\!\!
  {\textstyle \sum\limits_{j=2}^3} \beta_{k_j}^* {H^{n+m+s+5} \ov \dot\phi^3 2^3 k_1^3 k_2^3 k_3^3} 
    k_a^{n+1} k_b^{m+1} k_c^{s+1} \!\!\!\!\int^{\simeq 0}_{\eta_c}
    \!\!\!\!\!\!d\eta' (\text{-}\eta')^{\text{-}1+n+m+s}
    \Bigl(1+(\text{-}ik_1\eta')^{n^{{\text{{\sstiny$(\!1\!)$}}}}_{_{t}}\text{-}1}\Bigr) \\
    \times{e^{ik_1(1+\cos\theta_j)\eta'} \ov 2^{n^{{\text{{\sstiny$(\!1\!)$}}}}_{_{t}}}} 
    + \text{hermitian},
 \end{multline}
where we have used equations (\ref{khmomentumconservation}), (\ref{argKmdvsqueezed}) and $a, b, c$ run
over $1, 2, 3$ accounting for the permutations in
(\ref{genericcontrib}), from which we 
take the dominant ones~\footnote{\label{spatcontract} Recall that, since we are
  considering only scalar 
  perturbations, after the expansion the spatial derivatives indexes
  are contracted by 
the background metric, leading thus to contributions proportional, by
permutation, to $\vk_1\cdot \vk_2$, $\vk_1\cdot \vk_3$, 
$\vk_2\cdot \vk_3$. But then, in the squeezed limit, the
contributions proportional to $\vk_1\cdot \vk_2$ and $\vk_1\cdot \vk_3$
give an overall contribution proportional to 
$\vk_1\cdot (\vk_2 + \vk_3) \sim k_1^2$   
\cite{Creminelli:2011rh} that is subdominant with respect to the
contribution proportional to $\vk_2\cdot \vk_3$, which goes as 
$\vk_2\cdot \vk_3\to -k_2k_3 = -k_S^2$ in the limit.  
The latter is thus the leading one that we consider here and in the
following.}.
Finally, $n^{{\text{{\sstiny$(\!1\!)$}}}}_{_{t}}=\{0, 1\}$ accounts
for the presence or not of time 
derivatives acting on $G(k_1, \eta')$.

We concentrate now on the time-integral, which can be computed in
closed form. By writing 
$v\!\equiv
\!n\!+\!m\!+\!s(\!+n^{{\text{{\sstiny$(\!1\!)$}}}}_{_{t}}\!-\!1\!)$,
the integral is well defined
for the values attainable by $n, m, s$, discussed in section  
\ref{gencoupsec}, and gives ($j=2, 3$)
 \be
  \int^0_{\eta_c} d\eta' \; \eta'^{-1+v} \;
  e^{ik_1(1+\cos\theta_j)\eta'} 
   & = {(-1)^{v} \ov (i k_1)^{v}}
    \biggl(\sum_{r=0}^{v-1}\binom{v-1}{r}{r!(-ik_1\eta_c)^{v-1-r} \ov (1+\cos\theta_j)^{r+1}}
    e^{ik_1\eta_c(1+\cos\theta_j)} -{(v-1)! \ov (1+\cos\theta_j)^{v}}\biggr)
     \nonumber \\
  & \approx
  \begin{cases} 
     -{\eta_c^{v} \ov v}
    &   \text{if $|k_1\eta_c(\!1\!+\!\cos\theta_j\!)| \ll 1$} \\
    {(-1)^{v-1} \ov (i k_1)^{v}}
    {(v-1)! \ov (1+\cos\theta_j)^{v}}
    &   \text{if $|k_1\eta_c(\!1\!+\!\cos\theta_j\!)| \gg 1$}  \, ,
  \end{cases}
 \ee
where in the last passage we have written the leading contribution,
also considering the averaging to zero of large oscillations in the
final observables as the halo bias.
 
Inserting this result in equation (\ref{corrmodvacgen}), we obtain the
leading contribution (reinstating the delta function from conservation
of momentum)
 \beq \label{corrgenbispmv}
  \!\!\!\!\delta_\beta\langle\zeta_{\vk_1}\zeta_{\vk_2}\zeta_{\vk_3}\rangle^{^\text{mis}}_{_\lambda} \!\!
  = (2\pi)^3 \delta^{(3)}({\textstyle \sum\limits_i} \vk_i) \,\, \mathcal{B}^{^\text{mis}}_{_{\lambda_{\text{nms}}}}
  \, P_{_\text{st}}(k_1) P_{_\text{st}}(k_S) 
 \eeq
where, having substituted back $n+m+s$ in place of $v$,
 \begin{multline} \label{fNLcorrgenbispmv}
  \!\!\mathcal{B}^{^\text{mis}}_{_{\lambda_{\text{nms}}}} \! = \!
   \epsilon {\beta_{k_S}^* \ov 2^{n^{\!\!{\text{{\sstiny$(\!1\!)$}}}}_{_{t}}}} {\lambda_{_{nms}} \ov \sqrt{2\epsilon}}
   \,{\Lambda M_{_\text{Planck}} \ov H^2} 
    \bigl({k_1 \ov k_S}\bigr)^{^{\!\!1+\text{min}(n, m,s)}} \!\! 
   {\textstyle \sum\limits_{j=2}^3}
   \begin{cases} 
    ({H \ov \Lambda}|k_S\eta_c|)^{\text{{\tiny$n\!\!+\!\!m\!\!+\!\!s$}}}
    \bigl(\text{{\footnotesize$c_{_{nms}}(1)$}} \! + \! {\text{{\footnotesize$c_{_{nms}}(n^{{\text{{\sstiny$(\!1\!)$}}}}_{_{t}})$}} \ov (\text{-}i|k_1\eta_c|)^{1\text{-}n^{{\text{{\sstiny$(\!1\!)$}}}}_{_{t}}}}\bigr)
    & 
        \text{if $|k_1\eta_cv_{_{\theta_{\text{\stiny $j$}}}}| \ll 1$} \\
     \bigl(\text{-}i{H \ov \Lambda}{k_S \ov k_1}\bigr)^{\text{{\tiny$n\!\!+\!\!m\!\!+\!\!s$}}}
     \bigl({d_{_{nms}}(1) \ov v_{_{\theta_{\text{\stiny$j$}}}}^{n+m+s}} \! + \!
      {d_{_{nms}}(n^{{\text{{\sstiny$(\!1\!)$}}}}_{_{t}}) \ov v_{_{\theta_{\text{\stiny $j$}}}}^{n+m+s+n^{\!\!{\text{{\sstiny$(\!1\!)$}}}}_{_{t}}\text{-}1}}\bigr)
    & 
        \text{if $|k_1\eta_cv_{_{\theta_{\text{\stiny $j$}}}}| \gg 1$}
   \end{cases}  \\
    +\text{complex conjugate}
 \end{multline}
with 
 \beq
  {\textstyle v_{_{\theta_{\text{\stiny $j$}}}} \equiv (\!1\!+\!\cos\theta_j\!), \qquad
  c_{_{nms}}(x) \equiv {c_\lambda \ov (n+m+s-1+x)},
  \qquad
  d_{_{nms}}(x) \equiv (\text{-}1)^{x\text{-}1}\text{{\small$(n\!+\!m\!+\!s\!-\!2+x)!$}}d_\lambda}, 
 \eeq
and $c_\lambda, d_\lambda$ are numerical factors of order one, possibly
depending on $\theta_j$.

We observe that the generic contributions present both enhancement
factors (such as $|k_S\eta_c| \gg 1$, or ${k_S \ov k_1} \gg 1$) as
well as suppressing ones (such as 
$\bigl({H \ov \Lambda}\bigr)^{n+m+s-1}$ and
$\lambda_{_{nms}}$). We also observe that for
$|k_1\eta_cv_{_{\theta_{\text{\stiny $j$}}}}| \gg 1$ the 
contribution will not be in general of the local form. However, such
contributions could be suppressed and thus subleading. 
We will discuss these points in
details in section \ref{signaturessec}. As a check, note that when specialized to the couplings
(\ref{classdetailedexamplesmincub}),
(\ref{classdetailedexamplesquartder}) that we studied in details
previously, equation (\ref{fNLcorrgenbispmv}) leads to the results
(\ref{Bfactmodvacthreepoint}),  
(\ref{BfactmodvacthreepointHigDerSinglej}) that we 
found before, with the proper $c_\lambda, d_\lambda$.

\subsubsection{Modified dispersion relations}\label{gencoupmoddisprelsec}

As discussed in section \ref{corrbispmoddisprel} the leading
contribution to the bispectrum
arises when the Whightman functions have support mostly in the
interval of times in regions {\rm IV/III}, given the solution
(\ref{piecewisesolution}).
They go as, see appendix
\ref{Whightmanappendix},
 \begin{gather}
   \!\!\!\!(\partial_{\eta'} G^\pm)_k(0, \eta')\! =\! 
    \text{-}i{H^3 \ov \dot\phi^2} k{\gamma^{(*)}(k, \eta') \ov a(\eta')}
   {e^{\pm i{\Lambda \ov H}\Omega(k, \eta')} \ov 2k^2}, \quad
   (\partial_{i}^{r\text{{\tiny$\geq\!\!0$}}} G^\pm)_k(0, \eta')\! =\! {H^3 \ov \dot\phi^2}\,
    (\text{{\footnotesize$\pm$}} ik_i)^r \,
   {\chi^{(*)}(k, \eta') \ov a(\eta')}{e^{\pm i{\Lambda \ov H}\Omega(k, \eta')} \ov 2k^2}, 
 \end{gather}
where $ \chi^{(*)}(k, \eta')$ is defined in equation
(\ref{WhightmanNonStandardmdr}) and $\gamma^{(*)}(k, \eta')$ in
(\ref{DerWhightmanNonStandardmdr}). 

Once again, it is convenient to write the contribution to the bispectrum in
terms of the variables defined in equation (\ref{yxvariables}).
Inserting the Whightman functions  in (\ref{genericcontrib}), the
leading contribution linear in $\beta_{k_i}$ reads 
 \begin{multline} \label{corrmoddisprelgen} 
  \!\!\!\!\delta_\beta\langle\zeta_{\vk_1}\zeta_{\vk_2}\zeta_{\vk_3}\rangle^{^\text{mdr}}_{_\lambda}
  \propto  {\lambda_{_{nms}} \ov \Lambda^{n+m+s-1}}
  \sum_{j=2}^3 \beta_{k_j}^* {H^{n+m+s+5} \ov \dot\phi^3 2^3 k_1^3 k_2^3 k_3^3} 
    \biggl({\Lambda \ov H}\biggr)^{n+m+s} 
    {k_a^{n+1} k_b^{m+1} k_c^{s+1} \ov k_S^{n+m+s}} \\
   \times\,\int^{y\lesssim \text{-}{H \ov \Lambda}}_{y_{_{\rii II}} \sim -1} dy' \; (\text{-}y')^{n+m+s-1} \;
   \!\!\!\prod_{\text{{\sstiny$h\!\!=\!\!\{\!a, b, c\!\}$}}}\!\!\!
   (i\gamma^{(*)}(x_{h}, y'))^{n^{{\text{{\sstiny$(\!h\!)$}}}}_{_{t}}}
   (\chi^{(*)}(x_{_1}, y'))^{1\text{-}n^{{\text{{\sstiny$(\!1\!)$}}}}_{_{t}}} 
   e^{i{\Lambda \ov H}x_1  \tilde{v}_{_{\theta_{\text{\stiny $j$}}}}^{(n)}(y')} 
    + \text{hermitian},
 \end{multline}
where we have used equation (\ref{S0integrandsqueezlimit}), 
$n^{{\text{{\sstiny$(\!a,b,c\!)$}}}}_{_{t}}=\{0, 1\}$ accounts for
the presence or not of time derivatives in the coupling acting
on $\zeta_{k_{a, b, c}}$, and $a, b, c$ run
over $1, 2, 3$ taking into account the permutations, from which we
take the dominant ones according to footnote
(\ref{spatcontract}). Depending on the specific Lorentz-broken model, $n,
m, s$ can assume values larger than 
1 such that $n+m+s$ can be larger than 3~\footnote{However, time
  derivatives are at most of order 1, see 
  section \ref{gencoupsec} and \cite{Weinberg:2008hq}.}. 

The solution for the integral can be again obtained by asymptotic
techniques as those employed in sections \ref{corrbispmoddisprel},
\ref{corrbispmoddisprelhigderint}, 
so that the leading contribution reads
 \beq \label{corrgenbispmdr}
  \!\!\!\!\delta_\beta\langle\zeta_{\vk_1}\zeta_{\vk_2}\zeta_{\vk_3}\rangle^{^\text{mdr}}_{_\lambda} \!\!
  = (2\pi)^3 \delta^{(3)}({\textstyle \sum\limits_i} \vk_i) \,\, \mathcal{B}^{^\text{mdr}}_{_{\lambda_{\text{nms}}}}
  \, P_{_\text{st}}(k_1) P_{_\text{st}}(k_S) 
 \eeq
where, having written $n+m+s = v$ and
$n+m+s+n^{^{\text{{\sstiny$(\!1\!)$}}}}_{_{t}}\text{-}1 = v_{_{t}}$,
and used table \ref{behaviourtable} in 
appendix \ref{compbispmoddisprel}, 
\vspace{-0.2cm}
 \be \label{fNLcorrgenbispmdr}
  \mathcal{B}^{^\text{mdr}}_{_{\lambda_{\text{nms}}}} & =
   \epsilon \, \beta_{k_S}^*\,{\lambda_{_{nms}} \ov \sqrt{2\epsilon}}
    {\Lambda M_{_\text{Planck}} \ov H^2}\bigl({k_1 \ov k_S}\bigr)^{1+\text{min}(n, m,s)}
    i^{n^{{\text{{\sstiny$(\!1\!)$}}}}_{_{t}}}
     \\
   & \times{\textstyle\!\sum\limits_{j=2}^3}\!\!
   \left.
   \begin{cases} 
    ({1 \ov 2})^{n^{{\text{{\sstiny$(\!1\!)$}}}}_{_{t}}}
    \bigl(\mathcal{O}(1)+\bigl({\Lambda \ov H}{k_1 \ov k_S}\bigr)^{n^{{\text{{\sstiny$(\!1\!)$}}}}_{_{t}}-1}\mathcal{O}(1)\bigr)
    &   \text{if ${\Lambda \ov H}{k_1 \ov k_S} \ll 1$} \\
    & \\
   \!\!\left[\! 
   \begin{aligned} 
   & {\textstyle\!\bigl({H \ov \Lambda}{k_S \ov k_1}\bigr)^{v}
     \Bigl({(\text{-}i)^v\,\Gamma(v) \ov (1+\cos\theta_j)^{v}} 
     \bigl(\!1\!\!+\!\!C^{\text{{\tiny$(0)$}}}_{_{\theta_{\text{\stiny $j$}}}}({H \ov \Lambda}{k_S \ov k_1}) \!\bigr)\!+\!
     {(\text{-}i)^{v_{_{t}}}\,\Gamma(v_{_{t}}, iv_{_{\!\theta_{\!\text{\stiny $j$}}}}) \ov 2^{n^{{\text{{\sstiny$(\!1\!)$}}}}_{_{t}}}(1+\cos\theta_j)^{v_{_{\!t}}}}
    \Bigr)} 
    && {\textstyle\text{{\small if
     $v_{_{\theta_{\text{\stiny $j$}}}}\!\!>\!\!{H \ov \Lambda}{k_S \ov k_1}$ }}} \\
   & 
   {\textstyle\!\bigl({H \ov \Lambda}{k_S \ov k_1}\bigr)^{{v \ov \kappa+1}}
   {\Gamma({v \ov \kappa+1}) e^{i{\pi\text{{\tiny$v$}} \ov 2}{\text{{\tiny$\text{sign}(\!F^{\text{{\stiny(}}\!\kappa\!\text{{\stiny)}}}\!)$}} \ov \text{{\tiny$(\kappa\!\!+\!\!1)$}}}} \ov (\kappa+1) |F^{\text{{\stiny(}}\text{{\tiny$\kappa$}}\text{{\stiny)}}}|^{{v \ov \kappa+1}}}}
   \;{\textstyle\bigl(1\!+\!C^{\text{{\tiny$(\kappa)$}}}_{_{\theta_{\text{\stiny $j$}}\sim \pi}}({H \ov \Lambda}{k_S \ov k_1})
   \!\bigr)}
   && {\textstyle \text{{\small if
     $v_{_{\theta_{\text{\stiny $j$}}}}\!\!<\!\!{H \ov \Lambda}{k_S \ov k_1}$ }}} \\
   \end{aligned} \!\right]
    &   \text{if ${\Lambda \ov H}{k_1 \ov k_S} \gg 1$}
   \end{cases} \!\!\right\} \!+\! \! \! \! \!
    \begin{split}
    & \text{complex} \\ & \!\!\text{conjugate}
    \end{split} \nonumber
 \ee
with
 \beq
  C^{\text{{\tiny$(\rho)$}}}_{_{\theta_{\text{\stiny $j$}}}}({H \ov \Lambda}{k_S \ov k_1}) \equiv
   \sum\limits_{r, n=0}^\infty  
   c^{r, n}_{_{\theta_{\text{\stiny $j$}}}}\bigl({H \ov \Lambda}{k_S \ov k_1}\bigr)^{\!\!\text{{\tiny${\kappa+r+n \ov \rho+1}$}}}
   {\Gamma({v+\kappa+r+n \ov \rho+1})e^{i{\pi\text{{\tiny$(\kappa\!\!+\!\!r\!\!+\!\!n)$}} \ov \text{{\tiny$(\rho\!\!+\!\!1)$}}}{\text{{\tiny$\text{sign}(\!F^{\text{{\stiny(}}\!\rho\!\text{{\stiny)}}}\!)$}} \ov 2}} \ov   \Gamma({v \ov \rho+1}) |F^{\text{{\stiny(}}\text{{\tiny$\rho\!\!+\!\!n$}}\text{{\stiny)}}}|^{{\kappa+r+n \ov \rho+1}}}
 \eeq
containing all the subleading asymptotic contributions (the $c^{r, n}_{_{\theta_{\text{\stiny $j$}}}}$
are calculable order one coefficients, whose precise form is
irrelevant for us, obtained from the expansion of the
integrand in (\ref{corrmoddisprelgen})). Finally,
$v_{_{\theta_{\text{\stiny $j$}}}}$ has been defined in
(\ref{argKmdvsqueezed}), $\Gamma(v_{_{t}}, iv_{_{\!\theta_{\!\text{\stiny $j$}}}})$ 
is the lower incomplete gamma
function, and the $\mathcal{O}(1)$ factors in the second line of 
(\ref{fNLcorrgenbispmdr}) are written in 
details in
(\ref{Oonegeneralmdr}), appendix \ref{compbispmoddisprel}
.

This result is very similar to that obtained in the modified initial state
scenario, except for two differences. One is that now $n+m+s$ can also assume
values larger than 
3 in the case of couplings deriving from the
Lorentz-breaking terms in the action. The other is that for (nearly) folded 
configurations $\kappa$ is finite
and determined by the first correction to the
standard dispersion relations (see (\ref{kappafirstdispcorrec})), whereas
 in the modified initial state scenario it is $\kappa \to \infty$ for
the folded configuration. A
part from this,
the similar structure of the two results can again be seen
as showing that indeed the squeezed limit is dominated, at leading
order, by the generic features common to both scenarios (such as
particle content/creation, interference and time accumulation). 

As a check, one finds that when equation (\ref{fNLcorrgenbispmdr}) is
specialized to the couplings (\ref{classdetailedexamplesmincub}),
(\ref{classdetailedexamplesquartder}) that we studied previously as detailed
examples, 
it leads to the results we have found
before, see (\ref{BfactmdrII}),
(\ref{BfactmoddisrelHigDer}). Note, in particular, that since the one
in equation (\ref{hamilthigder}) is the sum of two elementary cubic
couplings (one with only time derivatives, the other with one time and
two space derivatives) such that for the
folded configurations the leading contribution is cancelled out, for
those configurations the result 
(\ref{BfactmoddisrelHigDer}) is given by the first subleading
correction in equation (\ref{fNLcorrgenbispmdr}).

\section{Signatures of very high energy physics in the squeezed
  limit}\label{signaturessec} 

In the previous sections we have studied the contributions
to the bispectrum in the squeezed limit in scenarios with modified
initial state or modified dispersion relations at high energies.
This has lead to the formulas
(\ref{corrgenbispmv})-(\ref{fNLcorrgenbispmv}) and 
(\ref{corrgenbispmdr})-(\ref{fNLcorrgenbispmdr}) for the leading
corrections to the standard result for all  
possible cubic couplings, see (\ref{gencubcoup}), in the effective Lagrangian.

Armed with these results, we are now going to 
\begin{itemize}
 \item discuss the leading features of the contributions to the
   bispectrum in the squeezed limit, individuating also which 
   cubic couplings yield the largest contributions;
  \item obtain the specific predictions for the modified scenarios
    reviewed in section \ref{revmodscen}. 
\end{itemize}
 
Concerning the analysis of the leading features of the bispectrum in
the squeezed limit, we will focus on the dependence on the probed large scale
$k_1^{-1}$ and on the magnitude of the non-Gaussianities. Indeed, those
are the most interesting features for observational purposes
(for example regarding the halo bias) \cite{Verde:2009hy,
  Schmidt:2010gw}. 

The results(\ref{corrgenbispmv})-(\ref{fNLcorrgenbispmv}) and 
(\ref{corrgenbispmdr})-(\ref{fNLcorrgenbispmdr}) show that the
bispectrum has different kinds of behaviour subject to the interplay
between the squeezed momentum scale $k_1$ and 
the scales/times $\Lambda$/$\eta_c$ of new physics. In particular, the
different behaviours occur depending on whether $|k_1\eta_c^{(k_S)}|$ or 
${k_1 \ov k_S}{\Lambda \ov H}$ are larger or smaller than 1. 
We will study the various cases separately.

\subsection{Enhancements}\label{enhancsignsec}
We investigate here the possibility of enhancements. 
As we have adopted a
phenomenological approach, we do not have the full-detailed knowledge
of the magnitude of the Bogoliubov coefficient $\beta_{k_S}$, which
the leading contributions to  
the bispectrum depend upon, see 
(\ref{corrgenbispmv})-(\ref{fNLcorrgenbispmv}) and 
(\ref{corrgenbispmdr})-(\ref{fNLcorrgenbispmdr}). However, we know
the constraints that it 
has to satisfy for phenomenological reasons and 
self-consistency of the theory (see section \ref{constraintsonBog}).
Therefore we can estimate the largest possible enhancements. 

Also, it is noteworthy that the leading contributions depend only on
$\beta_{k_S}$, which is independent of $k_1$,
see (\ref{fNLcorrgenbispmv}), (\ref{fNLcorrgenbispmdr}). This means that
our 
ignorance of the specific form of $\beta_{k_S}$ does not affect our
knowledge of
the $k_1$-dependence of the leading contributions.  

We turn now to the detailed analysis.

\paragraph*{{\em Cases $
    |k_1\eta_c^{(k_s)}|\ll 1 \;
    \text{and}\; {k_1 \ov k_S}{\Lambda \ov H} \ll 1$}} 
~~

In these cases the corrections to the bispectrum
(\ref{corrgenbispmv})-(\ref{fNLcorrgenbispmv}), 
(\ref{corrgenbispmdr})-(\ref{fNLcorrgenbispmdr})
for a generic cubic coupling contribute a 
parameter\footnote{As before, 'mis' indicates the modified initial state
  scenarios, while 'mdr' those with modified dispersion
  relations.}
$f_{NL} \sim \mathcal{B}^{^\text{mis, mdr}}_{_{\lambda_{\text{nms}}}}$, where, at leading order,
 \be \label{BsmallketacLovH} 
  \setlength{\jot}{12pt} 
  |\mathcal{B}^{^\text{mis, mdr}}_{_{\lambda_{\text{nms}}}}| & \sim
  \epsilon \, |\beta_{k_S}|\,{\lambda_{_{nms}} \ov \sqrt{2\epsilon}}
   {\Lambda M_{_\text{Planck}} \ov H^2}
    \bigl({k_1 \ov k_S}\bigr)^{1+\text{min}(n, m,s)}
  \,\Bigl({H\,D \ov \Lambda}\Bigr)^{n+m+s}
  \Bigl(\!1\!+\!\Bigl({k_1 D \ov k_S}\Bigr)^{n^{{\text{{\sstiny$(\!1\!)$}}}}_{_{t}}\text{-}1}\!\mathcal{O}(1)\!\!\Bigr)
  \\ 
  & \label{boundBsmallketacLovH} \leq
  \epsilon\sqrt{|\mu|} \, \,{\lambda_{_{nms}} \ov \sqrt{2}}
   \,   {M_{_\text{Planck}}^2 \ov H \Lambda}
    \bigl({k_1 \ov k_S}\bigr)^{1+\text{min}(n, m,s)}
   \,\Bigl({H\,D \ov \Lambda}\Bigr)^{n+m+s} 
  \Bigl(\!1\!+\!\Bigl({k_1 D \ov k_S}\Bigr)^{n^{{\text{{\sstiny$(\!1\!)$}}}}_{_{t}}\text{-}1}\!\mathcal{O}(1)\!\!\Bigr)
  \,
 \ee
and where $D = |k_S\eta_c^{(k_S)}|$ in the modified initial state case, while
$D ={\Lambda \ov H}$ in the scenario with modified dispersion
relations\footnote{However, in the NPHS approach   
 $|k_S\eta_c^{(k_S)}| ={\Lambda \ov H}$.}.
For the final inequality (\ref{boundBsmallketacLovH}) we have used the constraint
(\ref{betaconstraintnobackreaction}) on $|\beta_{k_S}|$.

Writing the above expressions using a common symbol $D$ for both kinds of
modified scenarios
is useful to make evident the similarity of
the results in the two different cases. As we remarked, this similarity
is an indication that the squeezed limit is dominated by the generic
features common to both of them (particle content/creation, interference
effects). 

We analyse now in details the features of
(\ref{boundBsmallketacLovH}), paying particular attention to the
possibility of enhancements (we will focus on the scale dependence in
section \ref{finalk1dependenceanalysis}). 

If we consider first $D =
|k_S\eta_c^{(k_S)}|$ (modified initial state), the inequality
(\ref{boundBsmallketacLovH}) reads   
  \beq \label{BforDketac}
  |\mathcal{B}^{^\text{$D\!\!=\!\!|k_S\eta_c|$}}_{_{\lambda_{\text{nms}}}}| \leq
  \epsilon\sqrt{|\mu|} \, \,{\lambda_{_{nms}} \ov \sqrt{2}}
   \,   {M_{_\text{Planck}}^2 \ov H \Lambda}
    \bigl({k_1 \ov k_S}\bigr)^{1+\text{min}(n, m,s)}
   \,\Bigl({|p_{_S}(\eta_c)| \ov \Lambda}\Bigr)^{n+m+s} 
  \Bigl(\!1\!+\!\Bigl({|p_{_1}(\eta_c)| \ov H}\Bigr)^{n^{{\text{{\sstiny$(\!1\!)$}}}}_{_{t}}\text{-}1}\!\mathcal{O}(1)\!\!\Bigr)
  \, ,
 \eeq
where $p_{_{S(1)}}(\eta_c) \equiv -k_{_{S(1)}}\eta_c^{(k_S)} H$ is the physical momentum associated with
the wavenumber $k_{_{S(1)}}$ at the boundary time $\eta_c^{(k_S)}$.

When instead $D = {\Lambda \ov H}$,
such as in the cases of modified dispersion relations or NPHS initial
state, the 
expression (\ref{boundBsmallketacLovH}) simplifies and we obtain 
  \beq \label{BforDLambdaH}
  |\mathcal{B}^{^\text{NPHS, mdr}}_{_{\lambda_{\text{nms}}}}| \leq
  \epsilon\sqrt{|\mu|} \, \,{\lambda_{_{nms}} \ov \sqrt{2}}
   \,   {M_{_\text{Planck}}^2 \ov H \Lambda}
    \bigl({k_1 \ov k_S}\bigr)^{1+\text{min}(n, m,s)} 
  \biggl(\!1\!+\!\Bigl({k_1 \ov k_S}{\Lambda \ov H}\Bigr)^{n^{{\text{{\sstiny$(\!1\!)$}}}}_{_{t}}\text{-}1}\!\mathcal{O}(1)\!\!\biggr)
  \, .
 \eeq

From (\ref{BforDketac}), (\ref{BforDLambdaH}), we can
understand various characteristics of the amplitude of the
non-Gaussianities when $|k_1\eta_{_c}^{(k_S)}|, \, {k_1 \ov k_S}{\Lambda \ov H} \ll 1$.
First of all, we see that (\ref{BforDketac}) is maximized for
$p_{_S}(\eta_c) = \Lambda$, that is, when the physical momentum
associated with $k_{_S}$ 
at $\eta_{_c}^{(k_S)}$ is really at the largest scale
that we can trust our effective theory at: the scale of new physics
$\Lambda$. 
This is precisely what occurs in the NPHS scenario of
initial state modification: the initial condition at
$\eta_{_c}^{(k_S)}$ is fixed by construction when the physical
momentum is at the cutoff scale $\Lambda$,
and we get (\ref{BforDLambdaH}). 

In the case of modified dispersion relations, equation
(\ref{BforDLambdaH}) can be re-written in a way that makes the similarity
with the modified initial case even more evident. It is
straightforward to obtain an equation which is analogous to
(\ref{BforDketac}), but where the place of
$p_{_S}(\eta_c)$ is taken by the physical momentum
$p_{_S}(\eta^{^{(k_S)}}_{\rii II})$ associated with $k_S$ 
at the particle creation/WKB breaking time 
$\eta^{^{(k_S)}}_{\rii II}$, which indeed has a magnitude of the order of 
$\Lambda$, see (\ref{BforDLambdaH}). This very simple exercise of
rewriting is not useless, as it is another indication of the universality
of the features of the squeezed limit in the modified scenarios.

Coming back to the question whether there can be actual overall enhancements,
potentially interesting for 
observations, one needs to take into account all the factors in
the equations (\ref{BforDketac}), (\ref{BforDLambdaH}) for
$\mathcal{B}^{\text{mis, mdr}}_\lambda$: both the 
enhancing (${M_{\text{Planck}}^2 \ov H \Lambda}$) and the suppressing
($\lambda_{_{nms}}\,\epsilon\sqrt{|\mu|} \, 
\bigl({k_1 \ov k_S}\bigr)^{1+\text{min}(n, m,s)}$). As for the latter
ones, it is immediately evident that terms with higher number 
of derivatives are more suppressed since $\text{min}(n, m,s)$
will be higher. This tells us that the contributions from  the
higher-derivative Lorentz-breaking couplings in the scenarios with
modified dispersion relations, where $n, m, s$ can assume values
also much larger that one, will be suppressed. 

Considering then the possible values of $n, m, s$, see section
\ref{gencoupsec}, it turns out, 
as it was in fact imaginable, that the cubic couplings 
(\ref{HIcubic}) and 
(\ref{hamilthigder}) that we have discussed as detailed examples
yield (among) the least suppressed contributions. Indeed, they
lead to a suppression respectively of order
$\bigl({k_1 \ov k_S}\bigr)^0$ and $\bigl({k_1 \ov k_S}\bigr)^{2}$, see
(\ref{classdetailedexamplesmincub}),
(\ref{classdetailedexamplesquartder}). In particular, the minimal
cubic coupling (\ref{classdetailedexamplesmincub}) is not actually
suppressed by factors ${k_1 \ov k_S}$. 
In fact, that coupling is leading also in the 
standard scenario, see \cite{Creminelli:2011rh}.

If we concentrate on these dominant (less suppressed) couplings, which
we had studied in details as examples in sections \ref{mincoupcubsec} and
   \ref{highdercubsec}, we find that there can indeed be actual overall 
enhancement of the non-Gaussianities with respect to the standard
scenario for certain values of the scale of new physics. 
Let us discuss a concrete example: take 
$\epsilon \sim |\mu|\sim 10^{-2}$, $H \sim 10^{-5}
M_{_\text{Planck}}$, compatible with the results of WMAP, and  
$\Lambda \sim 10^{-3} M_{_\text{Planck}}$, which 
corresponds to the supersymmetric GUT scale. Then, from
(\ref{BforDLambdaH}), if we can probe down to 
${k_1 \ov k_S} \lesssim 10^{-2}$,  
the largest possible enhanced amplitude 
that we could observe  
would be roughly of the order of  
 \beq \label{enahncementmincubcoupfinalan} 
  |\mathcal{B}^{^\text{NPHS, mdr}}_{_\text{min cub}}| \sim 10\,\epsilon \,
   \sim  10   \; (1-n_s) \, .  
 \eeq
in the case of the minimal coupling
(\ref{classdetailedexamplesmincub}), while it will be
 \beq
  |\mathcal{B}^{^\text{NPHS, mdr}}_{_\text{quart der}}| \sim 10^3 g \; (1-n_s) \sim 10  \; (1-n_s)
 \eeq
in the case of the quartic derivative interaction
(\ref{classdetailedexamplesquartder}) with $g \sim 10^{-2}$. 

It is noteworthy that 
for the higher (quartic)-derivative interaction the additional suppressing factor
$\bigl({k_1 \ov k_S}\bigr)^{2}$ and enhancing factor
${M_{\text{Planck}}^2 \ov \Lambda^2}$ (essentially due to the
power of $a^{-1}$ scaling) compensate each other to give a result of the same
magnitude as that of
the minimal cubic coupling. This result is quite different from the
one in the standard scenario, where the $a^{-1}$ scaling of the
interactions does not have any important effect since the
time integral in the bispectrum is dominated by late times.

Having obtained these field theory results, one should study the
implications for the actual experiments. This goes beyond the scope
of this work, and we leave it for future research.
However, we observe that, according to the analysis of \cite{Creminelli:2011rh}, 
values as ${k_1 \ov k_S} \sim 10^{-2}$ are indeed within the
reach of LSS  investigations, such as for example EUCLID, which in the
most optimistic estimates can probe down to ${k_1 \ov k_S} <
10^{-3}$ \cite{Creminelli:2011rh}. 

\paragraph*{{\em Cases $
  |k_1\eta_c^{(k_s)}| \gg 1 \;
  \text{and} \; {k_1 \ov k_S}{\Lambda \ov H} \gg 1$}} 
~~

In this case the correction 
(\ref{corrgenbispmv})-(\ref{fNLcorrgenbispmv}),
(\ref{corrgenbispmdr})-(\ref{fNLcorrgenbispmdr})
to the bispectrum for a
generic cubic coupling contributes a
parameter $f_{NL} \sim \mathcal{B}^{^\text{mis, mdr}}_{_{\lambda_{\text{nms}}}}$ 
with, at leading order, 
 \be \label{BbigketacLovH}
  |\mathcal{B}^{^\text{mis, mdr}}_{_{\lambda_{\text{nms}}}}| & \sim
  \epsilon \, |\beta_{k_S}|\,{\lambda_{_{nms}} \ov \sqrt{2\epsilon}}
   {\Lambda M_{_\text{Planck}} \ov H^2}
    \bigl({k_1 \ov k_S}\bigr)^{1+\text{min}(n, m,s)}E^{(n, m, s)}({H \ov \Lambda}{k_S \ov k_1})
  \\[0.12cm]
  & \label{boundBbigketacLovH} \leq
  \epsilon\sqrt{|\mu|} \, \,{\lambda_{_{nms}} \ov \sqrt{2}}
   \,   {M_{_\text{Planck}}^2 \ov H \Lambda}
    \bigl({k_1 \ov k_S}\bigr)^{1+\text{min}(n, m,s)}E^{(n, m, s)}({H \ov \Lambda}{k_S \ov k_1})
 \ee
where in the last inequality we have used the constraint
(\ref{betaconstraintnobackreaction}) on $|\beta_{k_S}|$.  
Here,
 \beq \label{integralresultlargeketacLambdaHx1}
  E^{(n, m, s)}({H \ov \Lambda}{k_S \ov k_1}) = {\textstyle\!\sum\limits_{j=2}^3}
  \begin{cases}
   \,\Bigl({H \ov \Lambda}{k_S \ov k_1}\Bigr)^{n+m+s} 
   {\textstyle\bigl(\!\mathcal{O}(1)\!\!+\!\! 
    \sum_{r=0}^\infty
    \mathcal{O}\bigl(({H \ov \Lambda}{k_S \ov k_1})^{^{\!\!\text{{\tiny$\kappa\!\!+\!\!r$}}}}\bigr)
   \bigr)_{_{[\theta_{\text{\stiny \!$j$}}\!]}}}
   & {\textstyle\text{{\small if $1\!+\!\cos(\theta_j)>{H \ov \Lambda}{k_S \ov k_1}$}}} \\[0.2cm]
   \left[\! 
   \begin{aligned} 
   & {\textstyle\!
   \,\bigl({H\,D \ov \Lambda}\bigr)^{n+m+s}
   \bigl(\!1\!+\!\bigl({k_1 D \ov k_S}\bigr)^{n^{{\text{{\sstiny$(\!1\!)$}}}}_{_{t}}\text{-}1}\!\mathcal{O}(1)\bigr)_{_{[\theta_{\text{\stiny \!$j$}}\text{\stiny $\sim$}\pi]}}} 
    && {\textstyle\text{{\small mis}}} \\
   & 
   {\textstyle\!
   \bigl({H \ov \Lambda}{k_S \ov k_1}\bigr)^{{n+m+s \ov \kappa+1}}
   }
   {\textstyle\bigl(\!1\!\!+\!\! 
    \textstyle{\sum_{r=0}^\infty}
    \mathcal{O}\bigl(({H \ov \Lambda}{k_S \ov k_1})^{^{\!\!\text{{\tiny${\kappa+r \ov \kappa+1}$}}}}\bigr)
    \!\bigr)_{_{[\theta_{\text{\stiny \!$j$}}\text{\stiny $\sim$}\pi]}}} 
    && {\textstyle \text{{\small mdr}}}
   \end{aligned} \!\right]
  & {\textstyle \text{{\small if $1\!+\!\cos(\theta_j) < {H \ov \Lambda}{k_S \ov k_1}$}}} 
  \end{cases},
 \eeq
where we have emphasized the orders of magnitude,
as we are interested in the enhancements --for the detailed
expressions see (\ref{fNLcorrgenbispmv}),(\ref{fNLcorrgenbispmdr}) -- 
(the $\theta_j$ dependence is indicated by the label
$_{[\theta_{\text{\stiny \!$j$}}]}$).  

A few features are evident at first sight from
(\ref{boundBbigketacLovH}), (\ref{integralresultlargeketacLambdaHx1}). 
First of all, the powers of 
$\bigl({H \ov \Lambda}{k_S \ov k_1}\bigr)$ in equations
(\ref{boundBbigketacLovH}), (\ref{integralresultlargeketacLambdaHx1})
are suppressing factors, because we are considering
the case ${k_1 \ov k_S}{\Lambda \ov H} \gg 1$. 

Second, the (nearly) folded
configurations (singled out by the condition 
$1\!+\!\cos(\theta_j) < {H \ov \Lambda}{k_S \ov k_1}\ll 1$) are enhanced
with respect to the other configurations ($D$ has been defined below
equation (\ref{BsmallketacLovH})), although less so in the
 scenario of modified dispersion
relations. It follows, as we see from
(\ref{integralresultlargeketacLambdaHx1}), 
that the result has a non-trivial
$\theta_j$ dependence
(different enhancements for different values of
$\theta_j$, recall also
that we consider a realistic squeezed
limit where $k_1$ is small but nonzero).

Finally, we see that the higher the number of derivatives, the
more the suppression because of the factors in
(\ref{boundBbigketacLovH}),
(\ref{integralresultlargeketacLambdaHx1}). In this respect,  
considering the possible values of $n, m, s$, the cubic
couplings (\ref{HIcubic}) and (\ref{hamilthigder}) (see their
schematic form in
(\ref{classdetailedexamplesmincub}), 
(\ref{classdetailedexamplesquartder})), which we have
discussed as detailed examples, turn out again to be the most dominant
ones.    

Equation (\ref{boundBbigketacLovH}) also includes enhancement
factors such as ${M_{\text{Planck}}^2 \ov H \Lambda}$, therefore the
presence of actual overall enhancements depends once more on the 
values of the scales and quantities at play. 
To give some pointers about the 
possibilities, let us again consider the
same example as before and the couplings (\ref{HIcubic}) and
(\ref{hamilthigder}), summarized in (\ref{classdetailedexamplesmincub}),
(\ref{classdetailedexamplesquartder}). Recall that we take 
$\epsilon \sim |\mu|\sim 10^{-2}$, $H \sim 10^{-5}
M_{_\text{Planck}}$, compatible with the results of WMAP, and  
$\Lambda \sim 10^{-3} M_{_\text{Planck}}$ (supersymmetric GUT scale).

Then, if we can only probe values down to ${k_1 \ov k_S} \gtrsim
10^{-2}$, we are in the case ${k_1 \ov k_S}{\Lambda \ov H} > 1$ and
from (\ref{boundBbigketacLovH}),
(\ref{integralresultlargeketacLambdaHx1}),
(\ref{classdetailedexamplesmincub}) we find the bound, roughly, 
 \beq \label{BbigketacLovHmincubcoup} 
  |\mathcal{B}^{^\text{mis, mdr}}_{_\text{min cub}}| \lesssim  
   {\textstyle\!\sum\limits_{j=2}^3}
   \begin{cases}
    10^{-1} {k_S \ov k_1} (1 -n_s)
   & {\textstyle\text{{\small if $1\!+\!\cos(\theta_j)>{H \ov \Lambda}{1 \ov x_1}$}}} \\[0.2cm]
   \left[\! 
   \begin{aligned} 
   & 10  (1 -n_s)
    && {\textstyle\text{{\small mis}}} \\[0.12cm]
   & 
   10^{{\kappa\text{-}1 \ov \kappa+1}}
   \bigl({k_S \ov k_1}\bigr)^{{1 \ov \kappa+1}}
    && {\textstyle \text{{\small mdr}}}
   \end{aligned} \!\right]
   & {\textstyle \text{{\small if $1\!+\!\cos(\theta_j) < {H \ov \Lambda}{1 \ov x_1}$}}}    
   \end{cases} \, .
 \eeq
We see that for ${k_1 \ov k_S} \simeq 10^{-2}$ the folded configurations,
both in the modified initial state and modified dispersion relations scenarios,
could be ten times enhanced with respect to the standard result,
whereas the other configurations would present only a
moderate enhancement compared to the result in the standard scenario.

In the case of the quartic-derivative interaction (\ref{hamilthigder}),
(\ref{classdetailedexamplesquartder}),
for a coupling $g \sim 10^{-2}$, we find from 
(\ref{fNLcorrgenbispmv}),(\ref{fNLcorrgenbispmdr})
-- taking into account the
comment at the end of section 
\ref{gencoupmoddisprelsec} --, or directly from
(\ref{BfactmodvacthreepointHigDerSinglej}),
(\ref{BfactmoddisrelHigDer}) that
 \beq \label{BbigketacLovHqurtdercoup}
  |\mathcal{B}^{^\text{mis, mdr}}_{_\text{quart der}}| \lesssim
  {\textstyle\!\sum\limits_{j=2}^3}
  \begin{cases} 
    10^{-1} {k_S \ov k_1} (1 -n_s)
   & {\textstyle\text{{\small if $1\!+\!\cos(\theta_j)>{H \ov \Lambda}{1 \ov x_1}$}}} \\[0.2cm]
   \left[\! 
   \begin{aligned} 
   &  10^3 \bigl({k_1 \ov k_S}\bigr)  (1 -n_s)
    && {\textstyle\text{{\small mis}}} \\[0.2cm]
   & 
   10^{3\text{-}{4 \ov \kappa+1}}
   \bigl({k_S \ov k_1}\bigr)^{{2 \ov \kappa+1}-1} (1 -n_s)
    && {\textstyle \text{{\small mdr}}}
   \end{aligned} \!\right]
   & {\textstyle \text{{\small if $1\!+\!\cos(\theta_j) < {H \ov \Lambda}{1 \ov x_1}$}}}    
   \end{cases} \, .
 \eeq
Note that for ${k_1 \ov k_S} \simeq 10^{-2}$,
(\ref{BbigketacLovHmincubcoup}) and (\ref{BbigketacLovHqurtdercoup})
are of the same magnitude, which again shows that in the
modified scenarios the different
$a^{-1}$ scaling of the higher-derivative interactions can compensate
for the stronger suppression by higher powers of ${k_1 \ov k_S}$ and
make the squeezed limit sensitive to higher-derivative interactions.

\subsection{Dependence on $k_1$}\label{finalk1dependenceanalysis}

We discuss now the dependence on the squeezed wavenumber $k_1$. Recall
that the leading contributions depend only on $\beta_{k_S}$, which is
independent of $k_1$, see
(\ref{corrgenbispmv})-(\ref{fNLcorrgenbispmv}) and
(\ref{corrgenbispmdr})-(\ref{fNLcorrgenbispmdr}). Thus, the
$k_1$-dependence of the  
leading corrections to the bispectrum in the squeezed limit is not
affected by our ignorance 
of the specific form of $\beta_{k_S}$.   

In the following, we will rewrite equations
(\ref{corrgenbispmv})-(\ref{fNLcorrgenbispmv}) and
(\ref{corrgenbispmdr})-(\ref{fNLcorrgenbispmdr}) highlighting the 
dependence on $k_1$. We will neglect the momenta conserving delta
function to avoid cluttering formulas.
Again, we find that the field theoretical
results for the modified
scenarios have two different kinds of behaviour, depending on the magnitude
of the scale of new 
physics as compared to the sensitivity of the observations (smallest
$k_1$ that can be probed). We will stress the differences with respect to the
standard scenario, 
where the bispectrum grows in the squeezed limit as  $\sim   
k_1^{-3}$ \cite{Maldacena:2002vr, Creminelli:2004yq,
  Creminelli:2011rh}, and also compare to the result obtained with the approximated
template proposed in \cite{Meerburg:2009ys} for scenarios with modified
initial state, which leads to a bispectrum growing as $\sim k_1^{-2}$,
see \cite{Verde:2009hy, Schmidt:2010gw}. 

\paragraph*{{\em Cases $|k_1\eta_c^{(k_s)}| \ll 1 \;
  \text{and}\; {k_1 \ov k_S}{\Lambda \ov H} \ll 1$}} 
~~
\vspace{0.08cm}

In this cases, equations (\ref{corrgenbispmv})-(\ref{fNLcorrgenbispmv}) and
(\ref{corrgenbispmdr})-(\ref{fNLcorrgenbispmdr}) yield at leading order
 \beq \label{boundBsmallketacLovHdepek1}  
   \!\!\delta_\beta\langle\zeta_{\vk_1}(\eta)\zeta_{\vk_2}(\eta)\zeta_{\vk_3}(\eta)\rangle_{_{\lambda_{\text{nms}}}} \!\!
  =  \epsilon \, |\beta_{k_S}|\,{\lambda_{_{nms}} \ov \sqrt{2\epsilon}}
   \,   {\Lambda M_{_\text{Planck}} \ov H^2}
    \!\,\Bigl(\!{H D \ov \Lambda}\!\Bigr)^{^{\!\!\!\!n+m+s}}\!\! 
   \Bigl({k_1 \ov k_S}\Bigr)^{\!\!^{\!\!1+\text{min}(n, m,s)}}\!\!
  \Bigl(\!1\!+\!c_{n^{{\text{{\sstiny$(\!1\!)$}}}}_{_{t}}}\!\Bigl({k_1 D \ov k_S}\Bigr)^{\!\!^{\!\!n^{{\text{{\sstiny$\!(\!1\!)\!$}}}}_{_{t}}\text{-}1}}\!\Bigr)
  P_{\!\!_\text{st}}(k_1) P_{\!\!_\text{st}}(k_S), 
 \eeq
where, as before, $D=|k_{_S}\eta_{_c}^{(k_s)}|$ or ${\Lambda \ov H}$, respectively for modified
initial state or modified dispersion relations.

The corrections (\ref{boundBsmallketacLovHdepek1}) scale as
$k_1^{-3+\text{min}(n,m,s)+n^{{\text{{\sstiny$(\!1\!)$}}}}_{_{t}}}$  
(recall $n^{{\text{{\sstiny$(\!1\!)$}}}}_{_{t}} = \{0, 1\}$, defined
below equations (\ref{corrmodvacgen}), (\ref{corrmoddisprelgen})). We thus find that
higher-derivative couplings (larger values for $n, m, s$) grow more
slowly for small $k_1$. The contribution that grows the fastest is of the form 
 \beq \label{boundBsmallketacLovHdepek1fastest}  
  \delta_\beta\langle\zeta_{\vk_1}(\eta)\zeta_{\vk_2}(\eta)\zeta_{\vk_3}(\eta)\rangle_{_{\text{min}(n, m,s)=\text{-}1}} \!\!
  = \epsilon \, |\beta_{k_S}|\,{\lambda_{_{nms}} \ov \sqrt{2\epsilon}}
   \,   {\Lambda M_{_\text{Planck}} \ov H^2}
    \,\biggl({H\,D \ov \Lambda}\biggr)^{n+m+s}
  \, P_{_\text{st}}(k_1) P_{_\text{st}}(k_S) \, , 
 \eeq
that is, it has a local shape $k_1^{-3}$, and can be found for couplings such as the
minimal coupling cubic, where min$(n, m, s) =-1$ (recall that this
value occurs due to the combination $(\vec\partial)^{-2}\dot\varphi$,
therefore it is always accompanied by
$n^{{\text{{\sstiny$(\!1\!)$}}}}_{_{t}} = 1$).

Although it has a local form, the result is still
different from the one in the standard scenario because
non-Gaussianities are/can be enhanced by the particle content/creation 
concerning the perturbations depending on $k_{2, 3}$ as  we have
discussed in section \ref{enhancsignsec}.

\paragraph*{{\em Cases $|k_1\eta_c^{(k_s)}| \gg 1 \;
   \text{and}\; {k_1 \ov k_S}{\Lambda \ov H} \gg 1$}} 
~~

In this case, equations (\ref{corrgenbispmv})-(\ref{fNLcorrgenbispmv}) and
(\ref{corrgenbispmdr})-(\ref{fNLcorrgenbispmdr}) yield, at leading order,
 \begin{multline} \label{boundBlargeketacLovHdepek1}  
  \!\!\!\!\delta_\beta\langle\zeta_{\vk_1}(\eta)\zeta_{\vk_2}(\eta)\zeta_{\vk_3}(\eta)\rangle_{_{\lambda_{\text{nms}}}} \!\!
  =  \epsilon \, |\beta_{k_S}|\,{\lambda_{_{nms}} \ov \sqrt{2\epsilon}}
   \,   {\Lambda M_{_\text{Planck}} \ov H^2}
  \;\;\bigl({k_S \ov k_1}\bigr)^{-1-\text{min}(n, m,s)} \\
  \qquad\times\left.\!\!
  {\textstyle\!\sum\limits_{j=2}^3}
  \begin{cases}
   \!\!\Bigl({H \ov \Lambda}{k_S \ov k_1}\Bigr)^{n+m+s} 
   & {\textstyle\text{{\small if $1\!+\!\cos(\theta_j)>{H \ov \Lambda}{1 \ov x_1}$}}} \\[0.2cm]
   \!\!\left[\! 
   \begin{aligned} 
   & {\textstyle\!
   \,\bigl({H\,D \ov \Lambda}\bigr)^{n+m+s}
   \bigl(\!1\!+\!\text{{\footnotesize$c_{n^{{\text{{\sstiny$(\!1\!)$}}}}_{_{t}}}$}}\bigl({k_1 D \ov k_S}\bigr)^{\!\!n^{{\text{{\sstiny$(\!1\!)$}}}}_{_{t}}\text{-}1}\!\bigr)
    }
    && {\textstyle\text{{\small mis}}} \\[0.12cm]
   & 
   {\textstyle\!
   \bigl({H \ov \Lambda}{k_S \ov k_1}\bigr)^{{n+m+s \ov \kappa+1}}
   }
    && {\textstyle \text{{\small mdr}}}
   \end{aligned} \!\right]
  & {\textstyle \text{{\small if $1\!+\!\cos(\theta_j) < {H \ov \Lambda}{1 \ov x_1}$}}} 
  \end{cases} 
  \right\}
  P_{\!\!_\text{st}}(k_1) P_{\!\!_\text{st}}(k_S) \, .
 \end{multline}
Except for the case of folded configurations in modified
initial state scenarios, it appears from
(\ref{boundBlargeketacLovHdepek1}) that  
higher-derivative couplings (larger values for $n, m, s$) would lead
to a more pronounced growth for small $k_1$ since its exponent will be
more negative. However,
these contributions are very suppressed 
because of the factors $\bigl({H \ov \Lambda}\bigr)^{n+m+s}, 
\bigl({H \ov \Lambda}\bigr)^{n+m+s \ov \kappa+1}$. The
most interesting cases are thus those that are least
suppressed. Considering the possible values for $n, m, s$, see
section \ref{gencoupsec},
this happens for instance for the minimal cubic coupling that we
studied as a detailed 
example, where $n+m+s=1$ and $\text{min}(n, m,s)=-1$, see
(\ref{classdetailedexamplesmincub}), in which case (roughly, see
(\ref{consistrelmodvac})-(\ref{Bfactmodvacthreepoint}),  
(\ref{consistrelmoddisprelII})-(\ref{BfactmdrII})
for the details)
 \beq \label{boundBlargeketacLovHdepek1mincubcoup}  
  \!\!\!\!\delta_\beta\langle\zeta_{\vk_1}(\eta)\zeta_{\vk_2}(\eta)\zeta_{\vk_3}(\eta)\rangle_{_\text{min cub}} \!\!
  = \epsilon \, |\beta_{k_S}|
   \left.
  {\textstyle\!\sum\limits_{j=2}^3}
  \begin{cases}
   \,{k_S \ov k_1} 
   & {\textstyle\text{{\small if $1\!+\!\cos(\theta_j)>{H \ov \Lambda}{1 \ov x_1}$}}} \\[0.12cm]
   \!\!\left[\! 
   \begin{aligned} 
   & {\Lambda \ov H}
    && {\textstyle\text{{\small mis}}} \\
   & 
   {\textstyle\!
   \bigl({\Lambda \ov H}\bigr)^{{\kappa \ov \kappa+1}}
   \bigl({k_S \ov k_1}\bigr)^{{1 \ov \kappa+1}}
   }
    && {\textstyle \text{{\small mdr}}}
   \end{aligned} \!\right]
  & {\textstyle \text{{\small if $1\!+\!\cos(\theta_j) < {H \ov \Lambda}{1 \ov x_1}$}}} 
  \end{cases} 
  \right\}\!\!
  P_{\!\!_\text{st}}(k_1) P_{\!\!_\text{st}}(k_S) \, .
 \eeq
We see from (\ref{boundBlargeketacLovHdepek1mincubcoup}) that in this
case the form of the bispectrum in 
the squeezed limit has {\em i}) a non-local shape, with a
scaling as $k_1^{-4}$ and
with amplitude not too severely suppressed for non-folded
configurations, {\em ii}) a non-local shape and enhanced amplitude
(see also section  
\ref{enhancsignsec}) for (nearly) folded
configurations in the case of modified 
dispersion relations, and {\em iii}) a local shape
and greater enhancement
for folded configurations in the
modified initial state case. Interestingly, the scaling with $k_1$ 
of the contribution from (nearly) folded configurations in the case of modified
dispersion relation captures the power $\kappa$ of the first
momentum correction to the 
standard dispersion relation, see (\ref{boundBlargeketacLovHdepek1}). 

\subsection{Predictions for the different modified scenarios}\label{secpredicscena}

We have seen that, most interestingly, in modified scenarios the
leading features of the bispectrum 
in the squeezed limit depend on very general aspects of the theory
(exit from horizon, particle production, interference and accumulation
with time), due to the modification of the Whightman functions in
these scenarios. These aspects
cannot determine the bispectrum in the squeezed limit in all details, but, even
adopting a wide-range phenomenological approach, we have been able to determine
both the leading dependence on the squeezed scale $k_1$ and the
presence of enhancements of non-Gaussianities and their bounds.

We have seen that these features depend on the values (less or more
than 1) of the quantities
$|k_1\eta_{_c}^{(k_s)}|$ and ${k_1 \ov  k_S}{\Lambda \ov H}$. Note that
${k_1 \ov k_S}$ represents the sensitivity of the observation (the
smallest ratio of scales that can be probed). We would like now to
ask ourselves which kind of conditions or predictions the different
theoretical scenarios impose or make about the values of  
$|k_1\eta_{_c}^{(k_s)}|$ and ${k_1 \ov k_S}{\Lambda \ov H}$, and, thus, what
features of the bispectrum would be realized in the squeezed limit.
\begin{itemize}[leftmargin=0.4cm,itemsep=0.2cm,parsep=0.2cm]
\setlength{\parindent}{0.0cm}
 \item {\em BEFT modified initial state}. In the Boundary Effective
   Field Theory approach, the
initial condition for the fields is fixed at a time 
$\eta_c$ independently of the modes $k$ (``beginning
of inflation''). Thus $\eta_{_c}^{(k_S)}=\eta_{_c}^{(k_1)}=\eta_{_c}$
and since initial conditions are always fixed when modes were subhorizon,
then $|k_1\eta_{_c}^{(k_1)}| \gg 1 \Rightarrow
|k_1\eta_{_c}^{(k_S)}| \gg 1$.

Looking therefore at the analysis in sections \ref{enhancsignsec},
\ref{finalk1dependenceanalysis},  we are in the cases of equations
(\ref{BbigketacLovH}), (\ref{boundBlargeketacLovHdepek1}), and the
leading possible contribution is given by
(\ref{boundBlargeketacLovHdepek1mincubcoup}), see also
(\ref{BbigketacLovHmincubcoup}). 
This means that if the BEFT is the correct physical approach
to model the effects
of very high energy physics from a low-energy point of view,
then the largest possible contribution to the bispectrum
for non-folded configurations would grow at large scales as
$k_1^{-4}$, but with only moderately 
enhanced non-Gaussianities, whereas the contribution from folded
configurations would be more enhanced, but growing only as $k_1^{-3}$. 

\item {\em NPHS modified initial state}. In the NPHS approach
the boundary condition is mode-dependent, and 
imposed at the time
given by  $|k \, \eta_{_c}^{(k)}| = {\Lambda \ov H} \gg 1$.
The picture is 
that different modes are created at different times
in a certain condition that fixes the initial state of the
perturbations in the effective theory
at scales lower than $\Lambda$. 

When computing the 
three-point function, therefore, we cannot trace the interaction of
the modes backward in time for a time earlier than the ``creation time'' for
the largest of the considered modes: the prior evolution (and our
ignorance about it) is encoded in
the initial state. Thus, in the case of the squeezed
limit, since $\eta_{_c}^{(k_S)} \neq \eta_{_c}^{(k_1)}$,
it can happen that
$|k_1\eta_{_c}^{(k_{_S})}|={k_1 \ov k_{_S}}{\Lambda \ov H}$
is larger or smaller than 1 (super- or sub-horizon condition)
despite the fact that $|k_1\eta_{_c}^{(k_{1})}| \gg 1$, 
depending on the values of the scales at play in the specific model.

Therefore the bispectrum will behave in different ways for the various
possibilities. Looking at the analysis in sections \ref{enhancsignsec},
\ref{finalk1dependenceanalysis}, we see that the leading correction to
the bispectrum would grow at large scales as $k_1^{-4}$ for non-folded
configurations if the scale of
new physics is such that $\Lambda > {k_S \ov k_1}H$, where ${k_1 \ov k_S}$ is
the smallest ratio of scales that can be probed, see
(\ref{boundBlargeketacLovHdepek1}), 
(\ref{boundBlargeketacLovHdepek1mincubcoup}). On the other hand, for folded
configurations, or when $\Lambda < {k_S \ov k_1}H$ (superhorizon condition), the
contribution would be of the local type, growing at most as
$k_1^{-3}$ as in the standard case, but will be enhanced, see 
(\ref{boundBsmallketacLovHdepek1}),
(\ref{boundBsmallketacLovHdepek1fastest}),
(\ref{boundBlargeketacLovHdepek1}), 
(\ref{boundBlargeketacLovHdepek1mincubcoup}),
(\ref{enahncementmincubcoupfinalan}).

An interesting outcome of this result is that since we have different
signatures depending on whether ${k_1 \ov k_{_S}}{\Lambda \ov H}$ is
larger or smaller than 1, in case we could
detect one or the other, say in the halo bias, and separate
them from other kinds of non-Gaussianities (such as the secondary
ones), we would obtain 
information on the magnitude of $\Lambda$ given our knowledge of the
observation sensitivity ${k_1 \ov k_S}$. 

\item {\em Modified dispersion relations}. In the scenarios with
  modified dispersion relations violating WKB at early times, 
  ${k_1 \ov k_S}{\Lambda \ov H}$ can be either smaller or greater than 1,
  depending on the value of the Lorentz-breaking scale $\Lambda$ and
  the sensitivity ${k_1 \ov k_S}$ of the observation (smallest ratio that can be probed).
  In the case when
  ${k_1 \ov k_S}{\Lambda \ov H} > 1$,
  there would be contributions to the bispectrum growing as
  fast as $k_1^{-3-N}$ with $N$ only bound by the specific model, and in
  principle also possibly large, due to the presence of
  higher-derivative Lorentz-breaking 
  couplings, see
  (\ref{boundBlargeketacLovHdepek1}). However, we have found that those
  contributions are severely suppressed by powers 
  $\bigl({H \ov \Lambda}\bigr)^M$, $M \geq N$. 

  The least-suppressed possible
  contributions (which could actually be 
  enhanced) show at leading order a growth as $k_1^{-4}$ for non-folded
  configurations, and as slower powers $k_1^{-r}$, with $r$ fractional, for the
  (nearly) folded ones, see (\ref{boundBlargeketacLovHdepek1}), 
  (\ref{boundBlargeketacLovHdepek1mincubcoup}). Remarkably, in the
  latter case, the exponent of $k_1$ is determined by the momentum power of
  the lowest high-energy correction to the standard dispersion
  relation, see (\ref{boundBlargeketacLovHdepek1mincubcoup}),   
  (\ref{kappafirstdispcorrec}),
  and it would thus be of great observational interest.

  On the other hand, in the case 
  $\Lambda < {k_S \ov k_1}H$, the leading contribution 
  would be of the local type growing at most as $k_1^{-3}$,
  see (\ref{boundBsmallketacLovHdepek1fastest}), but it could be
  enhanced, see (\ref{enahncementmincubcoupfinalan}).  

  Once again, if we
  could observe one or the other behaviour of the bispectrum, we could
  infer whether 
  ${k_1 \ov k_S}{\Lambda \ov H}$ is either larger or smaller than 1,
  and thus obtain
  information on the magnitude of the Lorentz-breaking scale from our
  knowledge of the probed $k_1$. 
\end{itemize}

\section{Comments on the halo bias}\label{halobiassec}

We will briefly comment here on the implications of
our results for the halo bias, leaving a more
detailed analysis for future research.
The halo bias depends on the bispectrum and the spectrum as
\cite{Verde:2009hy}\footnote{The formula (\ref{halobias}), obtained in 
\cite{Verde:2009hy}, is calculated in a local-Lagrangian-biasing
scheme. Reference \cite{Schmidt:2010gw} has a similar
but not identical formula calculated using a point-break splitting
method. As discussed in \cite{Schmidt:2010gw},
the formulas agree for large scales, but differ for
intermediate scale, where the point-break splitting method involves
more important approximations. It seems however that the formula
(\ref{halobias}) is the one that is better in agreement 
with the simulations \cite{Wagner:2010me}, see also
\cite{Creminelli:2011rh}.}  
 \begin{multline} \label{halobias}
  {\Delta b_h(k_1,R) \ov b_h} = {\delta_c \ov D(z)\mathcal{M}_R(k_1)} 
   {1 \ov 8\pi^2\sigma_R^2}\int_0^\infty dk_2 \, k_2^2 \mathcal{M}_R(k_2) \\ 
   \times\int_{-1}^1d\xi\, \mathcal{M}_R\bigl(\sqrt{k_1^2+k_2^2+2 k_1 k_2 \xi}\bigr) 
    {B(k_2,\sqrt{k_1^2+k_2^2+2 k_1 k_2 \xi},k_1) \ov P(k_1)} \, ,
\end{multline}
where $\sigma_R^2$ is the variance of the dark matter density
perturbations smoothed on a scale of Lagrangian radius $R$, $\delta_c$
is the critical threshold for the collapse of a spherical object
(for a matter-dominated universe $\delta_c = 1.686$), $D(z)$ is the
linear growth factor normalized to be $D(z) = (1+z)^{-1}$ during
matter domination, $P$ and $B$ are the 
two-point correlator\footnote{In the literature on the halo bias, the
  two-point correlator is often called the spectrum \cite{Verde:2009hy,
    Schmidt:2010gw}, not to be confused with the quantity 
    $\mathcal{P} = {k^3 \ov 2\pi^2}P$, also called spectrum, usual in the CMBR
literature.} and bispectrum for $\zeta$, see equations
(\ref{twopointdefgen})-(\ref{shape-func}), and 
$\mathcal{M}_R$ is the linear relation between the dark matter density
perturbations smoothed on a scale $R$ and the primordial curvature
perturbation: 
 \beq \label{darkmatterrelprimcurvpert}
  \mathcal{M}_R(k) \equiv {2 k^2 \ov 5\Omega_m H_0^2}T(k) W_R(k) \, ,
 \eeq
where $\Omega_m$ is the present time fractional density of matter. Finally,
$H_0$ is the present Hubble rate, $T$ is the transfer function
normalized to one on large scales and $W_R$ is the filter function with
characteristic scale $R$.  
 
The halo bias depends on both the primordial bispectrum and the
subsequent gravitational processing which generates additional
non-Gaussianities. These General Relativity corrections have been
studied for example in \cite{Verde:2009hy} and it has been shown that
they give a typical contribution to the halo bias, which could
permit to distinguish them. We will therefore comment only the
primordial contributions. 

The behaviour of the halo bias at large scales has been computed for
the local, equilateral and folded {\em templates} for the primordial
non-Gaussianities in
\cite{Verde:2009hy} (see also \cite{Schmidt:2010gw, Shandera:2010ei,
Desjacques:2009jb, Wagner:2010me, Smith:2010gx}), finding that the
bias goes approximately as $k_1^{-2}$ with the 
local bispectrum template, as a constant in $k_1$ with the
equilateral, and as $k_1^{-1}$ with the folded. 

The latter result is the one that can be compared with those we have
found for the scenario of the modified initial state scenario, since
that template was indeed proposed
as an approximation to model the bispectrum in that case
\cite{Meerburg:2009ys}. In fact, we have shown that it leads to the
wrong result for the squeezed limit:  the folded bispectrum
template goes as $k_1^{-2}$, which is very different from the true
bispectrum behaviour found in sections
\ref{bispsqueezlimsec} and \ref{signaturessec}. The reason for this is that
the template does not
depend on the additional 
scale(s) $\eta_c, \Lambda$ and thus its squeezed limit is
different than the one that actually occurs in the rigorous field theory
computation. 

We have indeed found in sections
\ref{bispsqueezlimsec} and \ref{signaturessec}
that in the squeezed limit the leading contribution
to the bispectrum grows as $k_1^{-4}$ or $k_1^{-3}$, depending on the
compared magnitudes of the scale of new physics and the sensitivity of
the observation (the smallest observable squeezed momentum
scale). 
Therefore, from (\ref{halobias}) and  the results in sections
\ref{bispsqueezlimsec} and \ref{signaturessec},
the halo bias for modified initial states would approximately
grow either as $k_1^{-3}$ or 
$k_1^{-2}$ for small $k_1$, and not as $k_1^{-1}$ as predicted by the
template.   

Similarly, in the scenario of
modified dispersion relations violating WKB at early times, equation
(\ref{halobias}) and the results in sections
\ref{bispsqueezlimsec} and \ref{signaturessec},
would also lead to a bias that approximately grows either as 
$k_1^{-3}$ or $k_1^{-2}$ for small $k_1$ for non-folded
configurations, again depending on the  
relative magnitudes of the high-energy scales and the sensitivity of 
the observations. 
(Nearly) folded configurations would 
lead to different, subleading 
scalings with $k_1$, determined by the 
exponent of the first correction to the standard dispersion relation
for small momenta. 

As for the magnitude of the halo bias, a feature relevant
for the observations of the bias is that, as we have seen, 
the high-energy modifications could enhance the single-field
non-Gaussianities in the squeezed limit, compared to the practically 
undetectable level that is present in the standard scenarios of single-field
models with Bunch-Davies vacuum and standard kinetic terms. However,
the final magnitude of the non-Gaussianities depends also on the
specific model, which determines the precise form and magnitude of the
coefficients $\beta_{k_S}$ entering the equations.  

\section{Final discussion and conclusion}\label{conclusions}

In this article we have analysed the squeezed limit of the bispectrum in
single-fields models of inflation when modifications of the theory at
very high energy are present. Due to the large number of inflationary models
available, it is useful to look at general
features that can be obtained when these modifications are
rather generic and simple. 

In particular, we have considered 1) the
presence of terms with higher derivatives, which are not suppressed any more
beyond a certain momentum scale and modify the dispersion relations
of the fields, and 2) the possibility of initial 
condition for the solution of the field equation that parametrize in a
simple and generic way our ignorance on the physics at very high
energies. In the case 1) the most significant results occur for
dispersion relation that violate WKB (for a short time) at early
times; in the case 2) we have considered both NPHS and BEFT
approaches, where the initial condition is imposed respectively in a
scale-invariant and non-scale-invariant way. 

The leading contribution to the squeezed limit appears
dominated by very general features of the scenarios (particle
content/creation at early times, interference and cumulation with time) and
thus it is possible to obtain general results and bounds. From the
more mathematical point of view, the reason for the leading differences with the
standard result is the modifications of the Whightman function of the
theory.

The result and bounds that we have found are interesting in two
respects: on the one 
hand, they show that non-Gaussianities could be enhanced by
high energy modifications of the theory, which would improve the
possibility of detection of those phenomena. On the other, the behaviour of the
bispectrum at large scale in these scenarios is distinctive, and
differs from the one that has been 
found in the standard scenarios, with Bunch-Davies vacuum and
standard dispersion relation. It is also different from that found in
previous works using the approximate template proposed in  
\cite{Meerburg:2009ys} for the modified initial state approach. 

Our results show that it could be possible to obtain various different
pieces of information
about the features of the high energy physics theory from large scale
observations such as those of the halo bias, if the signatures we have
found were detected. In fact, the enhancements and the
dependence of the bispectrum on the small wavenumber $k_1$ in the
realistic (observable) squeezed limit change 
in relation with the magnitude of the scale of new physics. Hence, if able to
detect certain behaviours of the bispectrum, one can place bounds on
the new scale of physics.

Of course, other sources could give origin to sizable
non-Gaussianities, for example in multi-field
models. An interesting question for future research would then be if
the effects due to high-energy modifications of the 
theory that we have considered  
can be distinguished from those. For example, the dependence on the
small mode $k_1$ of 
the halo bias could be different, or the enhancement factors could be
absent or of different magnitude. We leave this as an interesting
outlook for future research.

\section*{Acknowledgments}

The author is supported by a Postdoctoral F.R.S.-F.N.R.S. research
fellowship via the Ulysses Incentive Grant for the Mobility in Science
and is ``chercheur de recherche'' F.R.S.-F.N.R.S..

\appendix

\section{Appendices}

\subsection{Solution of the field equation in the case of modified
  dispersion relations}\label{solutionmoddisprel}

We parametrize the
modifications to the dispersion relation via a {\em generic} function 
$F\left({H \ov \Lambda}k\eta\right)$:
   \beq \label{powerexp}  
    \omega(\eta, k) =  a(\eta)\omega_{_\text{phys}}(p) 
      =  a(\eta) p \;  F\left(-{p \ov \Lambda}\right) 
      = k \;  F\left({H \ov \Lambda}k\eta\right) \,,
    \qquad F(x \to 0) \to 1 \, ,
    \qquad {H \ov \Lambda} \ll 1 \, ,
   \eeq
with the only request that the WKB conditions be violated for a short
period of time at early times. The generic shape of such a dispersion
relation is in figure \ref{INOUTdispersion}.
The approximate solution of the equation of motion (\ref{eqofmo}) in
these cases is given in equation (\ref{piecewisesolution}), where, see
\cite{Chialva:2011iz}: 
\begin{itemize}
 \item in regions {\rm I} and {\rm III}, in terms of the convenient
   variable ${H \ov \Lambda}k\eta \equiv y_k$:
   \beq \label{solutionWKB}
    u_1(y_k) =
    {e^{-i{\Lambda \ov H}\Omega_{\text{{\tiny F}}}(y_k)} \ov \sqrt{2 \, k \, U(y_k)}} \, , 
    \qquad
    u_2(y_k) = u_1^*(y_k) =  
    {e^{+i{\Lambda \ov H}\Omega_{\text{{\tiny F}}}(y_k)} \ov \sqrt{2 \, k \, U(y_k)}}
    \, ,
   \eeq
  with
   \beq \label{Omegaphase}
    \Omega_{\text{{\tiny F}}}(y_k)=\int^{y_k} U(y_k') dy_k'
    \, , \qquad 
    U(y_k) \equiv F + \epsilon^2 \biggl(-{\partial_y^2 F \ov 4 F^2}
    + 3{(\partial_y F)^{2} \ov 8 F^3} - {1 \ov F y_k^2}\biggr) \, ,
   \eeq
 \item in region {\rm IV}, where $\omega(\eta, k) \sim k$ and
   $k\eta \lesssim 1$,
   \beq
    \mathcal{V}_{1, 2} = \sqrt{-\eta} H^{(1, 2)}_\nu(-k\eta)\bigr|_{k\eta \lesssim 1}, 
    \qquad \nu \sim {3 \ov 2}
   \eeq
 \item in region {\rm II}, the partial solution $\mathcal{U}_{1, 2}(\eta, k)$ can be 
   found either by solving in details the specific equation, if one
   has a favourite model, or, more
   generally, by expanding the frequency around the minimum 
   $\omega_0$ at $\eta = \eta_{\text{min}}$ as 
   \beq \label{omeganearmin}
    \omega^2 = a^2 \omega_{_\text{phys}}^2 \simeq 
     {1 \ov \eta^2 H^2}\biggl(\omega_0^2+  
     \omega_0\omega_{_\text{phys}}^{''}(\eta_{\text{min}})(\eta -\eta_{\text{min}})^2 \biggr)\, ,
   \eeq
  so that
   \begin{gather}
    f_k(\eta)_{\bigr|_{\rm II}} = \sum_{i = 1}^2 B_i \, \mathcal{U}_{i}(\eta, k) 
     = B_1 \; W(i \kappa \,\eta_{\text{min}} , \sigma, 2i\kappa\,\eta)
    + B_2 \; W(-i \kappa\, \eta_{\text{min}} , \sigma, -2i\kappa\,\eta)
    \nonumber \\  
    \kappa  \equiv {\sqrt{\omega_0 \omega_{_\text{phys}}^{''}(\eta_{\text{min}})} \ov H} \qquad  
    \sigma \equiv {\sqrt{9H^2-4\omega_0^2-4 H^2 \kappa^2 \eta_{\text{min}}^2} \ov 2H} \, ,
   \end{gather}
  where $W(a, b, z)$ is the Whittacker function. As explained
  in \cite{Chialva:2011iz}, this approximation is well-justified also because
  backreaction constraints the interval $[\eta_{\ri I}, \eta_{\rii II}]$
  to be very small ($\Delta < 1$), so that $\eta \sim \eta_{\text{min}}$ for $\eta, \eta_{\text{min}}
  \in [\eta_{\ri I}, \eta_{\rii II}]$.
  In any case, it actually turns out that the details of the solution in region
  {\rm II} are not important for what concerns the leading
  contribution to the bispectrum \cite{Chialva:2011iz}. 
\end{itemize}

By asking for the continuity of the
function and its first derivative, and imposing the Wronskian condition $\mathcal{W}\{f, f^*\}
= -i$ to have the standard commutation relations in the quantum theory,
we obtain in full generality  
 \be \label{Dcoef}
  D_1 & = {\sqrt{\pi} \ov 2} e^{i{\pi \ov 2}\nu+i{\pi \ov 4}}\alpha_k &
  D_2 & = {\sqrt{\pi} \ov 2} e^{-i{\pi \ov 2}\nu-i{\pi \ov 4}}\beta_k \\
  \label{BogolBcoeffgener}
  B_1 & = {\mathcal{W}\{\varsigma_k \, u_1, \mathcal{U}_2\} \ov \mathcal{W}\{\mathcal{U}_1, \mathcal{U}_2\}}\biggr|_{\eta_{\ri I}} &
  B_2 & = -{\mathcal{W}\{\varsigma_k \, u_1, \mathcal{U}_1\} \ov \mathcal{W}\{\mathcal{U}_1, \mathcal{U}_2\}}\biggr|_{\eta_{\ri I}} \\
  \label{Bogolalpcoeffgener}
  \alpha_k & = {\mathcal{W}\{B_1 \, \mathcal{U}_1 + B_2 \, \mathcal{U}_2, u_2\} \ov \mathcal{W}\{u_1, u_2\}}\biggr|_{\eta_{\rii II}} &
  \beta_k & = -{\mathcal{W}\{B_1 \, \mathcal{U}_1 + B_2 \, \mathcal{U}_2, u_1\} \ov \mathcal{W}\{u_1, u_2\}}\biggr|_{\eta_{\rii II}} \, ,
 \ee
where $\mathcal{W}$ is the Wronskian. We choose $\varsigma_k =1$
picking up the usual adiabatic vacuum. The Wronskian condition also
imposes $|\alpha_k|^2 -|\beta_k|^2 =1$.
By expanding for small $\Delta$ around $\eta_{\ri I}$, we obtain
equation (\ref{alphabetaIII}) in the text.

Finally, the parameter signalling the WKB violation is \cite{Chialva:2011iz}
 \beq \label{WKBparameteratorderfour} 
  \mathcal{Q}  = -\biggl(-{\omega'' \ov 4 \omega^2}
  + 3{\omega^{'2} \ov 8 \omega^3} - {1 \ov \omega \eta^2}\biggr)^2 
  + {\omega'' \ov 2 \omega} - 3{\omega^{'2} \ov 4 \omega^2}
  -{U'' \ov 2 U} + 3{U^{'2} \ov 4 U^2} \, .
 \eeq

\subsection{Formulas for Whightman functions and their time
  derivatives}\label{Whightmanappendix} 

The bispectrum is computed at a time $\eta$ when all momenta have exit the
horizon, therefore the Whightman functions have the general form
 \beq \label{Whightmangen}
   G_k(\eta\sim 0, \eta') = \lim_{\eta\to 0} {H^2 \ov \dot{\phi}^{2}}
  {f_k(\eta) \ov a(\eta)}{f_k^{\ast}(\eta') \ov a(\eta')}
 \eeq  

We list here the explicit formulas obtained by substituting in the above
equations the field modes functions $f_k$ for $\zeta$ in the various
scenarios we consider. Due to the presence of positive- and
negative-energy solutions, the 
Whightman function will also have positive- and negative-energy parts,
which we indicate with $G_k^{\pm}$. 
\begin{itemize}[leftmargin=0.2cm,itemsep=0.2cm,parsep=0.0cm]
 \item {\em Standard scenario and modified initial state}. From equations
   (\ref{basicstandard}), (\ref{solutionmodinitstat})
  \beq \label{WhightmanNonStandardmv} 
   G_k^\pm(0, \eta') = \pm{H^2 \ov \dot{\phi}^2}{H^2 \ov 2k^{3}} 
      \, (1 \mp i k\eta') e^{\pm ik\eta'},
   \qquad 
   \partial_{\eta'} G_k^\pm(0, \eta') = \pm {H^2 \ov \dot{\phi}^2}{H^2 \ov 2k^{3}} 
      \, k^2 \eta' e^{\pm ik\eta'},
  \eeq
 \item{\em Modified dispersion relations}. From equation
   (\ref{piecewisesolution}), we find that in region {\rm IV} the Whightman
   function is the same as the one in equation
   (\ref{WhightmanNonStandardmv}), while in region {\rm III}
   it is obtained from (\ref{solutionWKB}). Thus,
  \beq \label{WhightmanNonStandardmdr}
   \!\!\!\! G_{k}^\pm(\eta\sim 0, \eta') = 
   {H^2 \ov \dot{\phi}^2} {e^{i{\pi \ov 2}} \ov 2 k^2}
   {H \ov a(\eta')} \chi^{(*)}(k, \eta')
   \, e^{\pm i{\Lambda \ov H}\Omega(k, \eta')},
    \quad
   \chi(k, \eta') = 
   \begin{cases} 
   {1 \ov \sqrt{U(k, \eta')}}
    & \text{in region {\rm III}} \\
    (1-{i \ov k\eta'})
    & \text{in region {\rm IV}}
  \end{cases}
  \eeq  
where $\Omega(k, \eta')$ is equal to $\Omega_{\text{{\tiny F}}}(k, \eta')$ in
region {\rm III} and to $k\eta'$ in region {\rm IV}.
The derivative of the Whightman function is
  \beq \label{DerWhightmanNonStandardmdr}
   \!\!\!\!\!\!\!\!
   \partial_{\eta'} G_{k}^{\pm}(\eta\sim 0, \eta') =
   - {H^2 \ov \dot{\phi}^2} {e^{i{\pi \ov 2}}\, H \ov 2 k}
   {\gamma^{(*)}(k, \eta') \ov a(\eta')}e^{\pm i{\Lambda \ov H}\Omega(k, \eta')},
   \quad
   \gamma(k, \eta') = 
   \begin{cases} 
   {1 \ov \sqrt{U}}\bigl({1 \ov 2k}{U' \ov U}\!+\!iU\!+\!{\mathcal{H} \ov k}\bigr)
    & \text{in region {\rm III}} \\
    i 
    & \text{in region {\rm IV}}
  \end{cases}
  \eeq  
 where the suffix $^{(*)}$ means that the positive-energy
 solution is associated with $\gamma^*, \chi^*$, while the
 negative-energy with $\gamma, \chi$. Finally, $\mathcal{H}$ is the
 conformal Hubble scale. 
\end{itemize}

\subsection{Time integral in bispectrum computations with modified
  dispersion relations}\label{compbispmoddisprel} 

When computing the bispectrum, we have encountered various Laplace
integrals of the form
 \beq \label{Jvgenericintegral}
 \mathcal{J}_j^{(w)} =  
  \int_{y_{\rii II}}^{y}
      dy' y'^{w-1}   e^{i{\Lambda \ov H} x_1 \tilde{v}_{_{\theta_{\text{\stiny $j$}}}}\!(y')} \,  
  h(x_1, x_S, \theta_j, y') \, ,
 \eeq
see equations (\ref{threepointFourierinteg}), (\ref{enhanchigder}),
(\ref{corrmoddisprelgen}), where $w$ and $h(x_1, x_S, \theta_j, y')$ 
change in the different integrals and the variables are defined in
(\ref{yxvariables}).  

In particular, in the case of the minimal coupling interaction
(\ref{HIcubic}), the integral present in the correction of order
$\beta_k$ to the bispectrum, see equation
(\ref{threepointFourierinteg}), has the form of
(\ref{Jvgenericintegral}) with $w=1$ and 
$h(\{x\}, \theta_j, y') \to g(\{x\}, \theta_j, y')$, where \cite{Chialva:2011iz}
 \beq \label{gmoddisprel}
  g(\{x_{h \neq j}\}, x_j, y', m) \equiv
   \prod_{h\neq j}\gamma^*(x_h, y')\tilde \gamma(x_j, y')
  \,  e^{i{H \ov \Lambda}\int^{y'} \bigl(\sum_{h \neq j}S_2(x_h)+(-1)^m S_2(x_j)\bigr)}
 \eeq
with $\gamma$ defined in (\ref{DerWhightmanNonStandardmdr}) and
 \begin{gather} \label{S2def}
  S_2(x_i, y) =  -{\partial_y^2 \omega(x_iy) \ov 4 \omega(x_iy)^2}
    + 3{(\partial_y \omega(x_iy))^{2} \ov 8 \omega(x_iy)^3} - {1 \ov \omega(x_iy) y^2}
   \, ,
  \qquad
  \tilde \gamma(x, y') = 
  \begin{cases}    
   \gamma^*(x, y')  & \text{if} \; m = 0 \\
   \gamma(x, y')  & \text{if} \; m = 1
  \end{cases} \, .
 \end{gather}
In the squeezed limit, for $m=1$ and $j = 2, 3$, looking at
(\ref{Omegaphase}) and (\ref{DerWhightmanNonStandardmdr}), we obtain
 \be  \label{gmoddisprelsqueezlim}
   g(y) \; \equiv \;          
   g(x_1 \text{{\scriptsize$\ll$}} 1, x_{_{j = 2, 3}} \text{{\scriptsize$\simeq$}} x_S \text{{\scriptsize$\simeq$}} 1, y, 1)
   & \;\simeq\;
   -i \, |\gamma(x_{_S}, y)|^2 
  \,  e^{i{H \ov \Lambda} \mathcal{O}(x_1, {H \ov \Lambda})} 
  \;\simeq\; -i|\gamma(1, y)|^2
 \ee
In particular, the $\mathcal{O}(1)$ coefficient
appearing in the result for the contribution to the bispectrum for 
${\Lambda \ov H}x_1 \ll 1$ in the last line of
(\ref{mincubbispcaseLarge}) and of (\ref{BfactmdrII}), is a purely
numerical factor, which reads in details: 
 \beq \label{Oonemincub}
  \int_{-1}^0 dy' |\gamma(1, y')|^2 \sim \mathcal{O}(1).
 \eeq 

In the case, instead, of the higher derivative interaction in equation
(\ref{hamilthigder}), the integral appearing
in equation
(\ref{enhanchigder}) has the form of
(\ref{Jvgenericintegral}) with
$w=3$ and $h(\{x\}, \theta_j, y') \to q(\{x\}, \theta_j, y')$, where
\cite{Chialva:2011iz}
 \begin{multline} \label{qfactenhanchigder}
  \!\!\!\!q(\{x_h\}, x_j, y)\!\! =\!\!\biggl(\!6\!\prod_{h \neq j} \gamma^*(x_{_h})\gamma(x_{_j})
  + {2\vec x_{_{j+1}} \!\cdot \!\vec x_{_{j+2}} \ov x_{j+1}\,x_{j+2}} \, \chi^*(x_{_{j+1}})\chi^*(x_{_{j+2}})\gamma(x_{_j})
  +{2\vec x_{_{j+1}} \!\cdot \!\vec x_{_{j}} \ov x_{j+1}\,x_{j}} \, \chi^*(x_{_{j+1}})\chi(x_{_{j}})\gamma^*(x_{_{j+2}})
  \\
  \!
  +\!{2\vec x_{_{j+2}} \!\cdot \!\vec x_{_{j}} \ov x_{j+2}\,x_{j}} \, \chi^*(x_{_{j+2}})\chi(x_{_{j}})\gamma^*(x_{_{j+1}})
  \biggr)
  \,  e^{i{H \ov \Lambda}\int^{y} \bigl(\sum\limits_{h \neq j}S_2(x_h)-S_2(x_j)\bigr)} \, ,
 \end{multline}
and $S_2$ has been defined in equation (\ref{S2def}) while $\chi$ in
(\ref{WhightmanNonStandardmdr}). We write 
explicitly only the dependence on $x$ for $\gamma, \chi$ to avoid cluttering
the formula. The labels of the $x$'s are defined modulo 3, here.

In the squeezed limit,
for $j = 2, 3$, looking at (\ref{Omegaphase}) and
(\ref{DerWhightmanNonStandardmdr}), the integral of the leading
contribution has
 \be \label{qfactenhanchigdersqueezed}
   q(\theta_j, x_1, y) 
   \; \equiv q(x_1 \text{{\scriptsize$\ll$}} 1, x_{_{j = 2, 3}} \text{{\scriptsize$\simeq$}} x_S  \text{{\scriptsize$\simeq$}} 1, y) 
   \; & \; \simeq \;
   -6i |\gamma^*(1, y)|^2 
   \text{+}2 i |\chi(1, y)|^2 - 4 i \chi^*(x_1, y)\cos(\theta_{j}),
 \ee
so that, again, the $\mathcal{O}(1)$ coefficient
appearing in the result for the contribution to the bispectrum for 
${\Lambda \ov H}x_1 \ll 1$, see last line of
(\ref{consistrelmoddisprelII}), is a purely numerical factor
 \beq \label{Oonemodquart}
  \int_{-1}^0 dy' \biggl(3|\gamma(1, y')|^2-|\chi(1, y')|^2\biggr) \sim \mathcal{O}(1).
 \eeq
Similarly, we find that the 
 $\mathcal{O}(1)$ coefficients
appearing in the result for the contribution (\ref{corrmoddisprelgen})
to the bispectrum for  
${\Lambda \ov H}x_1 \ll 1$, see last line of
(\ref{fNLcorrgenbispmdr}), are given by
 \beq \label{Oonegeneralmdr}
  \int_{-1}^0 dy'  (y')^{u}\gamma^*(1, y')^{n^{^{\text{{\sstiny$(\!h\!)$}}}}_{_{t}}}\gamma^*(1, y')^{n^{^{\text{{\sstiny$(\!h\!+\!1\!)$}}}}_{_{t}}}\Big|_{h\neq j} 
  \sim \mathcal{O}(1).
 \eeq
where $n^{^{\text{{\sstiny$(\!h\!)$}}}}_{_{t}}$ has been defined below
(\ref{corrmoddisprelgen}) and $u$ is equal to $v$/$v_t$,
defined above (\ref{corrmoddisprelgen}), 
in the first/second coefficient.

\begin{table}[t!]
\vspace{0.2cm}
\centering
\renewcommand{\arraystretch}{2}
\begin{tabular*}{1.022\textwidth}{@{\extracolsep{\fill}}||c|c|c|c||}
\hline   
  &  $|y| \ll 1$  & $y= y_{\rii II}$  &  $y \simeq y_*$ \\
\hline \hline 
$\!F(x_{_S}, y)\!$ & $1+\,(-y)^{\kappa}F^{(k)} $ &  
   $F(-1)$ & 
   $F_* + F^{(\nu_*)}_* \,|y-y_*|^{\nu_*}$    \\ \hline
   $\!\tilde{v}_{_{\theta_{\text{\stiny $j$}}}}\!(y)\!$ & 
   $\!\!\!\!\!\!y\,v_{_{\theta_{\text{\stiny $j$}}}}- 
   (-y)^{\kappa+1}F^{(k)}\cos(\theta_j)
   \!\!$ & 
   $-1-F(-1)\cos(\theta_j)$\!\! &
   \!\!\!\!$\begin{gathered}\!\!\!\!
    \!\!\!\tilde{v}_{_{\theta_{\text{\stiny $j$}}}}(y_*)+
    \tilde{v}_{_{\theta_{\text{\stiny $j$}}}}^{(\nu_*)}\!(y_*) \,|y-y_*|^{\nu_*}  \\
    \\
    \text{or} \\
    \\
    \!\!\!\!\!\!\!\!\!\!\!\!\!\!\!\!\!\!\!\!\!\!\!\!\!\!\!\!\!\! 
   \tilde{v}_{_{\theta_{\text{\stiny $j$}}}}(y_*)+
    \tilde{v}_{_{\theta_{\text{\stiny $j$}}}}^{(\nu_*)}(y_*) \\
    \qquad \times|y\text{-}y_*|^{\nu_*}\text{sign}(y\text{-}y_*) 
   \end{gathered}$
    \\ \hline
  $g(y)$  & $-i(1+|y|^{\kappa}F^{(k)})$ & 
   $-iF(-1)$ &
   $(y-y_*)^{\mu_*-1}g^{^{(\mu_*\!-\!1)}}\!(y_*)$  \\ \hline
   $q(\theta_j, x_1, y)$  & 
   \!\!\!\!\!\!\!\!\!\!\!\!
   $\!\!\!\!\!\!\!\!\!\!\!\!\!\!
   \begin{gathered}
   -4i (v_{_{\theta_{\text{\stiny $j$}}}}\!+\!2|y|^{\kappa}F^{(k)})
   \\
   \qquad\qquad\text{-}4i(\chi^{^*}\!(x_{_1},\!\!y)\text{-}1)\cos(\theta_j)
   \end{gathered}
   $\!\! &    
   \!\!$\!\!\!\!\!\!
   \begin{gathered}
   \!\!-6i|\gamma^*\text{{\small$(1, \text{-}1)$}}|^2
   +2i U\text{{\small$(1, \text{-}1)$}}^{\text{-}1}
   \\
   \qquad\qquad
      \text{-}4i \cos(\theta_{j})\chi^*\text{{\small$(x_{_1}, \text{-}1)$}}
   \end{gathered}
   \!\!$ &
   $ (y-y_*)^{\tau_*}q^{(\tau_*)}(\theta_j, y_*)$  \\ \hline
\end{tabular*}
\captionsetup{format=hang,font=small}
\caption{Behaviour of functions entering the integrands in equations
(\ref{threepointFourierinteg}), (\ref{enhanchigder}),
(\ref{corrmoddisprelgen}) close to
 the boundary
points (limits of integration) $y_{\rii II} \sim \text{-}1$ and $y
\sim\text{-}{H \ov \Lambda} \ll \text{-}1$, and close to the possible
stationary point(s) 
 $y_*$ in the squeezed limit $x_1 \ll 1$, $x_S \simeq 1$.}
\label{behaviourtable}
\end{table}

For ${\Lambda \ov H} x_1 \gg 1$, the integrals like
(\ref{Jvgenericintegral}) are dominated by  
the contribution of boundary and stationary points, where the
oscillations due to the large phase of the integrand are
reduced.  It is therefore
important to study the behavior at leading order of the integrands in
(\ref{threepointFourierinteg}), (\ref{enhanchigder}), 
(\ref{corrmoddisprelgen}) close to the boundary
points (limits of integration) $y_{\rii II} \sim -1$ and $y
\sim -{H \ov \Lambda} \ll -1$ and to the possible stationary point(s)
 $\{y_*\}$. In table \ref{behaviourtable}, we
report the leading and next-to-leading behaviour.

In the table, we include also the function $F(x, y')$, defined
in (\ref{powerexp}), and now written in terms of
the variables (\ref{yxvariables}).
From its definition (\ref{powerexp}), it appears that its expansion in powers of  
${p \ov \Lambda}=-y$ for $|y|\ll 1$ is in fact the expansion for the effective
frequency from the effective action. In particular, $\kappa$+1 is thus the power of
the first correction to the standard physical dispersion relation (model
dependent):
 \beq
  \omega_{_\text{phys}}(p) \sim p(1 + \,
     \textstyle{\bigl({p \ov \Lambda}\bigr)}^{\kappa} F^{(\kappa)}+\cdots) \, .
 \eeq 
In expanding $g(y)$, $q(\theta_j, x_1, y)$ -- see
(\ref{gmoddisprelsqueezlim}), (\ref{qfactenhanchigdersqueezed}) --   
we have taken into account the WKB conditions. The powers $\nu_*,
\mu_*, \tau_*$
are model dependent. 
We wish to be generic, so $\kappa, \nu_*, \mu_*, \tau_*$ can
be fractional or integer\footnote{Given the WKB conditions, it is
  $\mu_* \geq 1$, while $\tau_* \geq 0$.}. Then, since 
$F = {\omega_{\text{phys}} \ov p}$ is
a real function, its expansion must be clearly in powers of $|y|,\,
|y\!-\!y_*|$, with real coefficients, for 
generic $\nu_*, \kappa$. Similarly, it must be so for 
$\tilde{v}_{_{\theta_{\text{\stiny $j$}}}}(x, y)\Lambda =-(p+\omega_{\text{phys}}(p))$, 
which is a real function as well (in table \ref{behaviourtable} we
have been very generic 
concerning its expansion around $y_*$).



\end{document}